\documentclass[a4paper,11pt]{article}
 \pdfoutput=1
 \usepackage{jheppub} 
 \usepackage{graphicx}
 \usepackage{amssymb}
\usepackage{dcolumn}
\usepackage{bm}
\usepackage{color}
\usepackage{multirow}
 
\newcommand{\met}{{/\!\!\! E_T}} 
\newcommand{\mpt}{{\;/\!\!\!\! \vec{P}_T}} 

 \newcommand{\lsim}{{\;\raise0.3ex\hbox{$<$\kern-0.75em\raise-1.1ex\hbox{$\sim$}}\;}}
\newcommand{\gsim}{{\;\raise0.3ex\hbox{$>$\kern-0.75em\raise-1.1ex\hbox{$\sim$}}\;}}
\newcommand{\beq}{\begin{equation}}
\newcommand{\eeq}{\end{equation}}
\newcommand{\bea}{\begin{eqnarray}}
\newcommand{\eea}{\end{eqnarray}}

\def\baa{\begin{array}}
\def\eaa{\end{array}}

\mathchardef\minus="002D

\def\met{E_T\hspace{-0.45cm}/\hspace{0.25cm}}

\title{\boldmath Testing invisible momentum ansatze in missing energy events at the LHC}
 
\author[a]{Doojin Kim,}
\author[b]{Konstantin T.~Matchev,} 
\author[c]{Filip Moortgat,}
\author[c]{Luc Pape}

\affiliation[a]{Theory Division, CERN, CH-1211 Geneva 23, Switzerland}
\affiliation[b]{Physics Department, University of Florida, Gainesville, FL 32611, USA}
\affiliation[c]{CERN, Geneva CH-1211, Switzerland}

\abstract{We consider SUSY-like events with two decay chains, each terminating in an invisible particle, 
whose true energy and momentum are not measured in the detector. Nevertheless, a useful educated guess
about the invisible momenta can still be obtained by optimizing a suitable invariant mass function.
We review and contrast several proposals in the literature for such ansatze: 
four versions of the $M_{T2}$-assisted on-shell reconstruction (MAOS), as well as
several variants of the on-shell constrained $M_2$ variables. 
We compare the performance of these methods with regards to the mass determination of a new particle resonance along the decay chain
from the {\em peak} of the reconstructed invariant mass distribution. For concreteness, we consider the event topology of
dilepton $t\bar{t}$ events and study each of the three possible subsystems, in both a $t\bar{t}$ and a SUSY example. 
We find that the $M_2$ variables generally provide sharper peaks and therefore better ansatze for the invisible momenta.
We show that the performance can be further improved by preselecting events near the kinematic endpoint of the corresponding 
variable from which the momentum ansatz originates.
}
 
\preprint{CERN-TH-2017-061} 
\date{March 17, 2017}

\begin{document} 
\maketitle
\flushbottom

\section{Introduction}
\label{sec:introduction}

The bread and butter method for discovering a new particle\footnote{As well as measuring its mass and lifetime.} 
in high energy physics is the ``bump hunt": one identifies and measures the momenta and energies of all
relevant decay products, and forms their total invariant mass. Signal events, which are due to the production of a new resonance, 
appear as a localized ``bump" feature over the relatively smooth background continuum. This technique
has led to many discoveries in the past, including the most recent discovery of the Standard Model Higgs boson, which   
was first observed as an invariant mass peak in the four-lepton and di-photon channels \cite{Chatrchyan:2012xdj,Aad:2012tfa}.

However, this tried and true method faces a major challenge when one (or more) of the decay products are
neutral, weakly interacting particles, which are invisible in the detector, and as a result their energies and momenta remain unknown.
Many well-motivated models of new physics Beyond the Standard Model (BSM) contain such particles, as they are the prototypical 
dark matter candidates. Consequently, one has to develop alternative methods for discovery (and mass measurement)
which are applicable to the case of such semi-invisible\footnote{The case of a fully invisible decay, 
i.e., when the new resonance decays to invisible particles only, is rather trivial and will not be considered in this paper.} 
resonance decays.

The situation is further complicated by the fact that most BSM models with dark matter candidates introduce
a conserved discrete symmetry, often a $Z_2$ parity, in order to protect the lifetime of the dark matter particle.  
The new particles which are charged under this symmetry are necessarily pair produced; 
therefore, each event contains not one, but at least two invisible (dark matter) particles whose 4-momenta\footnote{Throughout this paper
we shall employ the convention where the letter $p$ ($q$) is used to denote the measured (unmeasured) momentum of a particle which is
visible (invisible) in the detector. In addition, the true (hypothesized) mass of the $i$-th invisible particle will be denoted with $m_i$ ($\tilde m_i$). } 
$q_1^\mu$ and $q_2^\mu$
are not individually measured. At hadron colliders, it is only the sum $\vec{q}_{1T}+\vec{q}_{2T}$  of their transverse momenta 
which can be measured in the form of the missing transverse momentum $\mpt$ of the event:
\beq
\vec{q}_{1T}+\vec{q}_{2T}=\mpt. 
\label{eq:mpt}
\eeq
However, the partitioning of the measured $\mpt$ into $\vec{q}_{1T}$ and $\vec{q}_{2T}$ is a priori unknown, and furthermore,
the longitudinal components $q_{1z}$ and $q_{2z}$ remain arbitrary at this point as well. 

Over the last 15-20 years, a large number of methods have been proposed to deal with measurements 
in such ``SUSY-like" events, i.e., events with two decay chains, each terminating in an invisible particle 
(see Ref.~\cite{Barr:2010zj} for a recent review). One possibility is to try to calculate exactly the unknown 
individual 4-momenta $q_{i}^\mu$ of the invisible particles, which in turn would allow one to 
reconstruct an invariant mass peak again. Unfortunately, this idea can only be applied to very specific event topologies, 
where the decay chains are sufficiently long, yielding enough mass-shell constraints in addition to 
(\ref{eq:mpt})~\cite{Kawagoe:2004rz,Nojiri:2007pq,Cheng:2008mg,Cheng:2009fw}.
This is why the majority of the proposed methods have abandoned the idea of directly measuring a mass peak,
and instead focused on measuring a kinematic {\em endpoint} for a suitably defined variable.

Now, what constitutes a ``good" kinematic variable for a kinematic endpoint measurement? The answer to this question 
in principle depends on several factors, including the assumed event topology, 
the nature of the visible SM particles in the final state, the precision with which their momenta $p_j$ are measured, etc. 
Roughly speaking, we can divide the set of kinematic variables into two categories:
\begin{itemize}
\item {\em Variables built only from directly measured quantities, i.e., the momenta $p_j$ of the visible final state particles
and the missing transverse momentum $\mpt$.} 
The primary example of such a variable is the invariant mass of a collection of visible particles. This idea 
forms the basis of the classic method for mass determination in supersymmetry from kinematic endpoints
\cite{Hinchliffe:1996iu,Bachacou:1999zb,Allanach:2000kt,Gjelsten:2004ki,Gjelsten:2005aw,Matchev:2009iw,Kim:2015bnd}.
Other variables belonging to this class include the scalar sum $H_T$ of the transverse momenta 
of visible objects (jets or leptons), the effective mass $M_{eff}$ \cite{Hinchliffe:1996iu,Tovey:2000wk},
the contransverse mass variable $M_{CT}$~\cite{Tovey:2008ui,Polesello:2009rn} 
and its variants $M_{CT_\perp}$  and $M_{CT_\parallel}$~\cite{Matchev:2009ad},
the ratio of visible transverse energies~\cite{Nojiri:2000wq,Cheng:2011ya},
and the energy itself~\cite{Agashe:2012bn,Agashe:2012fs,Agashe:2013eba,Chen:2014oha,Agashe:2015wwa,Agashe:2015ike}.
The advantage of these variables is their simplicity, since one does not have to even face the question about the 
individual momenta $q_i$ or masses $\tilde m_i$ of the invisible particles in the event. 
In principle, these variables are very general and can be usefully applied in certain situations;
however, they also fail to take advantage of the specific characteristics of the event, 
and become suboptimal for more complex event topologies.

\item {\em Variables defined in terms of both the measured momenta $p_j$ and the invisible momenta $q_i$.}
Of course, since the individual invisible momenta $q_i$ are unknown, the 
definition of any such variable 
\beq
v\equiv v(p_j,q_i)
\label{vdef}
\eeq 
must be supplemented with a procedure for fixing the values 
of the invisible momenta $q_i$ through a suitable ansatz. More concretely, the ansatz should allow us to compute the invisible 4-momenta
$q_i^\mu$ in terms of the measured visible 4-momenta $p_j^\nu$ and a set of hypothesized masses $\tilde m_i$
for the invisible particles:
\beq
q_i^\mu=q_i^\mu (p_j^\nu,\tilde m_i),
\label{ansatz}
\eeq
so that at the end of the day, the kinematic variable (\ref{vdef}) 
can be equivalently expressed in terms of visible momenta $p_j$ and invisible masses $\tilde m_i$ only:
\beq
v=v(p_j,q_i(p_j,\tilde m_i)).
\label{variable}
\eeq
If one is solely interested in the kinematic variable $v$ itself and its properties (differential distribution, kinematic endpoints, etc.),
the intermediate step (\ref{ansatz}) of computing the individual invisible momenta $q_i$ is unimportant
and can be regarded simply as a convenient calculational tool. In fact, many of the computer codes on the market 
which are used to compute kinematic variables of the type (\ref{vdef}), by default do not even report the values for the
invisible momenta found from the ansatz (\ref{ansatz}). There are also some special cases, e.g., the
minimum partonic center-of-mass energy $\sqrt{\hat{s}}_{min}$~\cite{Konar:2008ei,Robens:2011zm},
the razor variables~\cite{Rogan:2010kb,Buckley:2013kua}, or the transverse mass $M_T$~\cite{Smith:1983aa,Barger:1983wf},
where one can solve for the ansatz (\ref{ansatz}) analytically, eliminate the invisible momenta, and derive an exact analytical expression 
for the variable $v$ in the form of (\ref{variable}), which can then serve as an alternative definition, without reference to any
invisible momenta at all.

Perhaps the two best known examples of variables of the type (\ref{vdef}) 
are the transverse mass~\cite{Smith:1983aa,Barger:1983wf} and the Cambridge $M_{T2}$ variable~\cite{Lester:1999tx,Barr:2003rg}.
Recently this set of variables was expanded significantly and now includes 
$M_{T2\perp}$ and $M_{T2\parallel}$ \cite{Konar:2009wn},
the asymmetric $M_{T2}$ \cite{Barr:2009jv,Konar:2009qr},
$M_{2C}$~\cite{Ross:2007rm,Barr:2008ba}, 
$M_{CT2}$ \cite{Cho:2009ve,Cho:2010vz},
$M_{T2}^{approx}$ \cite{Lally:2012uj}, and 
the constrained $M_{2}$ variables \cite{Mahbubani:2012kx,Cho:2014naa,Cho:2014yma,Kim:2014ana,Cho:2015laa}.
As the index $``2"$ suggests, all these variables were designed for the case of SUSY-like events 
with {\em two} decay chains, and they also carry an implicit dependence\footnote{At first, the dependence on the unknown 
masses $\tilde m_i$ was considered undesirable, which perhaps prevented the more widespread use of variables 
of the type (\ref{vdef}). Later on, it was realized that the $\tilde m_i$ dependence itself contains a large amount of useful information, 
e.g., a ``kink" develops at the {\em true} value $m_i$ of the invisible particle 
mass~\cite{Cho:2007qv,Gripaios:2007is,Barr:2007hy,Cho:2007dh,Barr:2009jv}
(related techniques for measuring the invisible particle masses by utilizing the $\tilde m_i$ dependence are described in 
\cite{Matchev:2009ad,Matchev:2009fh,Alwall:2009sv,Konar:2009wn}).} 
on the test masses $\tilde m_i$ of the invisible particles, as indicated in (\ref{variable}). 
Despite the large number of such variables on the market, they all share the same common idea \cite{Barr:2011xt}:
choose a suitable target function and minimize it over all possible values of the individual invisible momenta $q_i$
which are consistent with the $\mpt$ condition (\ref{eq:mpt}). The variations arise because one faces a menu of choices:
\begin{itemize}
\item {\em Partitioning of the event.} One groups the final state particles according to the assumed production process ---
single production, pair production, etc. Ideally, one should also have a separate category for jets which are suspected 
to come from initial state radiation \cite{Lester:2007fq,Papaefstathiou:2009hp,Alwall:2009zu,Konar:2010ma}.
\item {\em Choice of target function.} The target function can be a full (3+1)-dimensional invariant mass, as in the case of 
$\sqrt{\hat{s}}_{min}$~\cite{Konar:2008ei}, $M_{2C}$~\cite{Ross:2007rm,Barr:2008ba} and 
$M_2$~\cite{Mahbubani:2012kx,Cho:2014naa}; a (2+1)-dimensional transverse mass, 
e.g., $M_T$~\cite{Smith:1983aa,Barger:1983wf} or $M_{T2}$~\cite{Lester:1999tx},
and even a (1+1)-dimensional mass as in the case of $M_{T2\perp}$ and $M_{T2\parallel}$ \cite{Konar:2009wn}.
Note that the projection to lower dimensions in general does not commute with the partitioning, 
so by performing those two operations in different order, one obtains in principle different variables \cite{Barr:2011xt}.
\item {\em Imposing additional on-shell constraints.} The minimization of (3+1)-dimensional mass target functions over the invisible momenta 
can be performed by taking into account the $\mpt$ constraint (\ref{eq:mpt}) only, or by adding additional kinematic constraints 
which are motivated by the assumed event topology \cite{Mahbubani:2012kx,Cho:2014naa,Swain:2014dha,Konar:2015hea}, 
a prior kinematic endpoint measurement \cite{Ross:2007rm},
or by the presence of a known SM particle in the decay chain
(for example, a $W$ boson \cite{Barr:2011ux,Barr:2011si,Bai:2012gs,Swain:2014dha} or a $\tau$ lepton \cite{Barr:2011he,Swain:2014dha,Konar:2016wbh}).
The additional on-shell constraints further restrict the allowed domain of values for the components of the individual invisible momenta 
$q_i$ and in general lead to a different outcome from the minimization procedure.
\end{itemize}
Note that whenever the target function is a {\em transverse} mass in (2+1) dimensions, the minimization fixes only the transverse
components $\vec{q}_{iT}$ of the invisible momenta, and for the longitudinal components one must rely on additional
measurements or assumptions. For example, in the $M_{T2}$-assisted on-shell (MAOS) reconstruction 
method, one assumes knowledge of the mass of the mother particle and enforces its on-shell condition, 
which allows to solve for the longitudinal momenta~\cite{Cho:2008tj}. The method
was then tested in examples where the mothers are known SM particles, e.g. top quarks, $W$-bosons 
or $\tau$-leptons~\cite{Choi:2009hn,Cho:2009wh,Park:2011uz,Choi:2010dw,Choi:2011ys,Guadagnoli:2013xia}.
Since the on-shell constraints are nonlinear functions, the MAOS approach typically yields 
multiple solutions for the longitudinal momentum components, so one must also specify a prescription for handling
this multiplicity. In contrast, target functions defined in (3+1) dimensions automatically yield ansatze for the full
energy-momentum 4-vectors $q^\mu_i$, without any need for additional assumptions \cite{Barr:2011xt}.
Another benefit of the (3+1) formulation is that the obtained solutions for the longitudinal components $q_{iz}$
are typically unique \cite{Cho:2014naa,Cho:2015laa}.
\end{itemize}

In this paper, we would like to reemphasize the existence of various ansatze (\ref{ansatz}) for the individual 
invisible momenta in missing energy events, and demonstrate their utility in the context of a mass measurement through a ``bump hunt".
Following previous studies, we shall consider the general event topology of dilepton $t\bar{t}$ events, which already have very rich kinematics, as
one can define and study three different subsystems \cite{Burns:2008va}: one associated with the two $b$-jets, 
another associated with the two leptons, and a third one referring to the event as a whole (see Fig.~\ref{fig:DecaySubsystem} below).
After briefly introducing our notation and conventions in Section~\ref{sec:notation}, in the next Section~\ref{sec:ansatze}
we shall carefully define and contrast the different ansatze for invisible momenta which follow from some of the 
most commonly discussed in the literature variables of type (\ref{vdef}): $M_{T2}$, $M_2$, and $\sqrt{\hat s}_{min}$. 
The transverse variable $M_{T2}$ is already at the heart of (as well as in the name of) 
the MAOS method \cite{Cho:2008tj}. In addition to the traditional MAOS method described earlier, in Section~\ref{sec:maos}
we shall also consider two modified MAOS prescriptions \cite{Choi:2009hn,Cho:2009wh,Park:2011uz,Choi:2010dw}, 
which avoid using information about the mother particle mass, and instead rely on the calculated value of $M_{T2}$ in the event.
(There will also be a fourth variant of the MAOS method, which will assume a known mass for a particle other than the parent.)
Then in Section~\ref{sec:M2} we shall consider the case of (3+1)-dimensional target functions, 
since it automatically provides an ansatz for the longitudinal invisible momenta \cite{Konar:2008ei,Cho:2014naa}. 

Next we would like to compare the performance of the difference ansatze (\ref{ansatz}).
One possibility is to compare the momenta predicted by (\ref{ansatz}) to the true invisible momenta in the event.
However, the ultimate goal of any invisible momentum reconstruction is to perform some kind of physics measurement.
In particular, once we have a guess for the invisible momenta, we can revisit the original idea for a bump hunt, 
and compare the precision of mass measurements performed with different ansatze. 
This will be the subject of Section~\ref{sec:m2aos}, in which we shall study the position and the sharpness of the
corresponding reconstructed invariant mass peak. Our main result will be that the invisible momenta 
provided by $M_{2}$-type variables generally lead to the most accurate mass measurements. 

In Section~\ref{sec:BSM} we shall generalize our discussion to the case of BSM collider signals exhibiting the $t\bar{t}$ event topology.
In particular, we shall explore the general mass parameter space of the three particles in each decay chain, and
analyze the performance of the invisible momentum reconstruction from $M_2$-type variables as a function of parameter space.
In doing so, we shall identify the parameter space regions where the accuracy is degraded, and 
then propose a solution for recovering sensitivity by applying a preselection cut.
The same idea has already been used successfully in the case of MAOS \cite{Cho:2008tj}
and here we demonstrate its validity in a more general context.
Sec.~\ref{sec:conclusions} is reserved for our conclusions.

\section{Notations and setup}
\label{sec:notation}

In this paper we shall largely follow the notation and terminology of Ref.~\cite{Cho:2014naa}, which we briefly review here for 
the reader's convenience.

\subsection{The physics process}

\begin{figure}[t]
\centering
\includegraphics[scale=1]{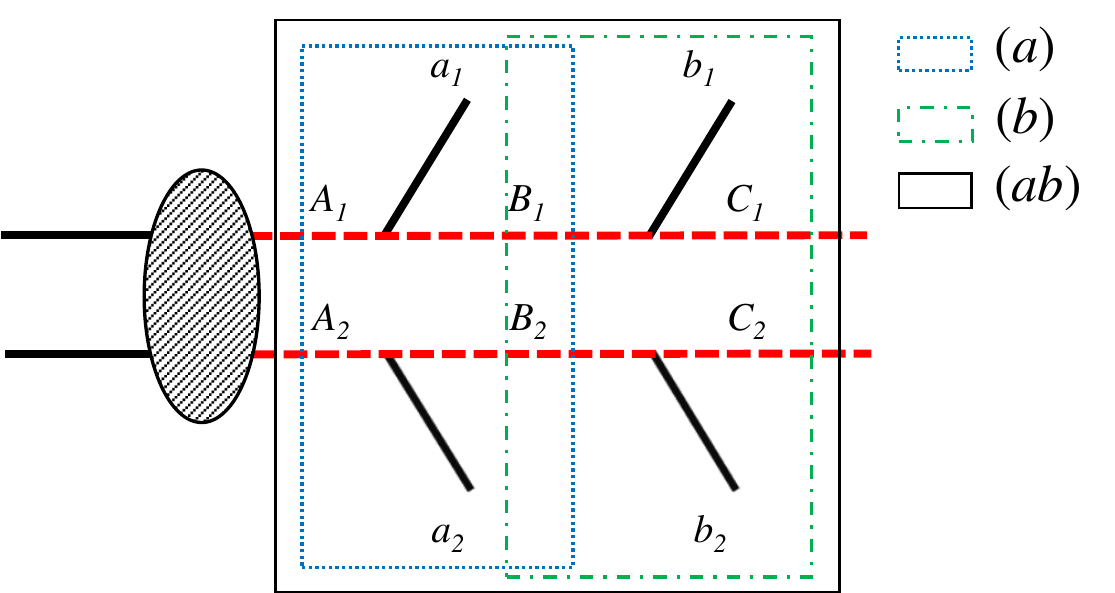}
\caption{\label{fig:DecaySubsystem} The decay topology under consideration in this paper
with the corresponding three subsystems explicitly delineated.
Each parent particle, $A_i$, ($i=1,2$) decays to two visible particles, $a_i$ and $b_i$,
and an invisible daughter particle, $C_i$, through an intermediate on-shell resonance, $B_i$.
The blue dotted, green dot-dashed, and black solid
lines indicate the subsystems $(a)$, $(b)$, and $(ab)$, respectively.}
\end{figure}

We focus on the generic event topology which is schematically depicted in Fig.~\ref{fig:DecaySubsystem}. 
We assume the pair production of two heavy particles, $A_1$ and $A_2$, whose subsequent decay chains
consist of two two-body decays:
\beq
pp\to A_1A_2, \quad A_i\to a_i B_i, \quad B_i\to b_iC_i, \quad (i=1,2).
\label{eq:process}
\eeq
Here the particles $a_i$ and $b_i$ are SM particles which are visible in the detector, so that their 4-momenta 
$p_{a_1}^\mu$, $p_{b_1}^\mu$, $p_{a_2}^\mu$, and $p_{b_2}^\mu$ are measured known quantities. 
In contrast, the particles $C_i$ are invisible in the detector --- they can be dark matter candidates or SM neutrinos ---
and their 4-momenta, $q_i^\mu$, are {\em a priori} unknown, being constrained only by the $\mpt$ measurement (\ref{eq:mpt})
and our conjectured values $\tilde m_{C_1}$ and $\tilde m_{C_2}$ for their masses:
\beq
q_i^2 = \tilde m_{C_i}^2, \quad (i=1,2).
\label{Conshell}
\eeq 
As usual, all visible particles are assumed massless (this is done merely for simplicity).
The masses of the intermediate resonances in Fig.~\ref{fig:DecaySubsystem}
are denoted by $m_{A_i}$ and $m_{B_i}$, with $m_{A_i}>m_{B_i}$. 
The process (\ref{eq:process}) depicted in Fig.~\ref{fig:DecaySubsystem} covers a large class of interesting and
motivated scenarios, including dilepton $t\bar{t}$ events in the SM, 
stop pair production in supersymmetry with $\tilde t\to b\tilde\chi^+$, followed by $\chi^+\to \ell^+\tilde\nu_\ell$,
gluino pair production in supersymmetry with $\tilde g \to \bar{q}\tilde q$, followed by $\tilde q\to q\tilde\chi^0$, 
and many more.

\subsection{Event subsystems and the particle family tree nomenclature}
 
As first discussed in the context of the $M_{T2}$ variable~\cite{Burns:2008va}, within the original event 
one can consider several useful subsystems which are delineated by the colored rectangles in Fig.~\ref{fig:DecaySubsystem}.
Each subsystem is defined by a choice of {\em parent} particles and a choice of {\em daughter} particles among the set of
three particles $\left\{ A_i, B_i, C_i\right\}$. Since the parents must be heavier than the daughters, there are only three possibilities,
and in each case, the remaining third type of particles will be referred to as {\em relatives}. 
Following the notation of \cite{Cho:2014naa}, we shall label each subsystem by the set of visible particles on each decay side
which are used to construct the kinematic variable:
\begin{itemize}
\item {\em The $(ab)$ subsystem.} This system refers to the event as a whole and is indicated by the solid black 
box in Fig.~\ref{fig:DecaySubsystem}. Here the $A_i$'s are the two parent particles and the $C_i$'s are the daughter particles,
leaving the intermediate resonances $B_i$ as the relative particles. The visible particles on each side, $a_i$ and $b_i$, are combined into
a composite visible particle with 4-momentum $p^\mu_{a_i}+p^\mu_{b_i}$.
\item {\em The $(b)$ subsystem.} This subsystem is outlined by the green dot-dashed box in Fig.~\ref{fig:DecaySubsystem}.
Now the parents are the $B_i$ particles, the daughters are the $C_i$ particles, and the relatives are the $A_i$ particles. 
The kinematic variables for this subsystem will be defined in terms of the 4-momenta $p^\mu_{b_i}$ of the visible particles $b_i$.
\item {\em The $(a)$ subsystem.} This subsystem is depicted by the blue dotted box in Fig.~\ref{fig:DecaySubsystem}.
The $A_i$ particles are again treated as parents, but the daughters are now the $B_i$ particles, while the relatives are 
the $C_i$ particles. The kinematic variables will use the 4-momenta $p^\mu_{a_i}$ of the visible particles $a_i$. 
\end{itemize}

\section{Ansatze for the invisible momenta}
\label{sec:ansatze}

We are now in position to define the different kinematic variables of interest, for each of the three subsystems: 
$(ab)$, $(b)$ and $(a)$. For each variable (\ref{vdef}), we first identify a target function, which is then minimized over all 
possible values of the individual invisible momenta $q^\mu_i$ consistent with the missing transverse momentum condition
(\ref{eq:mpt}). This minimization will yield the required ansatz for the missing momenta (\ref{ansatz}).
In Section~\ref{sec:maos} we begin our discussion with (2+1)-dimensional target functions defined on the transverse plane, 
where the minimization fixes only the transverse components $\vec{q}_{iT}$ of the invisible momenta.  
One then needs to impose an additional requirement in order to obtain a suitable value for the longitudinal components,  
and we shall review the different options discussed in the literature. Then in Section~\ref{sec:M2} we shall proceed to discuss
(3+1)-dimensional invariant mass target functions, where the minimization results in fully specified invisible momenta
$q^\mu_i$. In preparation for the numerical comparisons to follow in the next two sections,
we shall again review the different possibilities arising from applying various on-shell constraints 
on the parent and/or relative particles.

\subsection{Transverse mass target functions and MAOS reconstruction}
\label{sec:maos}

At hadron colliders, where the longitudinal momentum of the initial state is a priori unknown, 
transverse variables are attractive since they are invariant under longitudinal boosts.
When targeting an event topology with two separate decay chains like that of Fig.~\ref{fig:DecaySubsystem}, 
one should consider the two parent particles $P_i$ and their corresponding decay products $\{ a_i, b_i, C_i\}$.
In order to obtain a useful generalization of the canonical transverse mass variable for this case, 
one follows the prescription behind the Cambridge $M_{T2}$ variable~\cite{Lester:1999tx} ---
first form the individual transverse masses $M_{TP_i}$ of the two parents, then choose the larger of the two,
$\max\left(M_{TP_1},M_{TP_2}\right)$,
as our target function, and minimize it with respect to the transverse components of the momenta of the daughter particles, 
subject to the $\mpt$ constraint (\ref{eq:mpt}). We obtain three different versions of the $M_{T2}$ variable, depending on the 
subsystem under consideration \cite{Burns:2008va}. For subsystem $(ab)$, the parents are $A_i$ and the daughters are $C_i$, 
thus\footnote{In what follows, to simplify the notation we shall not indicate explicitly the parent mass dependence
on the visible momenta $p_{a_i}$ and $p_{b_i}$, which should be clear from the chosen subsystem. }
\bea
M_{T2} (ab) &\equiv& \min_{\vec{q}_{1T},\vec{q}_{2T}}\left\{\max\left[M_{TA_1}(\vec{q}_{1T},\tilde m_{C_1}),\;M_{TA_2} (\vec{q}_{2T},\tilde m_{C_2})\right] \right\}.  
\label{eq:mt2ab}\\
\vec{q}_{1T}+\vec{q}_{2T} &=& \mpt   \nonumber
\eea 
In subsystem $(b)$, the parents are the $B_i$ particles, and one gets
\bea
M_{T2} (b) &\equiv& \min_{\vec{q}_{1T},\vec{q}_{2T}}\left\{\max\left[M_{TB_1}(\vec{q}_{1T},\tilde m_{C_1}),\;M_{TB_2} (\vec{q}_{2T},\tilde m_{C_2})\right] \right\}.  
\label{eq:mt2b}\\
\vec{q}_{1T}+\vec{q}_{2T} &=& \mpt   \nonumber
\eea 
The case of subsystem $(a)$ is somewhat more complicated since the daughters are the $B_i$ particles 
and the minimization is performed in terms of {\em their} momenta as opposed to the momenta of the $C_i$ particles. 
If we introduce the 4-momenta of the $B_i$ particles,
\beq
Q_i\equiv q_i+p_{b_i},
\label{eq:Qidef}
\eeq
we can define 
\bea
M_{T2} (a) &\equiv& \min_{\vec{Q}_{1T},\vec{Q}_{2T}}\left\{\max\left[M_{TA_1}(\vec{Q}_{1T},\tilde m_{B_1}),\;M_{TA_2} (\vec{Q}_{2T},\tilde m_{B_2})\right] \right\}, 
\label{eq:mt2a}\\
\vec{Q}_{1T}+\vec{Q}_{2T} &=& \mpt + \vec{p}_{b_1T} + \vec{p}_{b_2T}  \nonumber
\eea 
where instead of (\ref{Conshell}) we have
\beq
Q_i^2 = \tilde m_{B_i}^2, \quad (i=1,2).
\label{Bonshell}
\eeq 

The three minimizations in (\ref{eq:mt2ab}), (\ref{eq:mt2b}) and (\ref{eq:mt2a}) in principle provide three independent 
ansatze for the {\em transverse} momenta $\vec{q}_{iT}$ of the $C_i$ particles\footnote{Strictly speaking,
in the case of subsystem (a), we initially obtain an ansatz for the transverse momenta $\vec{Q}_{iT}$ 
of the intermediate particles $B_i$, but they can be easily related to $\vec{q}_{iT}$ with 
the help of eq.~(\ref{eq:Qidef}).}, as shown in Table~\ref{tab:maos}. 
\begin{table}[t]
\centering
\begin{tabular}{||c|c||c|c|c||}
\hline
                       &                                    & \multicolumn{3}{c||}{Ansatz for the invisible momenta}\\
\cline{3-5}
Method           & required inputs           & longitudinal & No of    &transverse   \\ 
                       &                                    & components & solutions & components \\
\hline\hline
MAOS1(ab)    &         $m_A, m_C$     & eq.~(\ref{maos1ab}) & up to 4 &   \multirow{4}{*}{$M_{T2} (ab)$}                            \\ \cline{1-4} 
MAOS2(ab)    &        $m_C$               & eq.~(\ref{maos2ab}) & up to 2 &                              \\ \cline{1-4} 
MAOS3(ab)    &        $m_C$               & eq.~(\ref{maos3ab}) & unique &                                                       \\ \cline{1-4} 
MAOS4(ab)    &        $m_B, m_C$     & eq.~(\ref{maos4ab}) &  up to 4 &                                                       \\ \hline\hline
MAOS1(b)     &      $m_B, m_C$        &  eq.~(\ref{maos1b})  & up to 4 &   \multirow{4}{*}{$M_{T2} (b)$}                       \\ \cline{1-4} 
MAOS2(b)     &     \multirow{2}{*}{$m_{C}$}  & eq.~(\ref{maos2b}) &  \multirow{2}{*}{unique}  &                                    \\ \cline{1-1} \cline{3-3} 
MAOS3(b)     &                      &  eq.~(\ref{maos3b}) &   &                                            \\ \cline{1-4} 
MAOS4(b)     &   $m_A, m_C$          & eq.~(\ref{maos4b}) &  up to 4 &                                         \\ \hline\hline
MAOS1(a)    &    $m_A, m_B$         & eq.~(\ref{maos1a}) &  up to 4 &      \multirow{4}{*}{$M_{T2} (a)$}                                          \\ \cline{1-4} 
MAOS2(a)    &  \multirow{2}{*}{$m_{B}$}   & eq.~(\ref{maos2a}) & \multirow{2}{*}{unique} &                                     \\ \cline{1-1}   \cline{3-3} 
MAOS3(a)    &                        & eq.~(\ref{maos3a}) &         &                                              \\ \cline{1-4} 
MAOS4(a)    &    $m_C, m_B$          & eq.~(\ref{maos4a})&  up to 4 &                                           \\ \hline
\hline 
\end{tabular}
\caption{\label{tab:maos} A summary of the different possible MAOS schemes.
The transverse invisible momenta are fixed by the $M_{T2}$ calculation in one of the three
possible subsystems $(ab)$, $(a)$, and $(b)$, while the longitudinal invisible momenta 
can be computed from any one of the four conditions MAOS1, MAOS2, MAOS3 and MAOS4 
described in the text. The second column lists the required mass inputs for each case.}
\end{table}
As for the longitudinal components $q_{iz}$, one has to impose additional constraints
and compute $q_{iz}$ independently. There are several different options:
\begin{itemize}
\item {\em MAOS1: use the known mass of a parent particle.} This is the idea of the original MAOS method \cite{Cho:2008tj}.
If we imagine that the mass of the parent particle is already known from a prior measurement\footnote{It is important 
to distinguish the two different situations in which we can use such information about the parent mass. 
First, the parents can be SM particles, which decay semi-invisibly, e.g., top quarks, $W$-bosons 
or tau leptons. In this case the parent mass is known exactly. Second, the parents can be BSM particles, 
whose masses are a priori unknown, but some partial information can be obtained from the standard 
kinematic endpoint measurements, which typically establish a relationship between the mass of the daughter 
and the mass of the parent. In this case, the left-hand sides of eqs.~(\ref{maos1ab}-\ref{maos1a}) should
be thought of as functions of the test mass of the daughter particle, $\tilde m_{C_i}$ or $\tilde m_{B_i}$, 
depending on the subsystem. In other words, in practical applications of the MAOS method to BSM analyses,
one first introduces a value for the daughter test mass, after which the parent mass can be
computed from a kinematic endpoint measurement and substituted in (\ref{maos1ab}-\ref{maos1a}).},
we can enforce two mass shell conditions (one for each parent) in order to determine the 
longitudinal momentum of the respective invisible particle. 
Depending on the subsystem under considerations, the MAOS1 constraint reads
\bea
{\rm Subsystem~(ab):} \quad m_{A_i}^2 &=&(p_{a_i}+p_{b_i}+q_{i})^2,  \label{maos1ab} \\
{\rm Subsystem~(b):} \quad m_{B_i}^2 &=& (p_{b_i}+q_{i})^2, \label{maos1b} \\
{\rm Subsystem~(a):} \quad m_{A_i}^2  &=&  (p_{a_i}+Q_{i})^2. \label{maos1a}
\eea
The first two relations will provide an ansatz directly for $q_{iz}$, while the last one can be solved for $Q_{iz}$,
after which $q_{iz}$ will be obtained from (\ref{eq:Qidef}).
In all cases, we have to deal with a quadratic equation for each decay chain, thus we may end up with up to
four valid solutions, as indicated in Table~\ref{tab:maos}.
\item {\em MAOS2: use the value of $M_{T2}$ calculated in the event.} 
The main disadvantage of the original MAOS1 scheme is that one needs precise prior
knowledge of the mass of the parent particle, which may not be available immediately. 
In order to circumvent this difficulty, an alternative proposal, which 
does not require the parent mass as an input, was suggested in Refs.~\cite{Choi:2009hn,Cho:2009wh,Park:2011uz,Choi:2010dw}.
The idea is to use the {\em numerical value} of the event-wise $M_{T2}$ value 
in place of the parent mass. Depending on the subsystem, we have:
\bea
{\rm Subsystem~(ab):} \quad  M^2_{T2} (ab) &=&(p_{a_i}+p_{b_i}+q_{i})^2,  \label{maos2ab} \\
{\rm Subsystem~(b):} \quad M^2_{T2} (b)  &=& (p_{b_i}+q_{i})^2, \label{maos2b} \\
{\rm Subsystem~(a):} \quad M^2_{T2} (a)  &=&  (p_{a_i}+Q_{i})^2. \label{maos2a}
\eea
At first glance, these relations may look weird, since the left-hand side is a transverse quantity, while the 
right-hand side is a genuine (3+1)-dimensional invariant mass. This observation is the key to understanding the 
physical meaning of the ansatz: the invisible momentum is chosen so that its rapidity is the same as
the rapidity of the agglomerated visible decay products, which allows a longitudinal boost to a frame where 
the momenta are purely transverse, and the transverse mass becomes the same as the mass \cite{Barr:2011xt}. 
\item {\em MAOS3: use the individual parent transverse masses obtained in the $M_{T2}$ calculation.}
One remaining disadvantage of the MAOS2 method is that the obtained solution for the longitudinal momenta 
may not be unique. This occurs for the so-called ``unbalanced" events, where the minimum of the 
target function is at a point where the transverse masses of the two parents are not equal \cite{Cho:2014naa}.  
This motivated another choice, where one makes use of the individual parent transverse masses
in each branch \cite{Park:2011uz}, namely
\bea
{\rm Subsystem~(ab):} \quad  M^2_{TA_i} &=&(p_{a_i}+p_{b_i}+q_{i})^2,  \label{maos3ab} \\
{\rm Subsystem~(b):} \quad M^2_{TB_i} &=& (p_{b_i}+q_{i})^2, \label{maos3b} \\
{\rm Subsystem~(a):} \quad M^2_{TA_i} &=&  (p_{a_i}+Q_{i})^2. \label{maos3a}
\eea  
With this prescription, the obtained values for the longitudinal momenta are unique.
As shown in Table~\ref{tab:maos}, the distinction between MAOS2 and MAOS3 only arises 
in the case of subsystem $(ab)$, since subsystems $(a)$ and $(b)$ always lead to balanced events, 
for which MAOS2 and MAOS3 are identical procedures.
\item {\em MAOS4: use the known fixed mass of a relative particle.} This method is similar in spirit to MAOS1, 
only this time we use as an input the mass of a relative particle. In analogy to (\ref{maos1ab}-\ref{maos1a}), we get
\bea
{\rm Subsystem~(ab):} \quad m_{B_i}^2 &=&(p_{b_i}+q_{i})^2,  \label{maos4ab} \\
{\rm Subsystem~(b):} \quad m_{A_i}^2 &=& (p_{a_i}+p_{b_i}+q_{i})^2, \label{maos4b} \\
{\rm Subsystem~(a):} \quad m_{C_i}^2  &=&  (Q_{i}-p_{b_i})^2. \label{maos4a}
\eea
As shown in Table~\ref{tab:maos}, three different versions of MAOS4 are possible in dilepton $t\bar{t}$ events.
For example, MAOS4(ab) requires that the lepton and the neutrino on each side of the 
event reconstruct to the true $W$-boson mass, which makes it suitable for studying the
reconstructed top quark mass. On the other hand, in MAOS4(b) one demands that the two top quarks
have nominal masses, in which case the interesting variable to study would be the reconstructed $W$-boson mass. 
\end{itemize}

In principle, all twelve MAOS methods listed in Table~\ref{tab:maos} are valid procedures for
obtaining the invisible momenta and they will all be illustrated in Section~\ref{sec:reco_maos} below. 
To the best of our knowledge, only some of the options in Table~\ref{tab:maos} have been used in the literature so far.
The original proposal \cite{Cho:2008tj} focused on MAOS1(ab), while MAOS2(ab) and MAOS3(ab) were 
introduced later in \cite{Choi:2009hn,Cho:2009wh,Park:2011uz,Choi:2010dw}.
Ref.~\cite{Choi:2011ys} made use of MAOS1(ab) and MAOS4(ab) to tackle the two-fold combinatorial ambiguity in dilepton
$t\bar{t}$ events \cite{Rajaraman:2010hy,Baringer:2011nh}.
The possibility to use different subsystems for MAOS reconstruction was pointed out in Ref.~\cite{Guadagnoli:2013xia},
which performed a comparison of MAOS1(ab), MAOS1(a) and MAOS1(b) using dilepton $t\bar{t}$ events and 
concluded that the best ansatz for the momenta of the invisible particles is provided by MAOS1(ab), 
followed by MAOS1(b) and finally, MAOS1(a). Quite recently, the MAOS1(b) version was used by the CMS collaboration 
to measure the top mass in dilepton $t\bar{t}$ events \cite{CMS:2016kgk}.

\subsection{(3+1)-dimensional invariant mass target functions}
\label{sec:M2}

Following~\cite{Barr:2011xt,Mahbubani:2012kx,Cho:2014naa}, 
one could also consider target functions in (3+1)-dimensions.
Starting with the actual parent masses, $M_{P_i}$, 
we can schematically define the (3+1)-dimensional analogues of (\ref{eq:mt2ab}), (\ref{eq:mt2b}) and (\ref{eq:mt2a}) as
\bea
M_{2} (\tilde m) &\equiv& \min_{\vec{q}_{1},\vec{q}_{2}}\left\{\max\left[M_{P_1}(\vec{q}_{1},\tilde m),\;M_{P_2} (\vec{q}_{2},\tilde m)\right] \right\},  
\label{eq:m2def}\\
\vec{q}_{1T}+\vec{q}_{2T} &=& \mpt   \nonumber
\eea 
where $\tilde m$ is the daughter test mass for the corresponding subsystem and
the minimization is performed over all 3-components of the vectors $\vec{q}_{1}$ and $\vec{q}_{2}$.\footnote{Recall that in the case of
subsystem (a) we are actually using the momenta $Q_i$ which are related to $q_i$ by eq.~(\ref{eq:Qidef}).}
If (\ref{eq:m2def}) is left as is, we will obtain nothing new --- the result of the minimization will be equal to the corresponding
value of $M_{T2}$ \cite{Ross:2007rm,Barr:2011xt,Cho:2014naa}, and furthermore, we will derive the same
invisible momenta as with the MAOS2 method. This motivates us to modify the naive definition (\ref{eq:m2def})
appropriately, by taking into account the specific features of the event topology of Fig.~\ref{fig:DecaySubsystem}
\cite{Cho:2014naa}. For example, in many BSM realizations of Fig.~\ref{fig:DecaySubsystem}, the two decay chains are
symmetric in the sense that the original parent particles $A_i$ are identical (or at worst a particle-antiparticle pair)
and decay in the same fashion. As a result, the corresponding masses on the two sides of the event are the same:
\bea
m_{A_1}&=&m_{A_2}\equiv m_A, \label{eq:MAeq}\\
m_{B_1}&=&m_{B_2}\equiv m_B, \label{eq:MBeq}\\
m_{C_1}&=&m_{C_2}\equiv m_C, \label{eq:MCeq}
\eea 
and we can incorporate some number of these constraints into the definition of the kinematic variable.
Note that the first equal sign in eqs.~(\ref{eq:MAeq}-\ref{eq:MCeq}) refers to the symmetry of the event topology, while the second
additionally implies knowledge of the actual value\footnote{As in the case of MAOS, for BSM applications the parent 
mass may only be known as a function of the test daughter mass, as the latter is always a necessary input to the analysis.} 
of the mass, $m_A$, $m_B$ or $m_C$. Due to the freedom of choosing different sets among the constraints (\ref{eq:MAeq}-\ref{eq:MCeq}),
several classes of variables are possible.
\begin{itemize}
\item {\em Equality of the two parent masses.} In the absence of any knowledge of the actual masses of the parent particles, the best one can do is 
to apply the constraint of identical parents
\beq
M_{P_1} = M_{P_2}.
\label{eq:parents}
\eeq
Following the notation of \cite{Cho:2014naa}, variables for which this condition is enforced, will carry a {\em first} index $C$ for ``constrained".
\item {\em Equality of the two relative masses}. In analogy to (\ref{eq:parents}), 
we can demand that the two relative particles in each decay chain are the same:
\beq
M_{R_1} = M_{R_2}.
\label{eq:relatives}
\eeq
Following the notation of \cite{Cho:2014naa}, variables for which this condition is enforced, will carry a {\em second} index $C$.
\item {\em Fixed mass for the two relatives.} 
An even stronger constraint arises if we enforce the relative mass to be equal 
to some fixed value $M_R$ (compare to the MAOS4 method introduced above in Section~\ref{sec:maos}):
\beq
M_{R_1} = M_{R_2}\equiv M_R.
\label{eq:relativesfixed}
\eeq
Further expanding upon the notation of \cite{Cho:2014naa}, variables for which this condition is enforced, 
will carry a {\em second} index $R$ indicating the relative particle whose mass is known. 
For example, in the special case of the event topology of Fig.~\ref{fig:DecaySubsystem} 
applied to dilepton $t\bar{t}$ events, the index $R$ can take the values $R=t$ in subsystem (b),
$R=W$ in subsystem (ab), and $R=\nu$ in subsystem (a). 
\end{itemize}

In summary, the $M_2$ class of variables will be labelled by two\footnote{For simplicity, in this paper we shall always assume 
the masses of the two daughter particles in a given subsystem to be the same, otherwise we would need a third index for the daughter particles. 
This assumption is done only for simplicity and can be easily relaxed, see, e.g., \cite{Barr:2009jv,Konar:2009qr}.} subscripts.
The first refers to the parent hypothesis and takes a value $C$ if (\ref{eq:parents}) is applied, and $X$ otherwise.
The second subscript refers to the relative hypothesis and takes a value $C$ if (\ref{eq:relatives}) is applied,
a value $R$ if (\ref{eq:relativesfixed}) is applied, and $X$ otherwise.
Altogether, we have six\footnote{Additional variables can be obtained if we make further assumptions about the event topology.
For example, if we assume an ``antler" topology, where the two parents $A_i$ arise from the decay of a heavy resonance $G$
with a known mass $m_G$, one can further impose the constraint $(\sum_i (p_{a_i}+p_{b_i}+q_i))^2=m_G^2$ 
\cite{Konar:2015hea,Konar:2016wbh}. }
possible variables: $M_{2XX}$, $M_{2XC}$, $M_{2XR}$, $M_{2CX}$, $M_{2CC}$, and $M_{2CR}$.

\begin{table}[t]
\centering
\begin{tabular}{||c|c|c||c|c||}
\hline
   &   Subsystem   & Mass   &
\multicolumn{2}{c||}{Applied constraints for }  \\ \cline{4-5}
    Variable           &  type   &    inputs                &                  parents                        & relatives   \\ \hline  \hline
$M_{2XX}(ab)$   &      (ab)         &    $m_C$              &                      ---                            &   ---         \\
$M_{2XC}(ab)$   &      (ab)         &    $m_C$              &                      ---                            &   $m_{B_1}=m_{B_2}$    \\
$M_{2CX}(ab)$   &      (ab)         &    $m_C$             &   $m_{A_1}=m_{A_2}$                 &   ---    \\
$M_{2CC}(ab)$   &      (ab)         &    $m_C$             &   $m_{A_1}=m_{A_2}$                 &   $m_{B_1}=m_{B_2}$    \\  \hline
$M_{2XR}(ab)$   &      (ab)         &  $m_B, m_C$      &                      ---                            &   $m_{B_1}=m_{B_2}=m_B$    \\
$M_{2CR}(ab)$   &      (ab)         &    $m_B, m_C$    &   $m_{A_1}=m_{A_2}$                  &   $m_{B_1}=m_{B_2}=m_B$    \\
\hline \hline
$M_{2XX}(b)$    &      (b)         &    $m_C$              &                      ---                            &   ---         \\
$M_{2XC}(b)$    &      (b)         &    $m_C$              &                      ---                            &   $m_{A_1}=m_{A_2}$    \\
$M_{2CX}(b)$    &      (b)         &    $m_C$             &   $m_{B_1}=m_{B_2}$                 &   ---    \\
$M_{2CC}(b)$    &      (b)         &    $m_C$             &   $m_{B_1}=m_{B_2}$                 &   $m_{A_1}=m_{A_2}$    \\  \hline
$M_{2XR}(b)$    &      (b)         &  $m_A, m_C$      &                      ---                            &   $m_{A_1}=m_{A_2}=m_A$    \\
$M_{2CR}(b)$    &      (b)         &    $m_A, m_C$    &   $m_{B_1}=m_{B_2}$                  &   $m_{A_1}=m_{A_2}=m_A$    \\
\hline \hline
$M_{2XX}(a)$    &      (a)         &    $m_B$              &                      ---                            &   ---         \\
$M_{2XC}(a)$    &      (a)         &    $m_B$              &                      ---                            &   $m_{C_1}=m_{C_2}$    \\
$M_{2CX}(a)$    &      (a)         &    $m_B$             &   $m_{A_1}=m_{A_2}$                 &   ---    \\
$M_{2CC}(a)$    &      (a)         &    $m_B$             &   $m_{A_1}=m_{A_2}$                 &   $m_{C_1}=m_{C_2}$    \\  \hline
$M_{2XR}(a)$    &      (a)         &  $m_B, m_C$      &                      ---                            &   $m_{C_1}=m_{C_2}=m_C$    \\
$M_{2CR}(a)$    &      (a)         &    $m_B, m_C$    &   $m_{A_1}=m_{A_2}$                  &   $m_{C_1}=m_{C_2}=m_C$    \\
\hline \hline
\end{tabular}
\caption{\label{tab:M2variables} A summary of the $6\times 3=18$ variables of type $M_2$ defined in the text.
For each of the three subsystems $(ab)$, $(a)$, and $(b)$, one may choose to apply (or not) the
parent constraint (\ref{eq:parents}), and then choose to apply (or not) one of the
relative constraints (\ref{eq:relatives}) or (\ref{eq:relativesfixed}).}
\end{table}

In Table~\ref{tab:M2variables} we collect the full set of $6\times 3=18$ variables of type $M_2$.
The table is organized as follows. We group the variables by subsystem --- first (ab), then (b), and finally, subsystem (a).
Within each subsystem, we order the variables according to the amount of theoretical input --- variables with fewer (more)
constraints appear earlier (later) in the list. As indicated by the entries in the third column of Table~\ref{tab:M2variables},
four of the variables within each subsystem require a single input mass parameter, namely the hypothesized mass of the 
daughter particle for this subsystem. These 12 variables, of type $M_{2XX}$, $M_{2CX}$, $M_{2XC}$, and  $M_{2CC}$,  
are precisely the on-shell constrained $M_2$ variables discussed in \cite{Cho:2014naa}. The remaining 6 variables
in Table~\ref{tab:M2variables} require an additional mass input --- the mass of the relative particle.
In this sense, they are the analogues of the MAOS1 or MAOS4 schemes for invisible momentum reconstruction, 
which also required an additional mass input, see Table~\ref{tab:maos}. 

The pros and cons of the different types of $M_2$ variables from Table~\ref{tab:M2variables} will be discussed in our numerical
examples below (see Section~\ref{sec:reco_M2}). The exact definition for each variable should be clear from our earlier discussion
(see also \cite{Cho:2014naa}), but at this point it may still be instructive to give a few specific examples, particularly for 
the newly introduced variables $M_{2XR}$ and $M_{2CR}$ which employ the stricter constraint (\ref{eq:relativesfixed}).
 
For concreteness, let us consider the dilepton $t\bar{t}$ realization of the event topology of Fig.~\ref{fig:DecaySubsystem}, 
in which the visible particles are: a pair of b-quarks ($a_1=b$, $a_2=\bar{b}$) and a pair of leptons ($b_1=\ell^+$, $b_2=\ell^-$).  
One could imagine that the leptons are still the result of leptonic decays of SM $W$-bosons to neutrinos, so that $m_{B_i}=m_W$ and
$m_{C_i}=0$, while the parents $A_i$ are some new particles, e.g., 4th generation up-type quarks. 
Then, the physics process under consideration (\ref{eq:process}) becomes
\beq
pp\to t'\bar{t}', \quad t'\to b W^+, \quad W^+\to \ell^+ \nu_\ell.
\label{eq:4thgen}
\eeq
In this case, it makes sense to consider the variable $M_{2CW}(b\ell)$ defined as
\bea
M_{2CW}^2(b\ell) &\equiv& \min_{\vec{q}_{1},\vec{q}_{2}}\left\{\max\left[(p_{b}+p_{\ell^+}+q_1)^2,\;(p_{\bar{b}}+p_{\ell^-}+q_2)^2 \right] \right\},
\label{eq:m2CWab}\\
q_1^2&=& 0 \nonumber \\
q_2^2&=& 0 \nonumber \\
\vec{q}_{1T}+\vec{q}_{2T} &=& \mpt   \nonumber \\
(p_{\ell^+}+q_1)^2&=&m_W^2 \nonumber \\
(p_{\ell^-}+q_2)^2&=&m_W^2 \nonumber \\
(p_{b}+p_{\ell^+}+q_1)^2&=&(p_{\bar{b}}+p_{\ell^-}+q_2)^2 \nonumber 
\eea  
whose upper kinematic endpoint would be the mass of the top partner $t'$.

Another possibility is to consider stop production in SUSY, followed by sequential decays to charginos and sneutrinos:
\beq
pp\to \tilde t\, \tilde t^\ast, \quad \tilde t\to b \tilde \chi^+, \quad \tilde\chi^+\to \ell^+ \tilde\nu_\ell.
\label{eq:stop}
\eeq 
In this case, a prior measurement of the $M_{T2}(\ell)$ kinematic endpoint could provide knowledge of the chargino mass as a function 
of the sneutrino mass, $m_{\tilde\chi^\pm}(m_{\tilde \nu_\ell})$, which would allow us to consider the
maximally constrained kinematic variable $M_{2C\tilde\chi^\pm}(b\ell)$ defined as
\bea
M_{2C\tilde\chi^\pm}^2(b\ell) &\equiv& \min_{\vec{q}_{1},\vec{q}_{2}}\left\{\max\left[(p_{b}+p_{\ell^+}+q_1)^2,\;(p_{\bar{b}}+p_{\ell^-}+q_2)^2 \right] \right\},
\label{eq:m2Cchiab}\\
q_1^2&=& m^2_{\tilde \nu_\ell} \nonumber \\
q_2^2&=& m^2_{\tilde \nu_\ell}  \nonumber \\
\vec{q}_{1T}+\vec{q}_{2T} &=& \mpt   \nonumber \\
(p_{\ell^+}+q_1)^2&=& m^2_{\tilde\chi^\pm}(m_{\tilde \nu_\ell}) \nonumber \\
(p_{\ell^-}+q_2)^2&=& m^2_{\tilde\chi^\pm}(m_{\tilde \nu_\ell})  \nonumber \\
(p_{b}+p_{\ell^+}+q_1)^2&=&(p_{\bar{b}}+p_{\ell^-}+q_2)^2 \nonumber 
\eea  
The minimizations in (\ref{eq:m2CWab}) and (\ref{eq:m2Cchiab}) are essentially one-dimensional minimizations, since they 
involve a total of seven constraints for the eight unknown components $q_1^\mu$ and $q_2^\mu$.
 
 One could also consider situations where the masses for the $A_i$ particles are known instead. If we stick to the case where 
 $A_i$ is the SM top quark, we can imagine that the particles $B_i$ are not $W$ bosons, but some other charged scalars $H^\pm$.
Then the process under consideration becomes
\beq
pp\to t\bar{t}, \quad t\to b H^+, \quad H^+\to \ell^+ \nu_\ell.
\label{eq:H+}
\eeq
The relevant variable now is 
\bea
M_{2Ct}^2(\ell) &\equiv& \min_{\vec{q}_{1},\vec{q}_{2}}\left\{\max\left[(p_{\ell^+}+q_1)^2,\;(p_{\ell^-}+q_2)^2 \right] \right\},
\label{eq:m2Ctb}\\
q_1^2&=& 0 \nonumber \\
q_2^2&=& 0 \nonumber \\
\vec{q}_{1T}+\vec{q}_{2T} &=& \mpt   \nonumber \\
(p_{\ell^+}+q_1)^2 &=& (p_{\ell^-}+q_2)^2 \nonumber\\
(p_{b}+p_{\ell^+}+q_1)^2&=& m_t^2 \nonumber \\
(p_{\bar{b}}+p_{\ell^-}+q_2)^2&=& m_t^2\nonumber 
\eea 
whose upper kinematic endpoint is the mass of the charged boson $H^\pm$. Note that the first $M_2$ subscript ``$C$"
in (\ref{eq:m2Ctb}) refers to the presence of the parent mass constraint (\ref{eq:parents}) for the $H^\pm$ particles in the leptonic subsystem, 
while the second subscript ``$t$" identifies the relative particles $A_i$ as top quarks.

In conclusion of this section, we also mention the possibility to define a class of variables, $M_1$, where one minimizes a target mass function 
without any partitioning of the event \cite{Barr:2011xt}. If this minimization is performed in the absence of any additional kinematic constraints 
besides (\ref{eq:mpt}), one obtains the usual $\sqrt{s}_{min}$ variable \cite{Konar:2008ei,Konar:2010ma}. In the example of the $t\bar{t}$ event topology
we have
\bea
s_{min}(b\ell) \equiv M_{1XX}^2(b\ell) &\equiv& \min_{\vec{q}_{1},\vec{q}_{2}}\left\{(p_{b}+p_{\ell^+}+q_1+p_{\bar{b}}+p_{\ell^-}+q_2)^2 \right\},
\label{eq:m1XXbl}\\
q_1^2&=& 0 \nonumber \\
q_2^2&=& 0 \nonumber \\
\vec{q}_{1T}+\vec{q}_{2T} &=& \mpt   \nonumber 
\eea
where we have assumed zero test masses for the two invisible particles. However, one may also choose to 
partition the event post factum in order to define a suitable kinematic constraint of the type (\ref{eq:parents}), 
(\ref{eq:relatives}) or (\ref{eq:relativesfixed}). Consider, for example, the single production of a heavy Higgs boson, $H^0$, 
subsequently decaying to two on-shell $W$-bosons, which in turn decay leptonically:
\beq
pp\to H^0, \quad H^0\to W^+W^-, \quad W^+\to \ell^+ \nu_\ell, \quad W^-\to \ell^- \bar{\nu}_\ell.
\label{eq:HWW}
\eeq
The relevant variable to consider in this case would be
\bea
M_{1W}^2(\ell) &\equiv& \min_{\vec{q}_{1},\vec{q}_{2}}\left\{(p_{\ell^+}+q_1+p_{\ell^-}+q_2)^2 \right\},
\label{eq:m1W}\\
q_1^2&=& 0 \nonumber \\
q_2^2&=& 0 \nonumber \\
\vec{q}_{1T}+\vec{q}_{2T} &=& \mpt   \nonumber \\
(p_{\ell^+}+q_1)^2 &=& m_W^2 \nonumber\\
(p_{\ell^-}+q_2)^2 &=& m_W^2 \nonumber
\eea
which was called $m_T^{bound}$ in \cite{Barr:2011si} and $\hat{s}_{min}^{cons}$ in \cite{Swain:2014dha}.
In all those cases, the minimization again results in an ansatz for the invisible 3-momenta $\vec{q}_1$ and $\vec{q}_2$, 
so that the $M_1$ class of variables can in principle also be used for fixing the momenta of the invisible particles.

\section{$M_{T2}$-assisted and $M_2$-assisted mass reconstructions of mass peaks}
\label{sec:m2aos}

In the previous section, we identified a number of different ways in which one can obtain an ansatz for the 
unknown momenta of the invisible particles in the event. The main purpose of this section is to compare the usefulness of
these different ansatze with regards to mass measurements through bump hunting. To be specific, we shall focus on the
dilepton $t\bar{t}$ event topology from Fig.~\ref{fig:DecaySubsystem} and we shall consider the three subsystems,
$(ab)$, $(a)$ and $(b)$. In subsection~\ref{sec:reco_maos} we shall first discuss the twelve versions of the traditional MAOS method
which are listed in Table~\ref{tab:maos}, while in subsection~\ref{sec:reco_M2} we shall compare the different types of
$M_2$-based reconstructions from Table~\ref{tab:M2variables}. Depending on the procedure, one expects to obtain an invariant 
mass  bump for one of the three particles involved --- the top quark, the $W$-boson or the neutrino, as the case may be. 
The sensitivity of the mass measurement will be judged by the width of the obtained invariant mass distribution --- 
a narrow (broad) peak will indicate high (reduced) sensitivity. Finally, in subsection~\ref{sec:reco_all} we shall contrast 
the MAOS methods from Sec.~\ref{sec:reco_maos} to the $M_2$-based methods from Sec.~\ref{sec:reco_M2}.

\subsection{Comparison of the different MAOS methods}
\label{sec:reco_maos}

\begin{figure}[t]
\centering
\includegraphics[width=4.9cm]{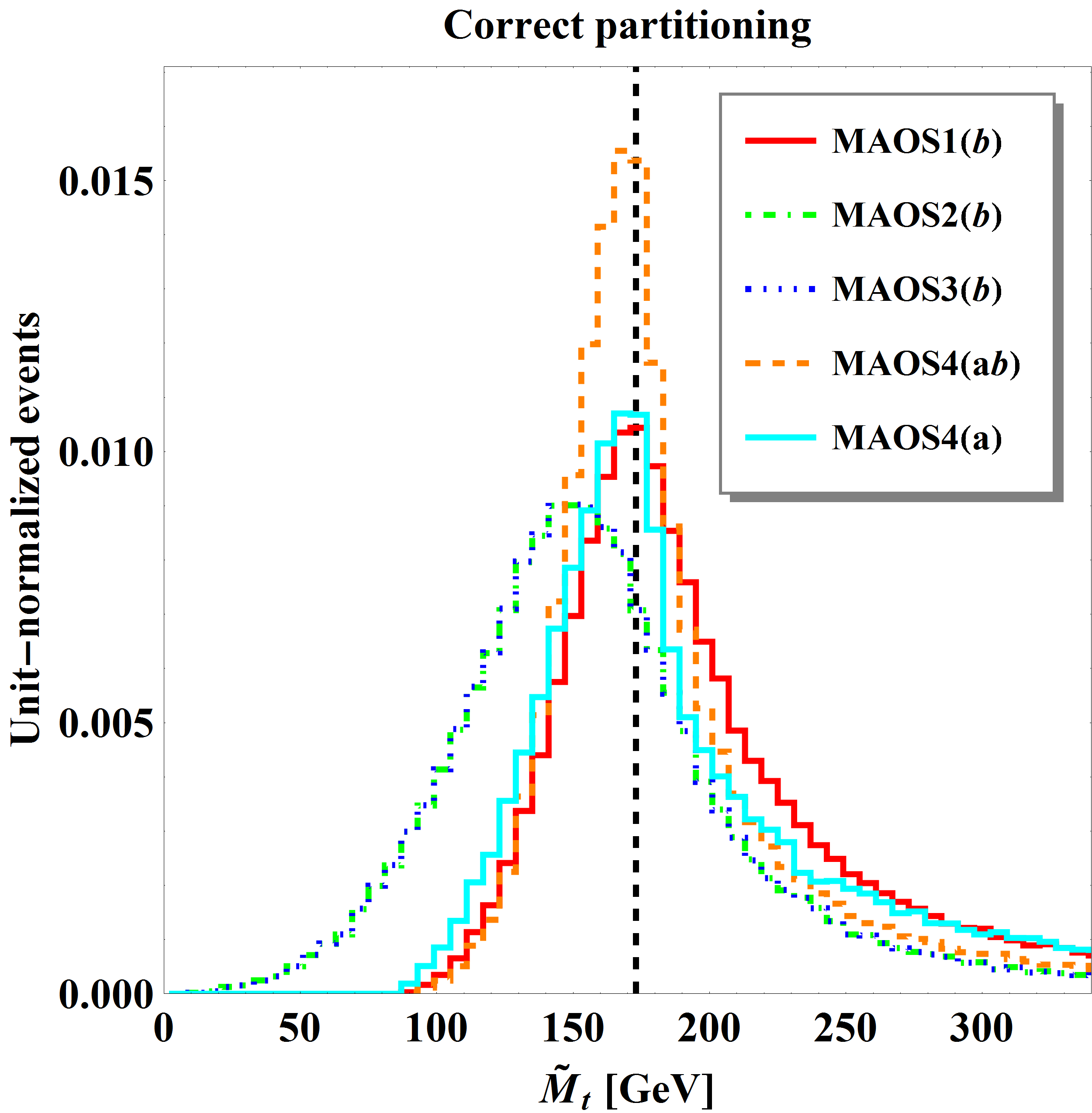}
\includegraphics[width=4.9cm]{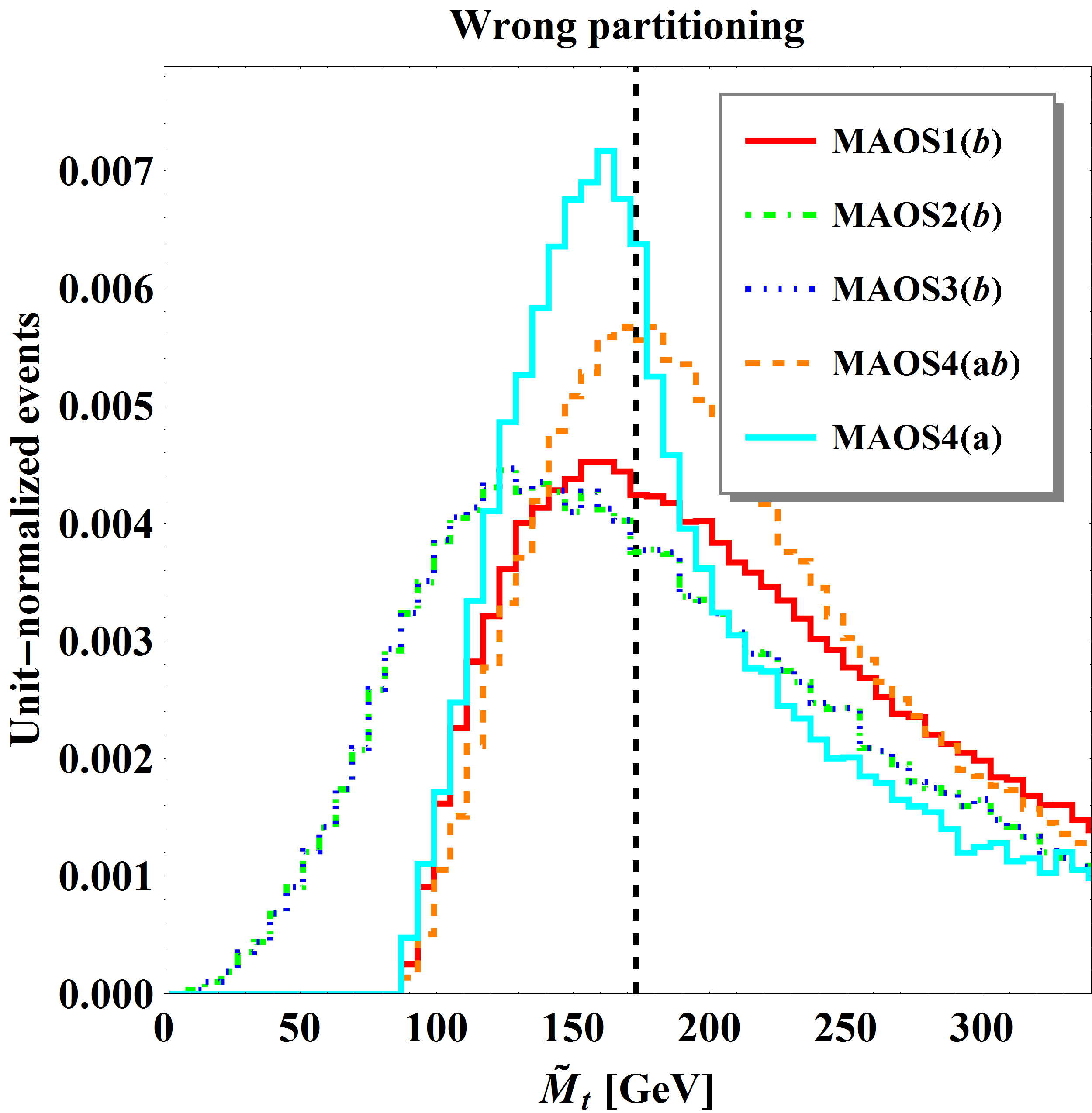}
\includegraphics[width=4.9cm]{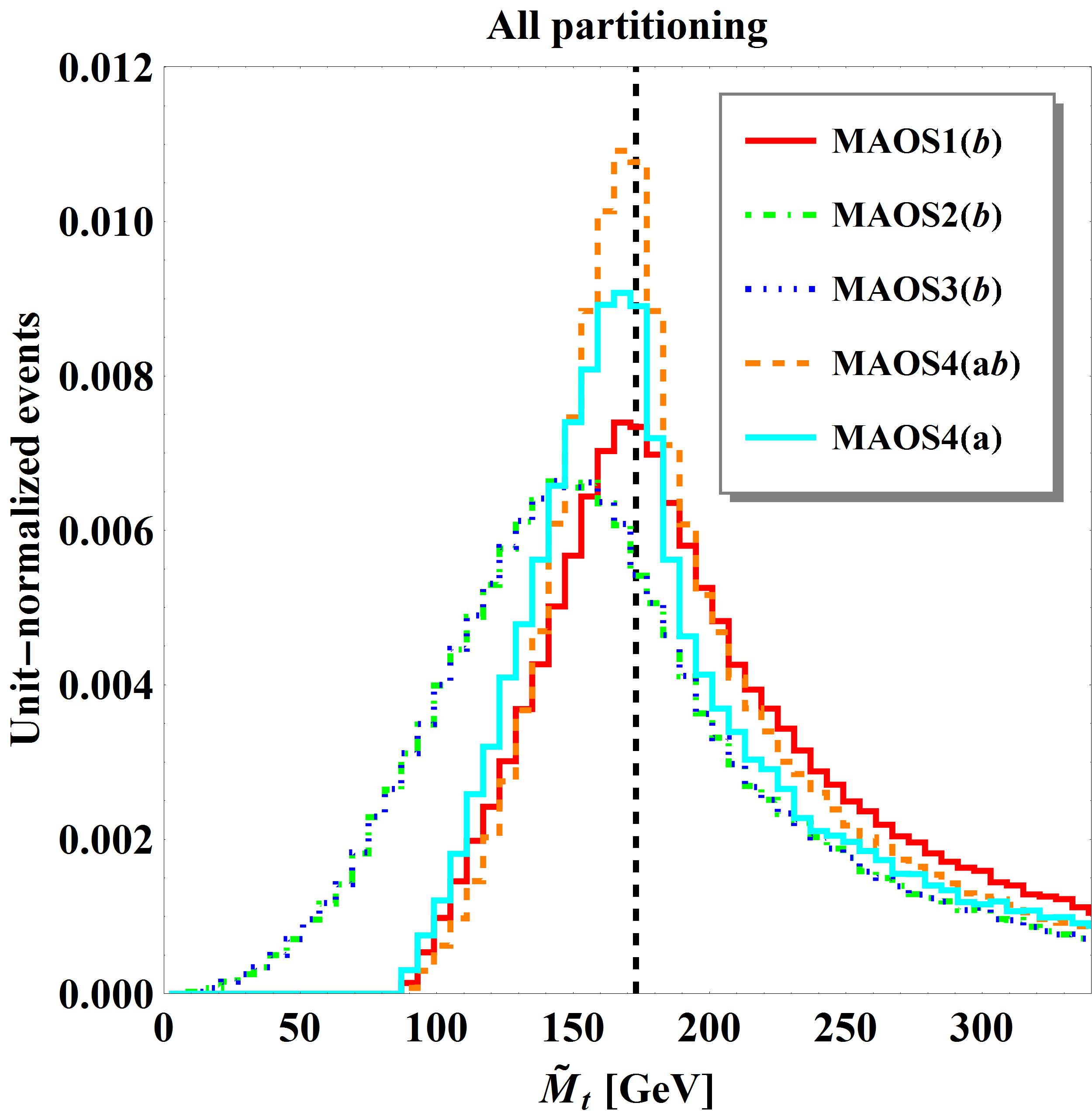}
\caption{\label{fig:reco_maos_top} Comparison of the performance of different MAOS schemes in reconstructing the top mass
in dilepton $t\bar{t}$ events. Distributions of the reconstructed top mass $\tilde M_t$ are shown for the case of the correct 
lepton-jet pairing (left panel), the wrong lepton-jet pairing (middle panel) and both pairings (right panel).
The top quark is treated as a relative particle in the case of MAOS1(b) (red solid line), MAOS2(b) (green dot-dashed line)
and MAOS3(b) (blue dotted line), and as a parent particle in the case of MAOS4(ab) (orange dashed line) and MAOS4(a) (cyan solid line).}
\end{figure}

\begin{figure}[t]
\centering
\includegraphics[width=4.9cm]{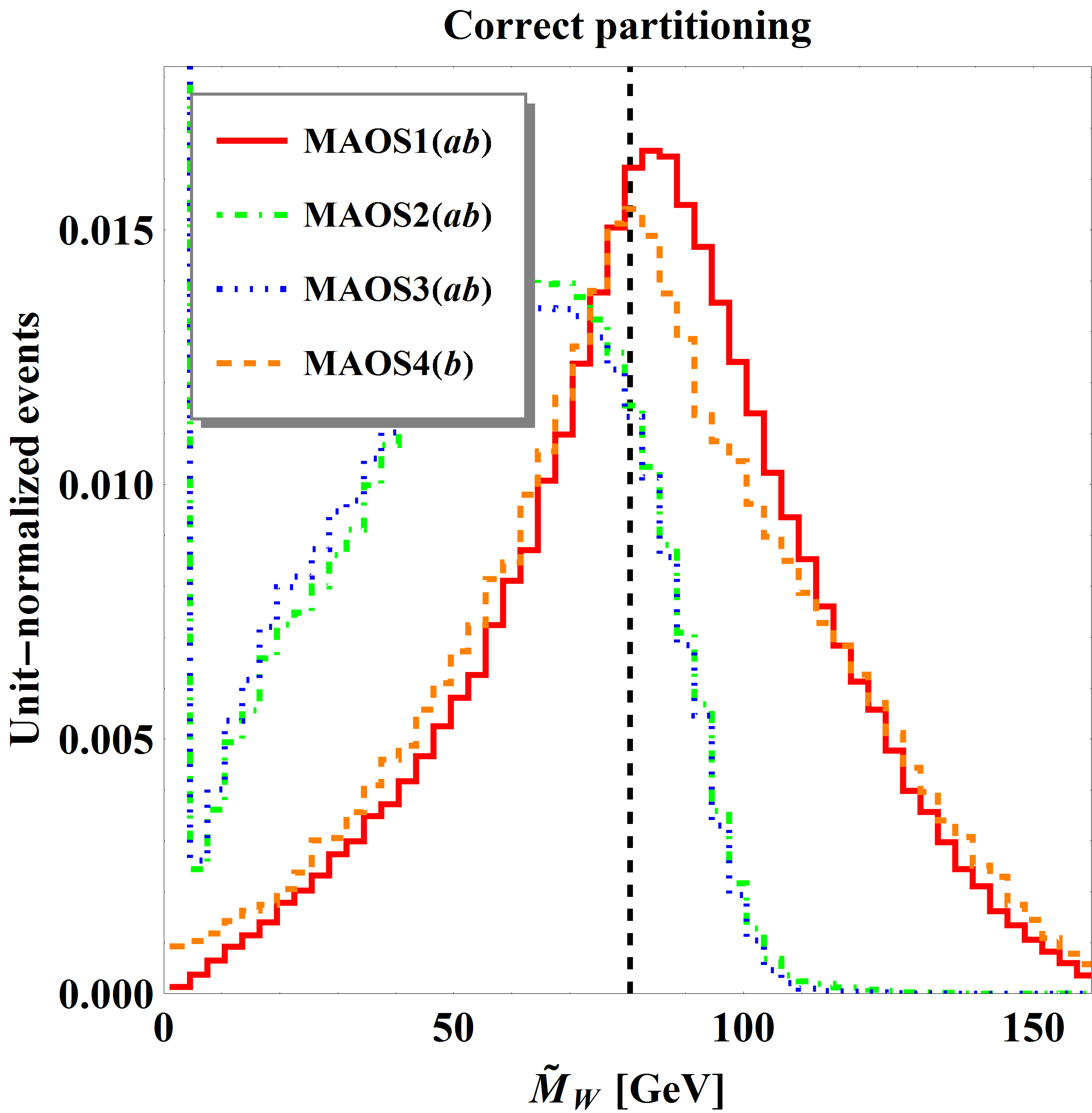}
\includegraphics[width=4.9cm]{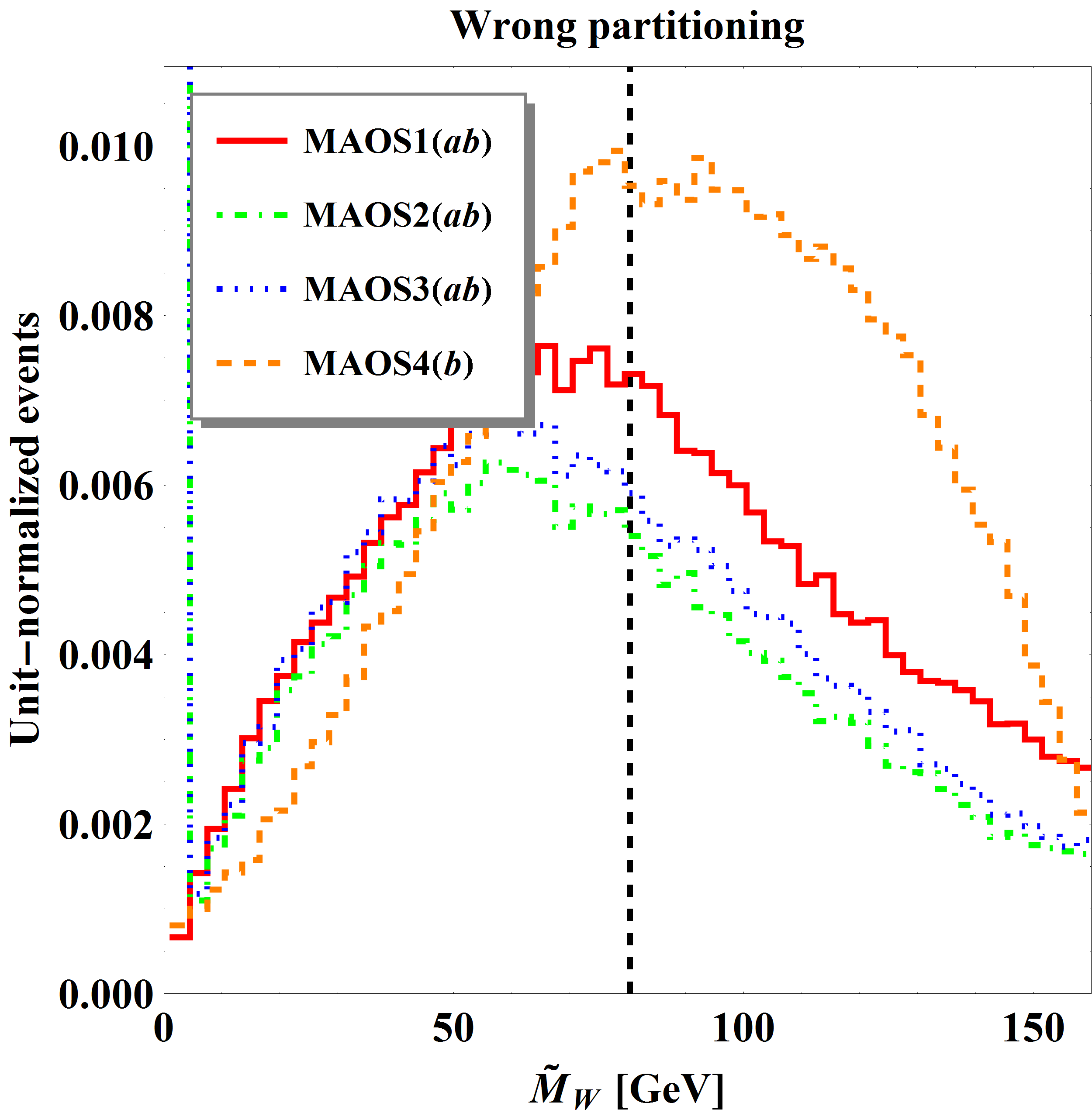}
\includegraphics[width=4.9cm]{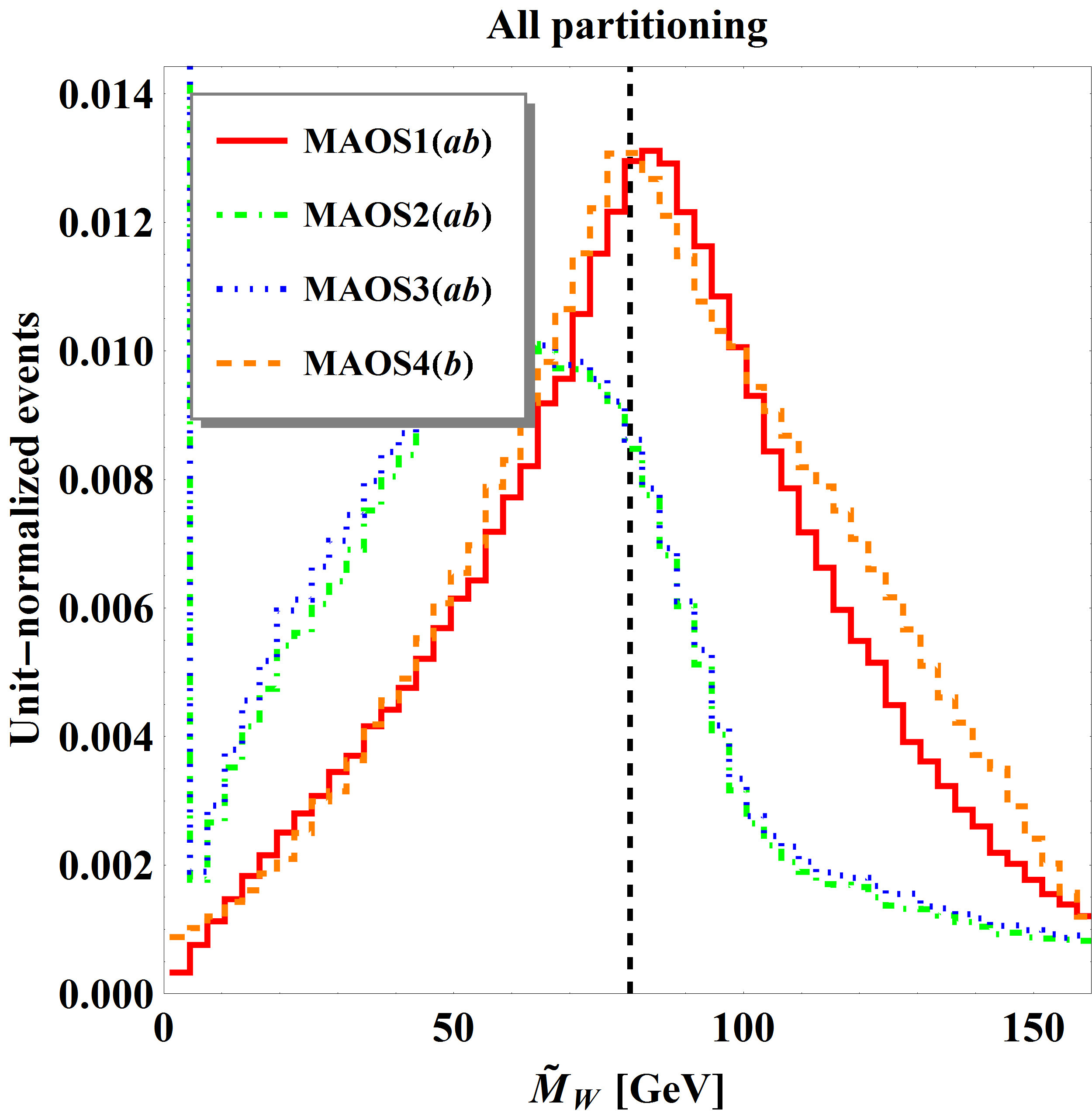}
\caption{\label{fig:reco_maos_W} Comparison of the performance of different MAOS schemes in reconstructing the $W$-boson mass
in dilepton $t\bar{t}$ events. Distributions of the reconstructed $W$-boson mass $\tilde M_W$ are shown for the case of the correct 
lepton-jet pairing (left panel), the wrong lepton-jet pairing (middle panel) and both pairings (right panel).
The $W$-boson is treated as a relative particle in the case of MAOS1(ab) (red solid line), MAOS2(ab) (green dot-dashed line)
and MAOS3(ab) (blue dotted line), and as a parent particle in the case of MAOS4(b) (orange dashed line).}
\end{figure}

\begin{figure}[t]
\centering
\includegraphics[width=4.9cm]{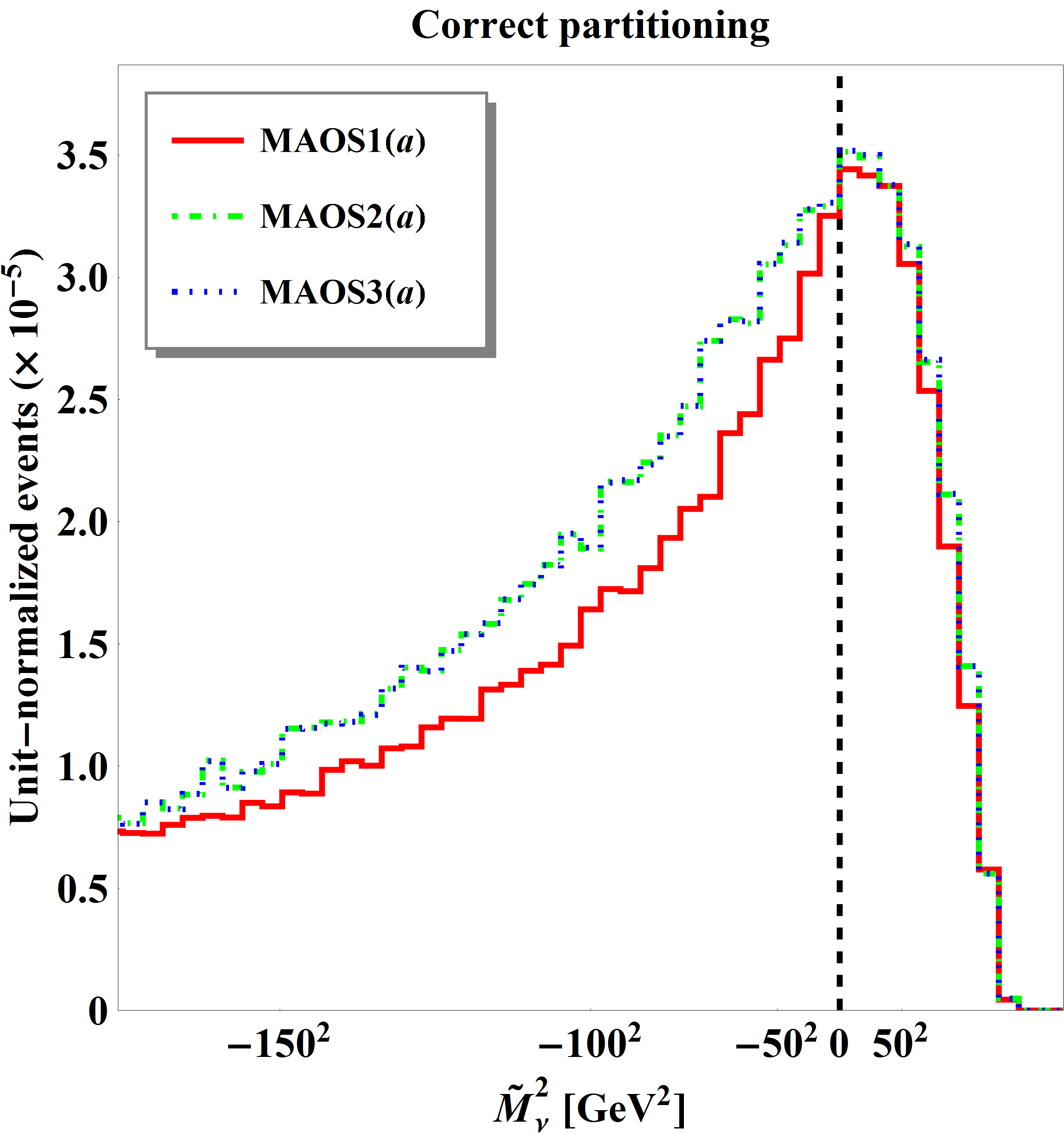}
\includegraphics[width=4.9cm]{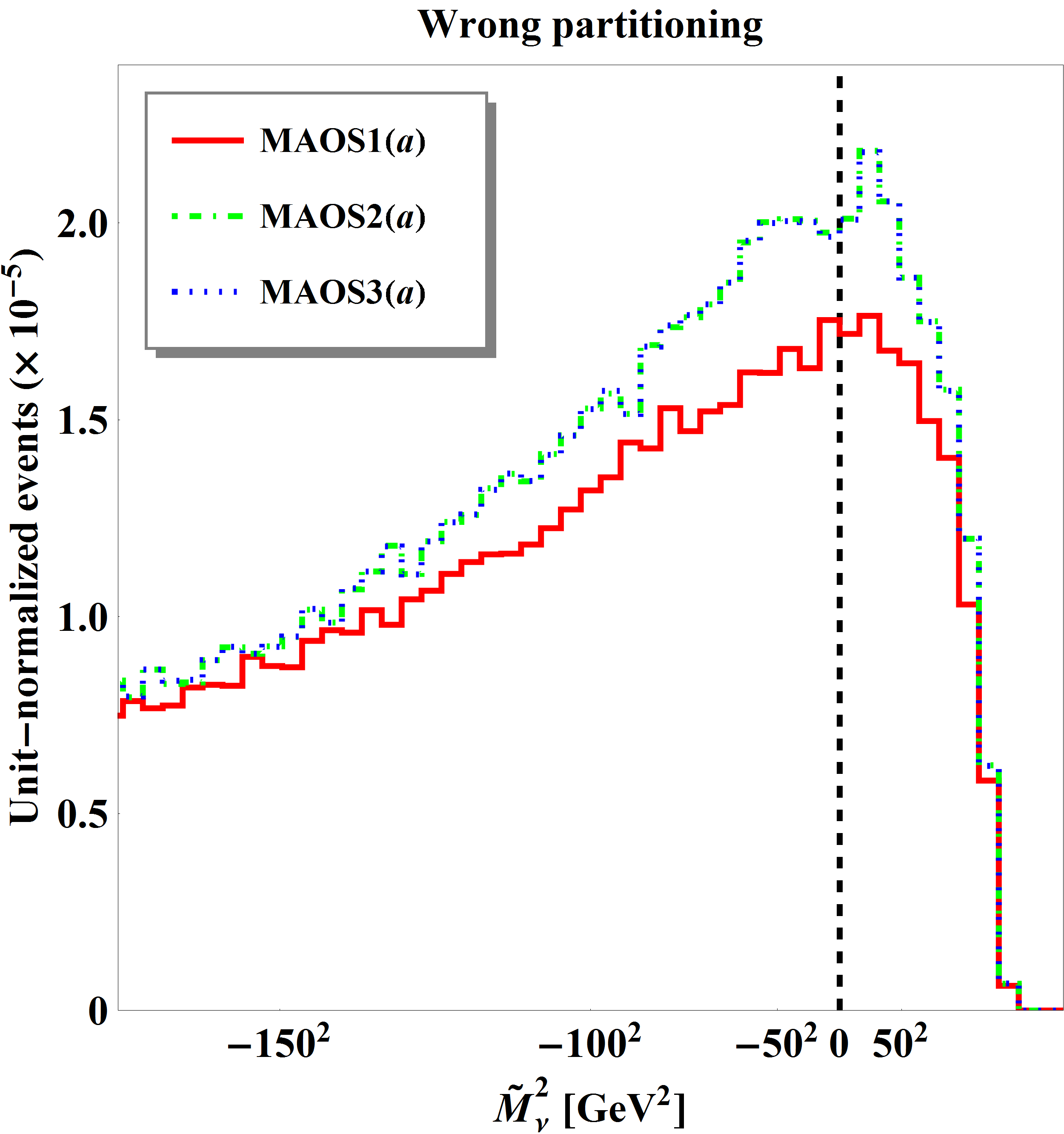}
\includegraphics[width=4.9cm]{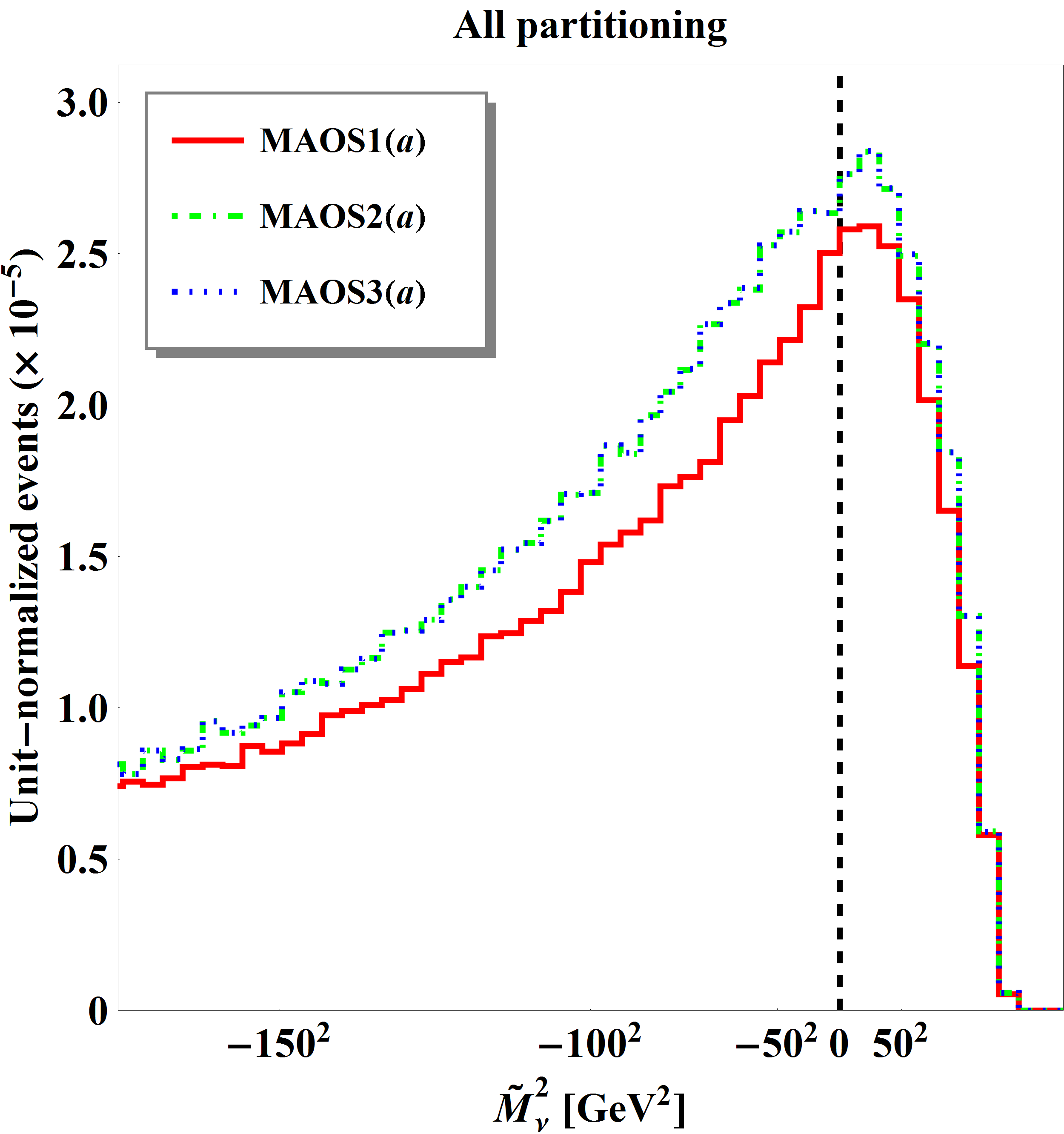}
\caption{\label{fig:reco_maos_nu} Comparison of the performance of different MAOS schemes in reconstructing the neutrino mass-squared
in dilepton $t\bar{t}$ events. Distributions of the reconstructed neutrino mass-squared $\tilde M^2_\nu$ are shown for the case of the correct 
lepton-jet pairing (left panel), the wrong lepton-jet pairing (middle panel) and both pairings (right panel).
Here the neutrino is always treated as a relative particle in the case of MAOS1(a) (red solid line), MAOS2(a) (green dot-dashed line)
and MAOS3(a) (blue dotted line).}
\end{figure}

First we compare the performance of the twelve different MAOS schemes introduced in Sec.~\ref{sec:maos}.
Generally, we will be reconstructing the mass of the relative particle --- the $W$ boson mass $\tilde M_W$ in subsystem $(ab)$,
the top quark mass $\tilde M_t$ in subsystem $(b)$ and the neutrino mass $\tilde M_\nu$ in subsystem $(a)$. 
However, in the case of MAOS4, the result would be trivial since the mass of the relative particle itself is used as one of the constraints. 
This is why in the case of MAOS4 only we shall instead plot the mass of the parent particle, i.e., $\tilde M_W$ for MAOS4(b)
and $\tilde M_t$ for MAOS4(ab) and MAOS4(a). Our results are presented in Figs.~\ref{fig:reco_maos_top}-\ref{fig:reco_maos_nu},
where events were generated with {\sc Madgraph} \cite{Alwall:2014hca} for the LHC with energy 14 TeV.
Since it is difficult to distinguish a $b$-jet from a $\bar{b}$-jet  in practice, 
there is a two-fold combinatorial ambiguity which may occur at different stages --- in forming the $M_{T2}$ variable, in using
the top mass to solve for $q_{iz}$, or in forming $\tilde M_t$. Either way, this combinatorial ambiguity inevitably affects the results, 
which is why in the figures we show separately results for the correct lepton-jet pairing (left panels), 
the wrong lepton-jet pairing (middle panels) and combining both pairings (right panels).

Fig.~\ref{fig:reco_maos_top} shows results from reconstructing the top quark mass $\tilde M_t$
with the five relevant MAOS methods: MAOS1(b) (red solid lines), MAOS2(b) (green dot-dashed lines), MAOS3(b) (blue dotted lines), 
MAOS4(ab) (orange dashed lines), and MAOS4(a) (cyan solid lines).
In all cases, we use the correct test mass when calculating $M_{T2}$: the true neutrino mass $m_\nu=0$ in subsystems (ab) and (b),
and the true $W$-boson mass $m_W=80$ GeV for subsystem (a). In the case of MAOS2(b) and MAOS3(b), this is the only mass input 
needed to reconstruct $\tilde M_t$, see Table~\ref{tab:maos}. Unfortunately, this theoretical advantage seems to be offset by 
the inferior performance of these two methods: even for the correct lepton-jet combination, the MAOS2(b) and MAOS3(b) 
distributions in the left panel in Fig.~\ref{fig:reco_maos_top} peak below the true top mass $m_t$, so that a bump hunt will
systematically underestimate the value of $m_t$. The remaining three MAOS methods illustrated in the figure, 
MAOS1(b), MAOS4(ab) and MAOS4(a), use an additional mass input, and are thus expected to perform better.\footnote{In MAOS1 and MAOS4, 
the additional mass input is used to solve for the longitudinal momenta. Since the relevant equations are non-linear, 
one may end up with multiple solutions. In such cases, we plot the result for each solution with a corresponding weight 
factor so that each event has weight 1. Similar comments apply to the case of MAOS2, where for unbalanced events
one may find two solutions for $q_{iz}$.} 
This is confirmed by Fig.~\ref{fig:reco_maos_top}, which suggests that MAOS4(ab) slightly outperforms the other other two
methods, MAOS4(a) and  MAOS1(b), which are utilizing the smaller individual subsystems (a) and (b).
There are two effects which contribute to this. First, for the correct lepton-jet combination (the left panel in 
Fig.~\ref{fig:reco_maos_top}) the distributions for all three methods, MAOS1(b), MAOS4(ab) and MAOS4(a), have their peaks
very close to the true mass $m_t$, but the peak for MAOS4(ab) is more narrow than the other two. Second, for the 
wrong lepton-jet combination (the middle panel in Fig.~\ref{fig:reco_maos_top}), the MAOS4(ab) distribution is relatively
broad, but happens to peak right around the top quark mass again, while the distributions for MAOS4(a) and  MAOS1(b)
peak at slightly lower values. If one does not attempt to resolve the combinatorics \cite{Rajaraman:2010hy,Baringer:2011nh,Choi:2011ys} 
and instead does the simplest thing, namely, combine the two distributions from the left and middle panels of Fig.~\ref{fig:reco_maos_top}, 
one would obtain the combined distributions shown in the right panel of Fig.~\ref{fig:reco_maos_top}.
We see that among the methods using two mass inputs, MAOS4(ab) appears to be the best, followed by 
MAOS4(a) and MAOS1(b). The remaining two procedures, MAOS2(b) and MAOS3(b), rely on a single mass input, 
and give identical answers, in accordance with our expectations for subsystem $(b)$.

Fig.~\ref{fig:reco_maos_W} shows the analogous results for the reconstruction of the mass $\tilde M_W$ of the $W$-boson,
using MAOS1(ab) (red solid line), MAOS2(ab) (green dot-dashed line), MAOS3(ab) (blue dotted line), and MAOS4(b) (orange dashed line).
In all cases we use the correct test mass as an input to the $M_{T2}$ calculation, and then the correct value of the 
additional mass input required for MAOS1(ab) and MAOS4(b). The left panel of Fig.~\ref{fig:reco_maos_W} clearly demonstrates
the benefit of the additional mass input, as MAOS1(ab) and MAOS4(b) greatly outperform MAOS2(ab) and MAOS3(ab).
Since the corresponding wrong-combination distributions in the middle panel have similar shapes, this advantage is preserved in the
combined distributions shown in the right panel. Upon closer inspection, MAOS4(b) (orange dashed line) appears slightly better 
than MAOS1(ab) (red solid line). However, in new physics applications of the MAOS methods, the knowledge of the 
additional mass input is not always guaranteed, and one would have to do with MAOS2(ab) or MAOS3(ab), which perform
very similarly. Among the two, MAOS3(ab) has a slight theoretical advantage in the sense that its invisible momentum ansatz 
is always unique and well-defined.

Our third and final mass reconstruction for the dilepton $t\bar{t}$ topology is shown in Fig.~\ref{fig:reco_maos_nu}, where we plot
in analogous fashion the reconstructed neutrino mass-squared $\tilde M^2_\nu$ when the neutrino is treated as a relative particle
in subsystem (a): MAOS1(a) (red solid line), MAOS2(a) (green dot-dashed line) and MAOS3(a) (blue dotted line). 
This time the benefit of the additional mass input $m_t$ in the case of MAOS1(a) is not so clear --- all three distributions 
have similar shapes (the distributions for MAOS2(a) and MAOS3(a) are in fact identical, since 
subsystem $(a)$ has only balanced events) and peak near the origin.

\subsection{Comparison of the different $M_2$-based methods}
\label{sec:reco_M2}

We shall now use the dilepton $t\bar{t}$ example to test the accuracy of the invisible momentum reconstruction 
from the different $M_2$-based methods listed in Table~\ref{tab:M2variables}. We shall not consider all 18 possibilities 
in Table~\ref{tab:M2variables}, since some are closely related. For example, it is known that for any subsystem,
the $M_{2XX}$ and $M_{2CX}$ variables are identical, and furthermore, equal to the value of the Cambridge 
transverse mass variable $M_{T2}$ \cite{Cho:2014naa}:
\beq
M_{2XX} = M_{2CX} = M_{T2}.
\label{M2XXMT2equality}
\eeq
In spite of this relation, the corresponding three ansatze for the invisible momenta are not necessarily the same. 
First of all, $M_{T2}$ is a transverse variable and it only fixes the transverse components $\vec{q}_{1T}$ and $\vec{q}_{2T}$,
while $M_{2XX}$ and $M_{2CX}$ in addition provide values for the longitudinal components $q_{1z}$ and $q_{2z}$.
In the case of balanced events, those predictions are unique and the same for $M_{2XX}$ and $M_{2CX}$, while
for unbalanced events, there is a two-fold ambiguity for $q_{1z}$ and $q_{2z}$ in the case of $M_{2CX}$ 
and a flat direction in the case of $M_{2XX}$ \cite{Cho:2014naa}. In what follows, we shall therefore prefer to 
consider the invisible momentum reconstruction from the $M_{2CX}$ variable instead of $M_{2XX}$. 

Similar considerations apply in the case of the pair of variables $M_{2XC}$ and $M_{2CC}$, as well as for
$M_{2XR}$ and $M_{2CR}$. In each case, the variables are equal for balanced events and only differ for unbalanced events, 
where this time the obtained invisible momentum configurations are unique. This is why we shall also not consider
$M_{2XC}$ and $M_{2XR}$, and instead focus on $M_{2CC}$ and $M_{2CR}$, respectively.

\begin{figure}[t]
\centering
\includegraphics[width=4.9cm]{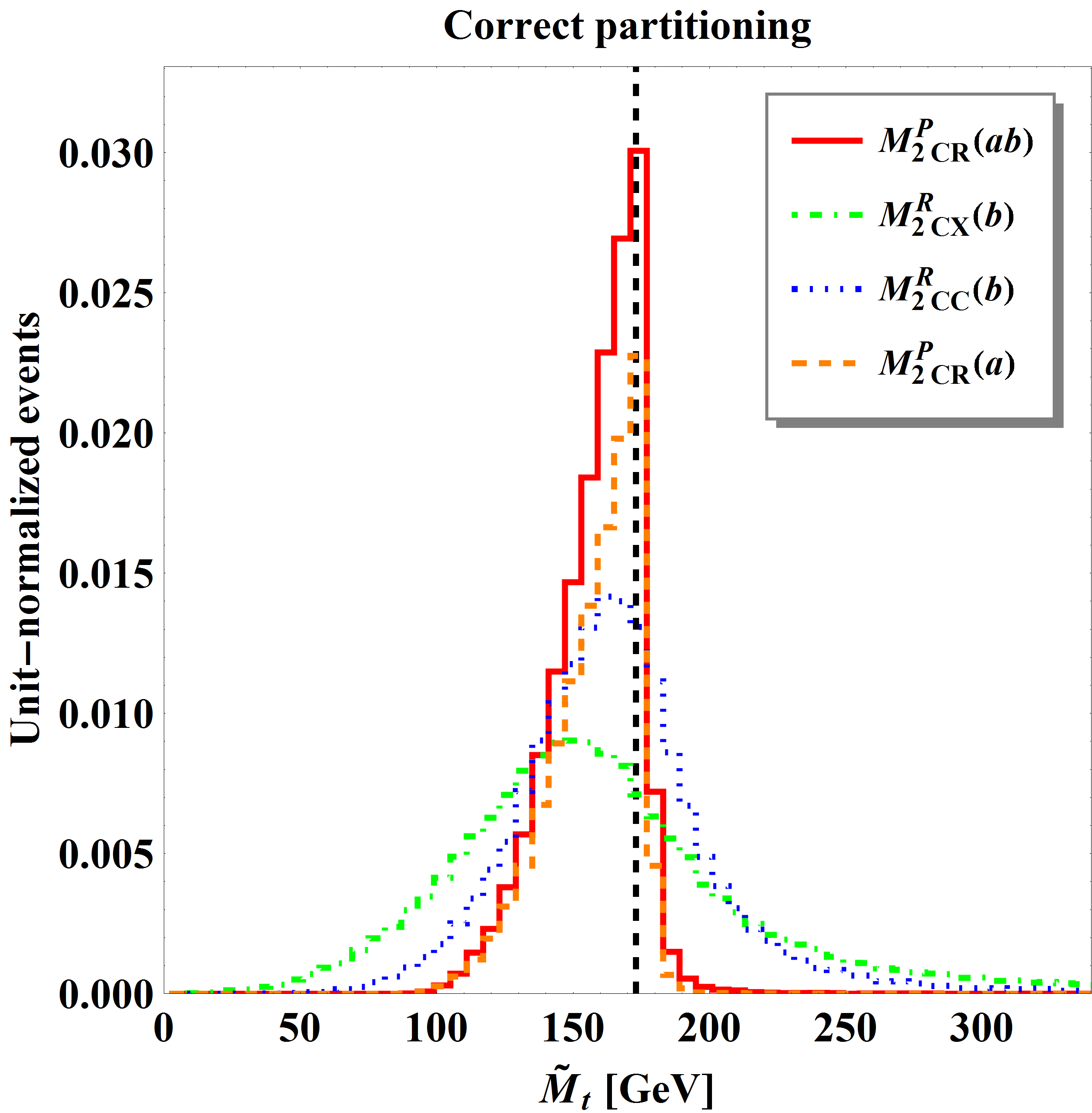}
\includegraphics[width=4.9cm]{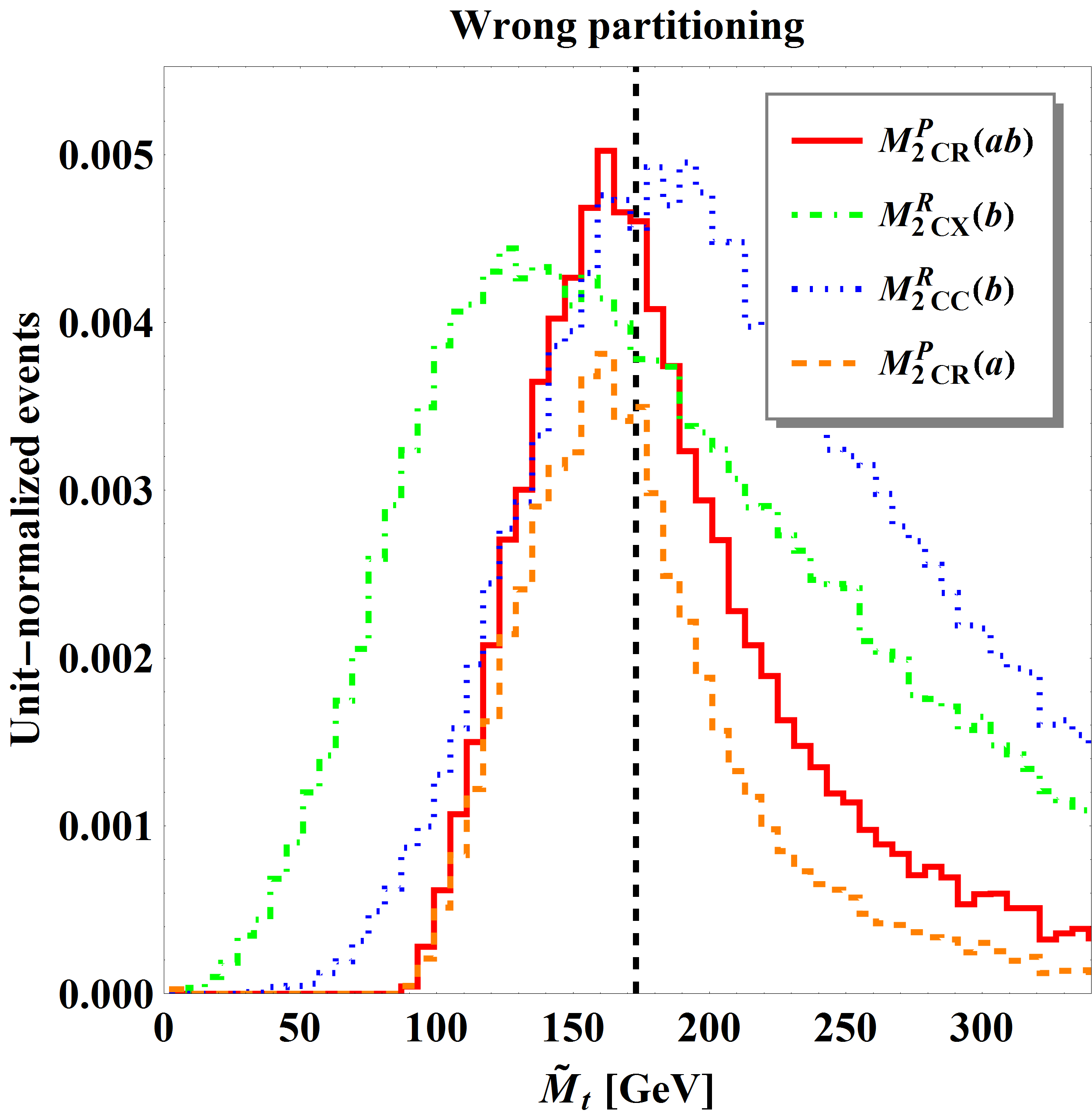}
\includegraphics[width=4.9cm]{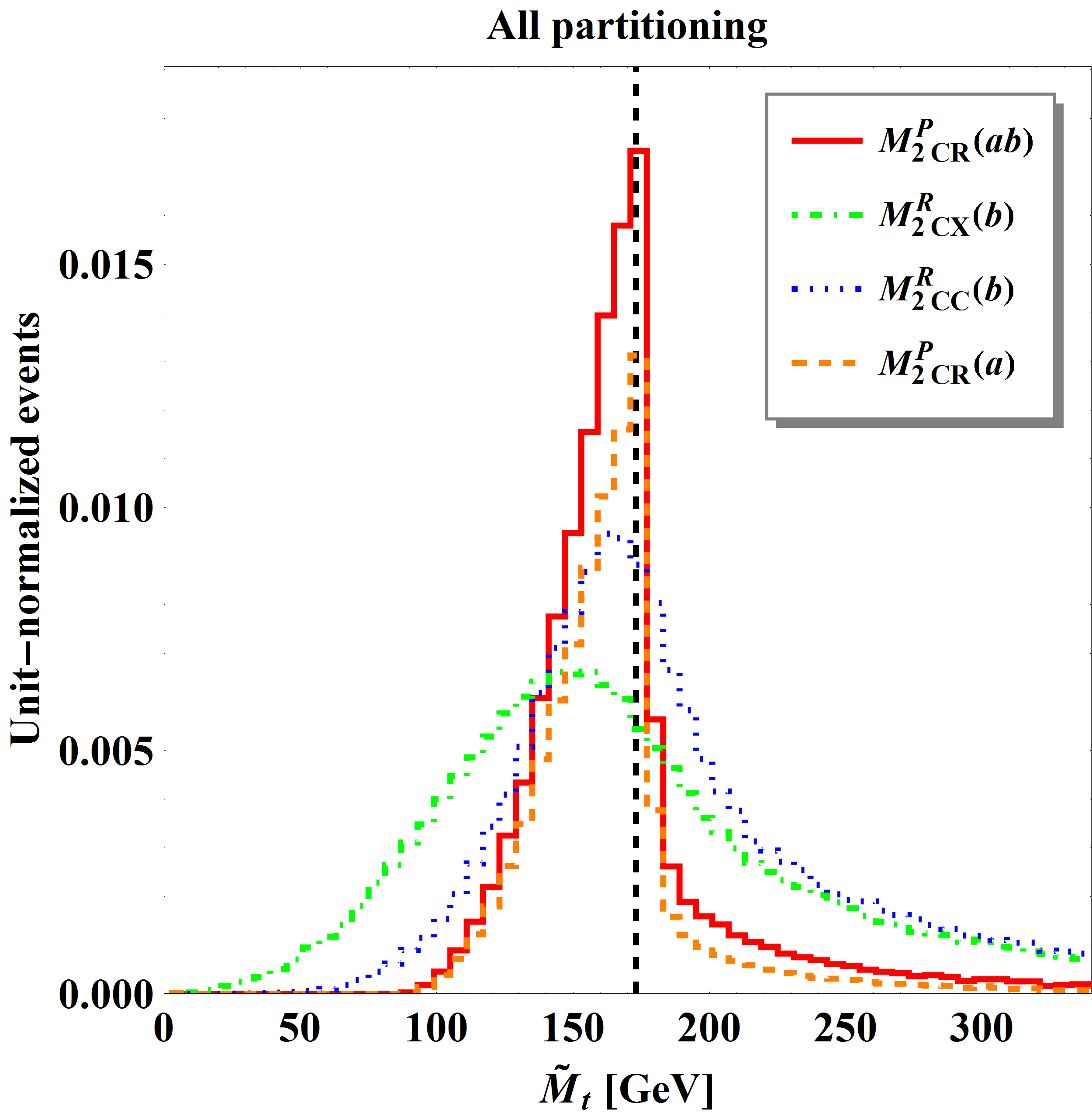}
\caption{\label{fig:reco_M2_top} The same as Fig.~\ref{fig:reco_maos_top}, but using the appropriate $M_2$ variables for 
fixing the invisible momenta. Distributions of $\tilde M_t$ are shown for the case of 
$M_{2CR}(ab)$ (red solid line), 
$M_{2CX}(b)$ (green dot-dashed line),
$M_{2CC}(b)$ (blue dotted line), and
$M_{2CR}(a)$ (orange dashed line).}
\end{figure}

Fig.~\ref{fig:reco_M2_top} shows distributions of the reconstructed top quark mass $\tilde M_t$ with the four relevant 
$M_2$ methods from Table~\ref{tab:M2variables}: $M_{2CR}(ab)$ (red solid line), $M_{2CX}(b)$ (green dot-dashed line),
$M_{2CC}(b)$ (blue dotted line), and $M_{2CR}(a)$ (orange dashed line). In analogy to Fig.~\ref{fig:reco_maos_top}, 
we show separately the distributions obtained for the correct lepton-jet pairing (left panel), the wrong lepton-jet pairing 
(middle panel), and both pairings (right panel). Note that some distributions have fewer events, since the constraints 
cannot be simultaneously satisfied. This is most notable for the case of $M_{2CR}(a)$, and is typically due to events in which
an intermediate resonance (a top quark or a $W$-boson) is rather off-shell (we expect this effect to be further
amplified once we account for the finite detector resolution). Also note that in subsystems $(ab)$
and $(a)$ the top quark is a parent particle, while in subsystem $(b)$ it is a relative particle. This distinction is indicated
in the legend of Fig.~\ref{fig:reco_M2_top} with a superscript $P$ or $R$, respectively.

Fig.~\ref{fig:reco_M2_top} confirms that the more constrained variables generally provide better guesses for the invisible momenta,
as measured by the location and width of the reconstructed mass peak in $\tilde M_t$. The most constrained version of the $M_2$ variable 
is $M_{2CR}$, which has one parent constraint and two relative constraints, leaving a single momentum degree of freedom to be minimized over.
In Fig.~\ref{fig:reco_M2_top}, both $M_{2CR}(ab)$ and $M_{2CR}(a)$ seem to work very well --- for the correct lepton-jet pairing,
the reconstructed top mass peak is very well defined and located at the correct position (marked with the vertical dashed line).
However, the disadvantage of $M_{2CR}(ab)$ and $M_{2CR}(a)$ is that one uses both the $W$-boson mass and the neutrino mass
as inputs to the calculation, which restricts their applicability to BSM scenarios. Under those circumstances, the single-input variables
$M_{2CX}(b)$ and $M_{2CC}(b)$ will be more useful for momentum reconstruction --- in Fig.~\ref{fig:reco_M2_top} the corresponding 
$\tilde M_t$ distributions are shown with the green dot-dashed and the blue dotted line, respectively. We see that even with 
the lack of knowledge of the precise value of the $W$-boson mass, the $M_{2CC}(b)$ variable still provides a good momentum ansatz,
as judged by the location of the peak of its $\tilde M_t$ distribution.

\begin{figure}[t]
\centering
\includegraphics[width=4.9cm]{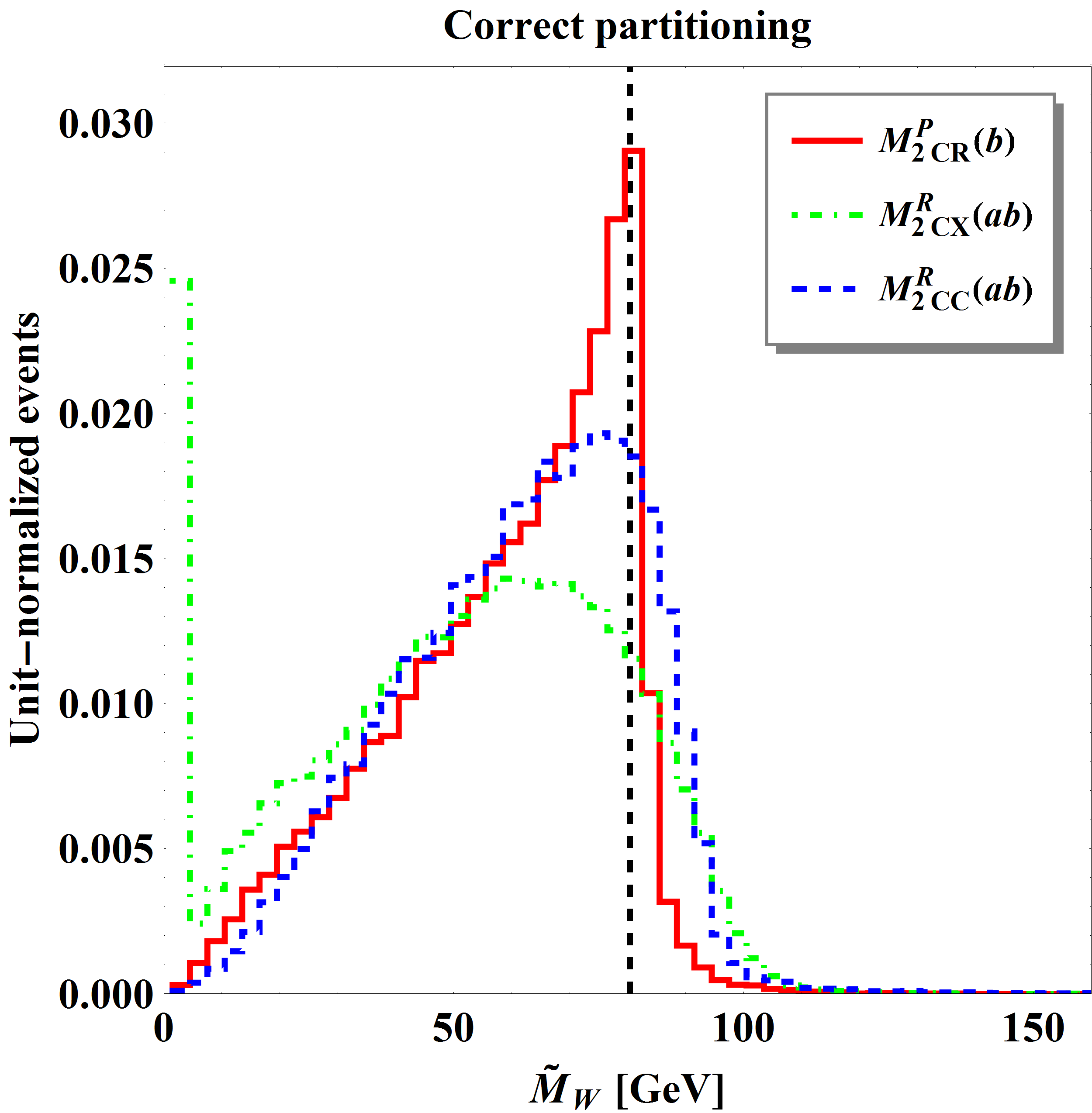}
\includegraphics[width=4.9cm]{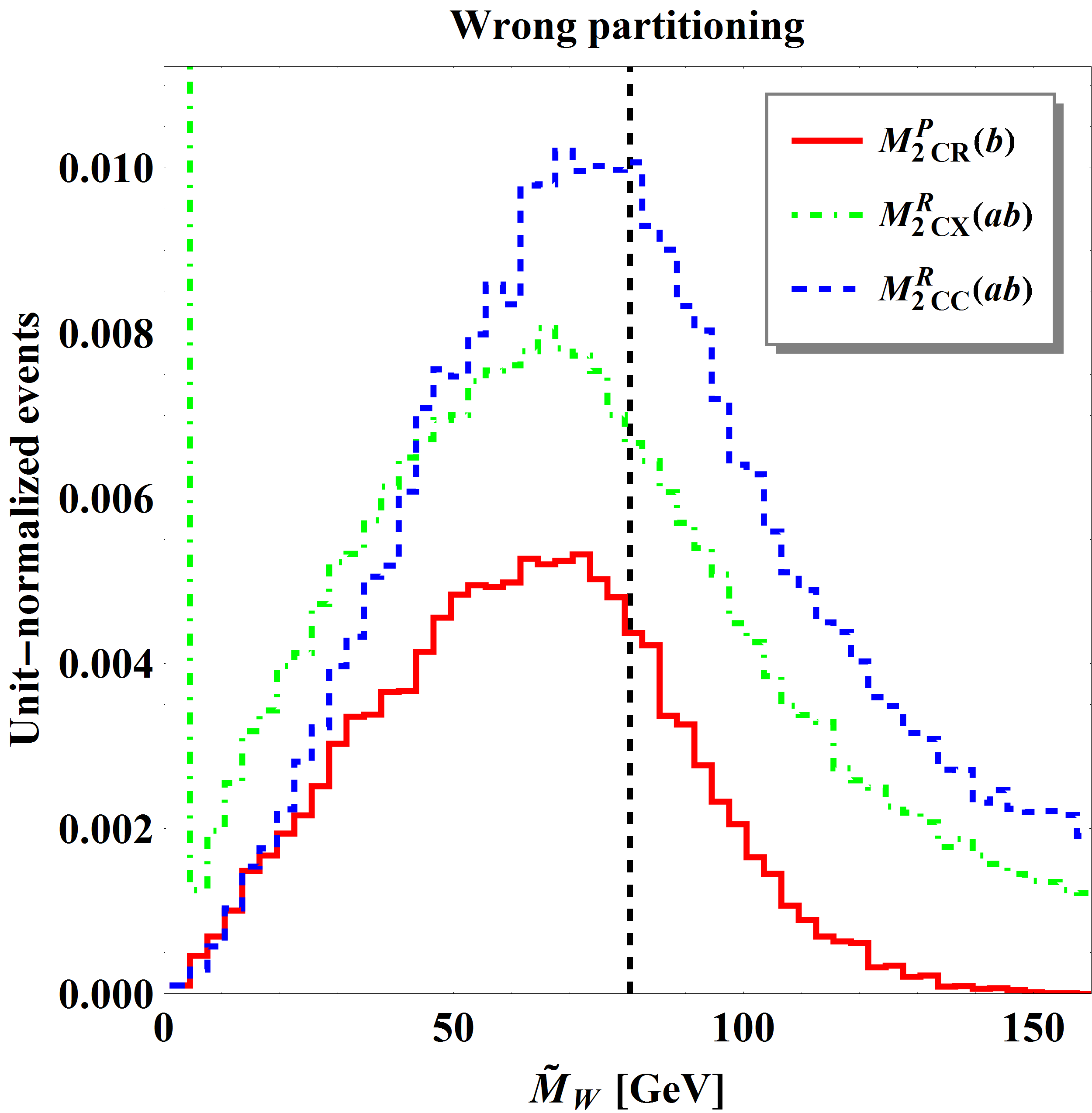}
\includegraphics[width=4.9cm]{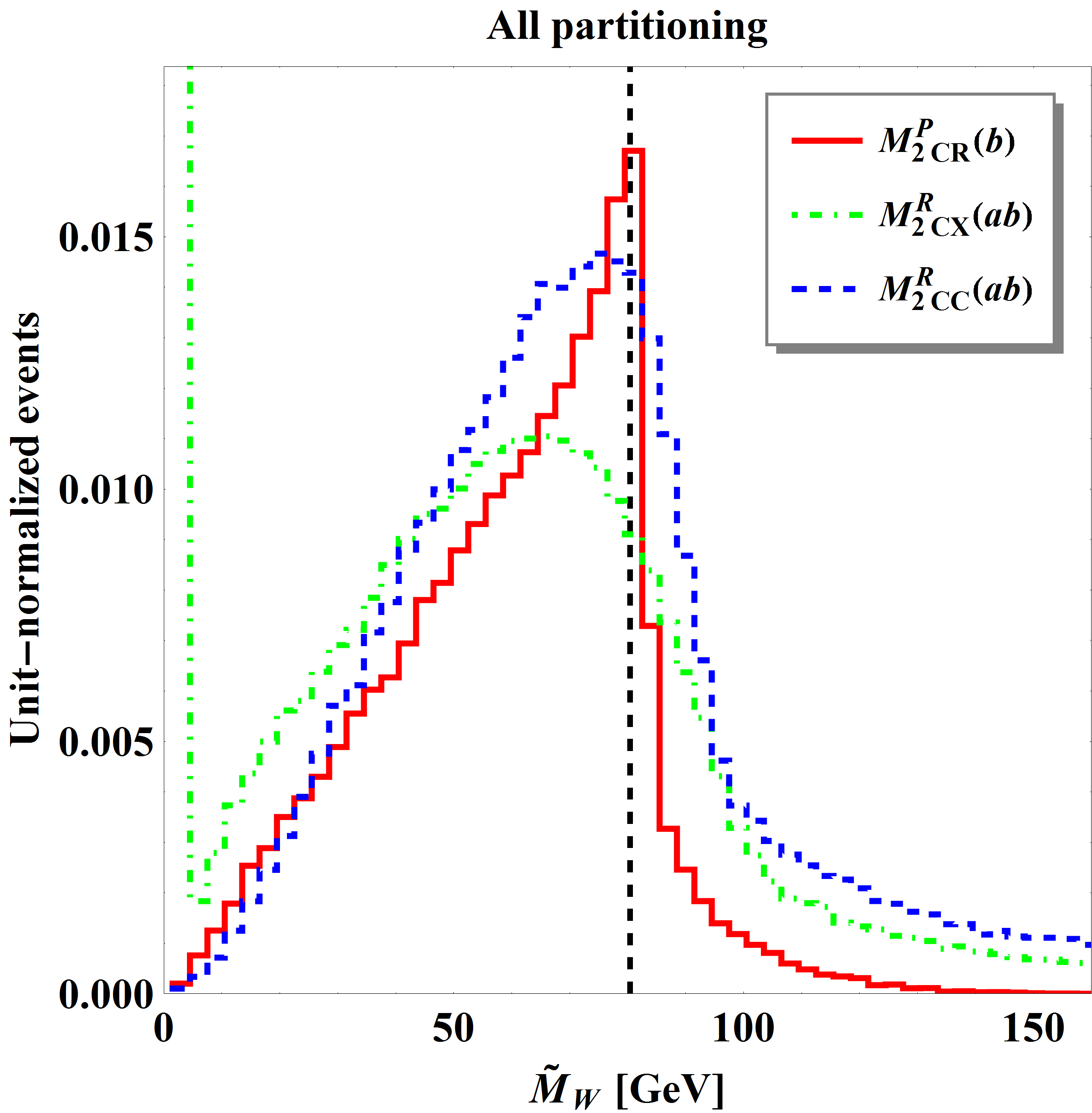}
\caption{\label{fig:reco_M2_W} The same as Fig.~\ref{fig:reco_maos_W}, but using the appropriate $M_2$ variables for 
fixing the invisible momenta. Distributions of $\tilde M_W$ are shown for the case of 
$M_{2CX}(ab)$ (green dot-dashed line), 
$M_{2CC}(ab)$ (blue dotted line), and
$M_{2CR}(b)$ (red solid line).}
\end{figure}
\begin{figure}[t]
\centering
\includegraphics[width=4.9cm]{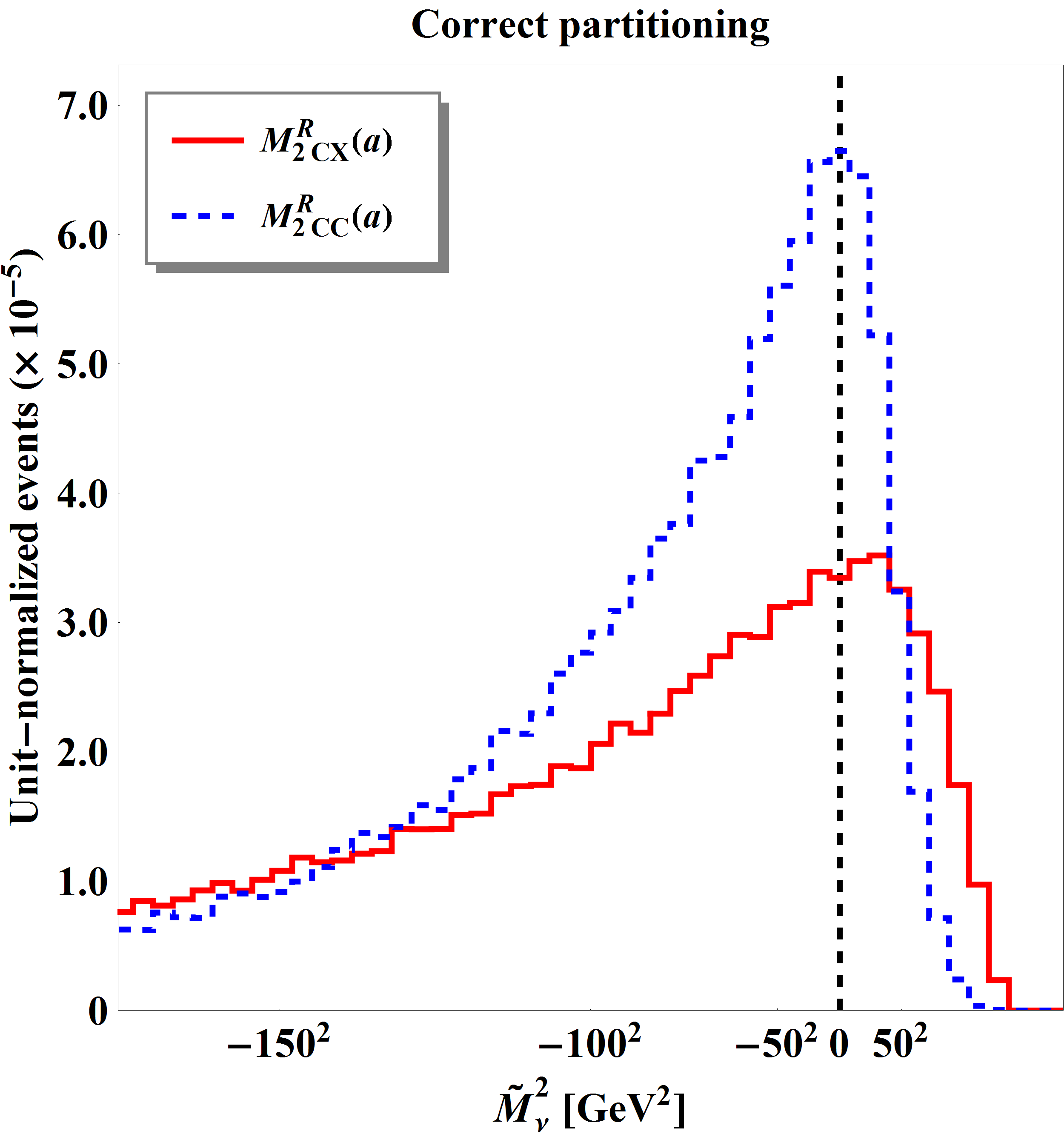}
\includegraphics[width=4.9cm]{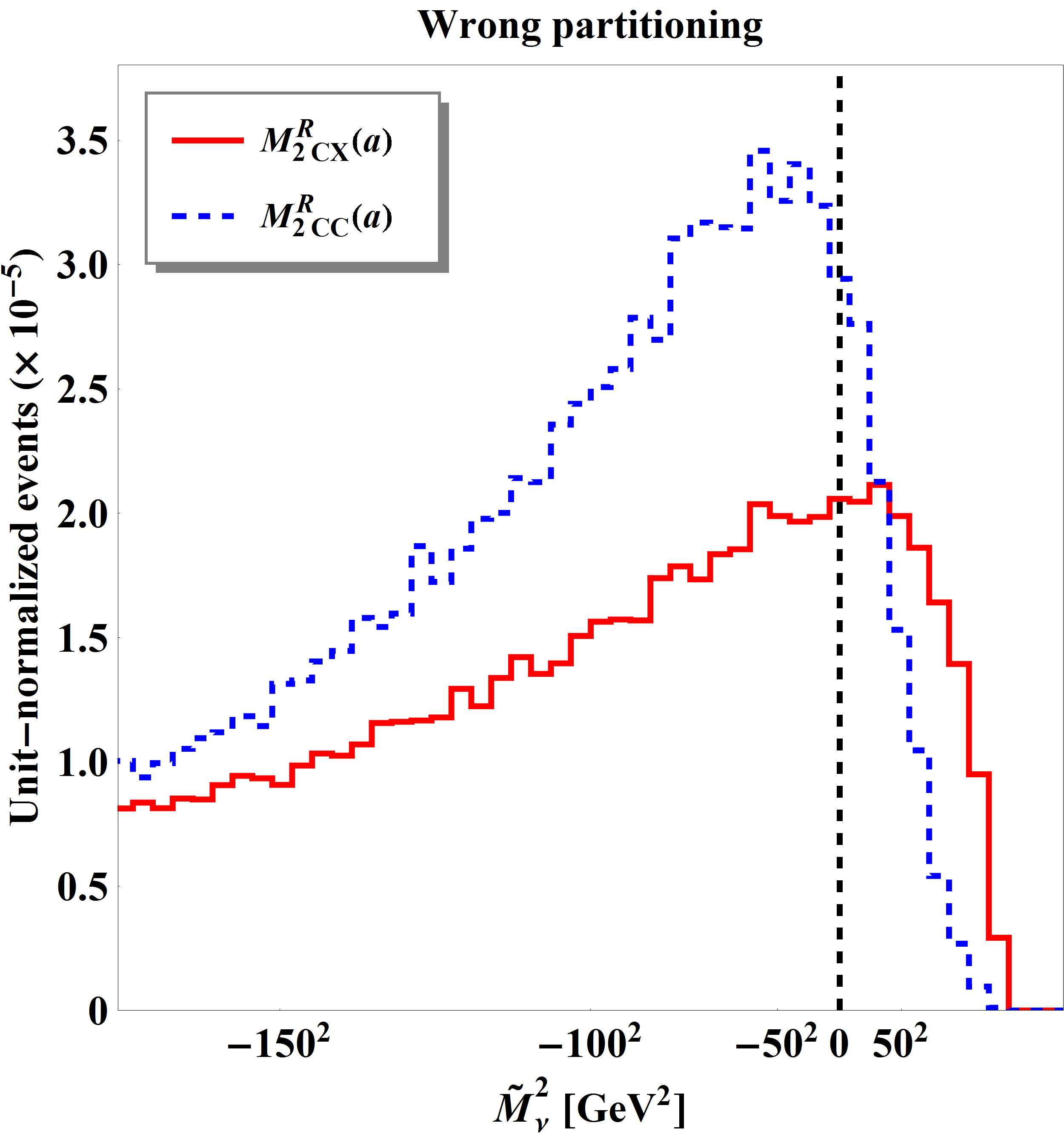}
\includegraphics[width=4.9cm]{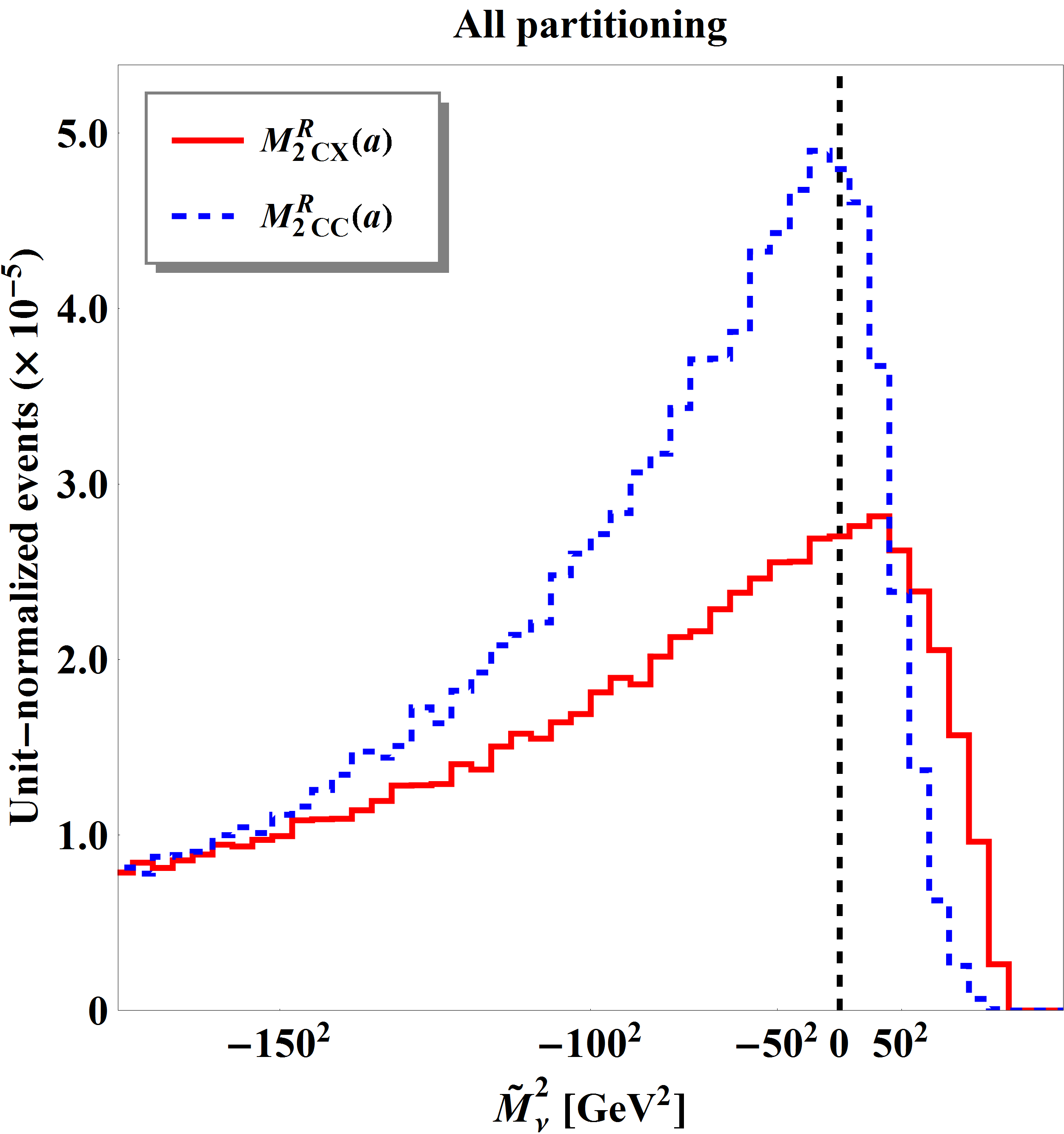}
\caption{\label{fig:reco_M2_nu} The same as Fig.~\ref{fig:reco_maos_nu}, but using the appropriate $M_2$ variables for 
fixing the invisible momenta. Distributions of $\tilde M^2_\nu$ are shown for the case of 
$M_{2CX}(a)$ (red solid line) and
$M_{2CC}(a)$ (blue dashed line).}
\end{figure}

In Figs.~\ref{fig:reco_M2_W} and \ref{fig:reco_M2_nu} we similarly show distributions of the reconstructed
$W$-boson mass $\tilde M_W$ and the reconstructed neutrino mass squared $\tilde M^2_\nu$, respectively. 
(These two figures are the analogues of Figs.~\ref{fig:reco_maos_W} and \ref{fig:reco_maos_nu} for the MAOS case.)
The $\tilde M_W$ distributions in Fig.~\ref{fig:reco_M2_W} use invisible momentum reconstruction from 
$M_{2CX}(ab)$ (green dot-dashed line), $M_{2CC}(ab)$ (blue dotted line), and $M_{2CR}(b)$ (red solid line),
while the $\tilde M^2_\nu$ distributions in Fig.~\ref{fig:reco_M2_nu} use the invisible momenta obtained 
by $M_{2CX}(a)$ (red solid line) and $M_{2CC}(a)$ (blue dashed line). We again observe that the maximally constrained
variable, $M_{2CR}(b)$, which uses as inputs the neutrino and top quark masses, is able to provide us with 
a very good ansatz for the invisible momenta, and the $\tilde M_W$ distribution in Fig.~\ref{fig:reco_M2_W} 
exhibits a very narrow peak at the proper location (80 GeV). The remaining four distributions in Figs.~\ref{fig:reco_M2_W} 
and \ref{fig:reco_M2_nu} are derived from single-input variables, where we again observe that $M_{2CC}$ 
performs slightly better than $M_{2CX}$.

In the above discussion of Figs.~\ref{fig:reco_M2_top} and \ref{fig:reco_M2_W} we have been focusing 
on measuring the top quark mass and the $W$-boson mass from the peaks of the respective $\tilde M_t$ and $\tilde M_W$ distributions.
However, one should keep in mind that whenever we reconstruct a {\em parent} mass, we always have the option 
of measuring it from a kinematic endpoint as well. This is clearly evident in the left panel of Fig.~\ref{fig:reco_M2_top} 
for the case of $M_{2CR}(ab)$ and $M_{2CR}(a)$, and in the left panel of Fig.~\ref{fig:reco_M2_W} for the
case of $M_{2CR}(b)$. Even with the pollution from the wrong combinatorics in the middle panels of 
Figs.~\ref{fig:reco_M2_top} and \ref{fig:reco_M2_W}, the endpoint structures are still preserved in the 
corresponding combined distributions shown in the right panels. 

\subsection{Comparison of $M_{T2}$-assisted and $M_2$-assisted reconstruction schemes}
\label{sec:reco_all}

\begin{figure}[t]
\centering
\includegraphics[width=6.5cm]{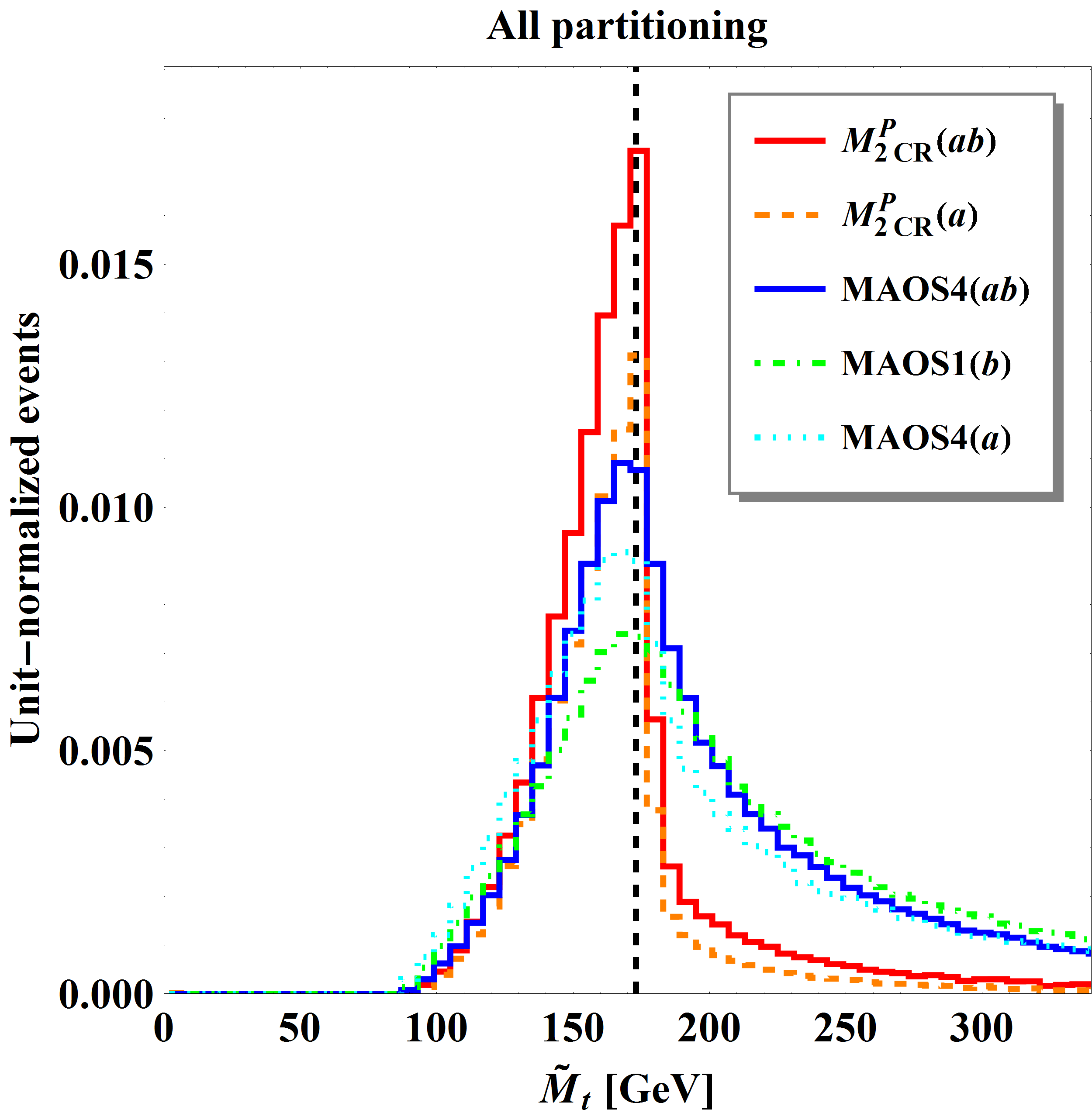}~~
\includegraphics[width=6.5cm]{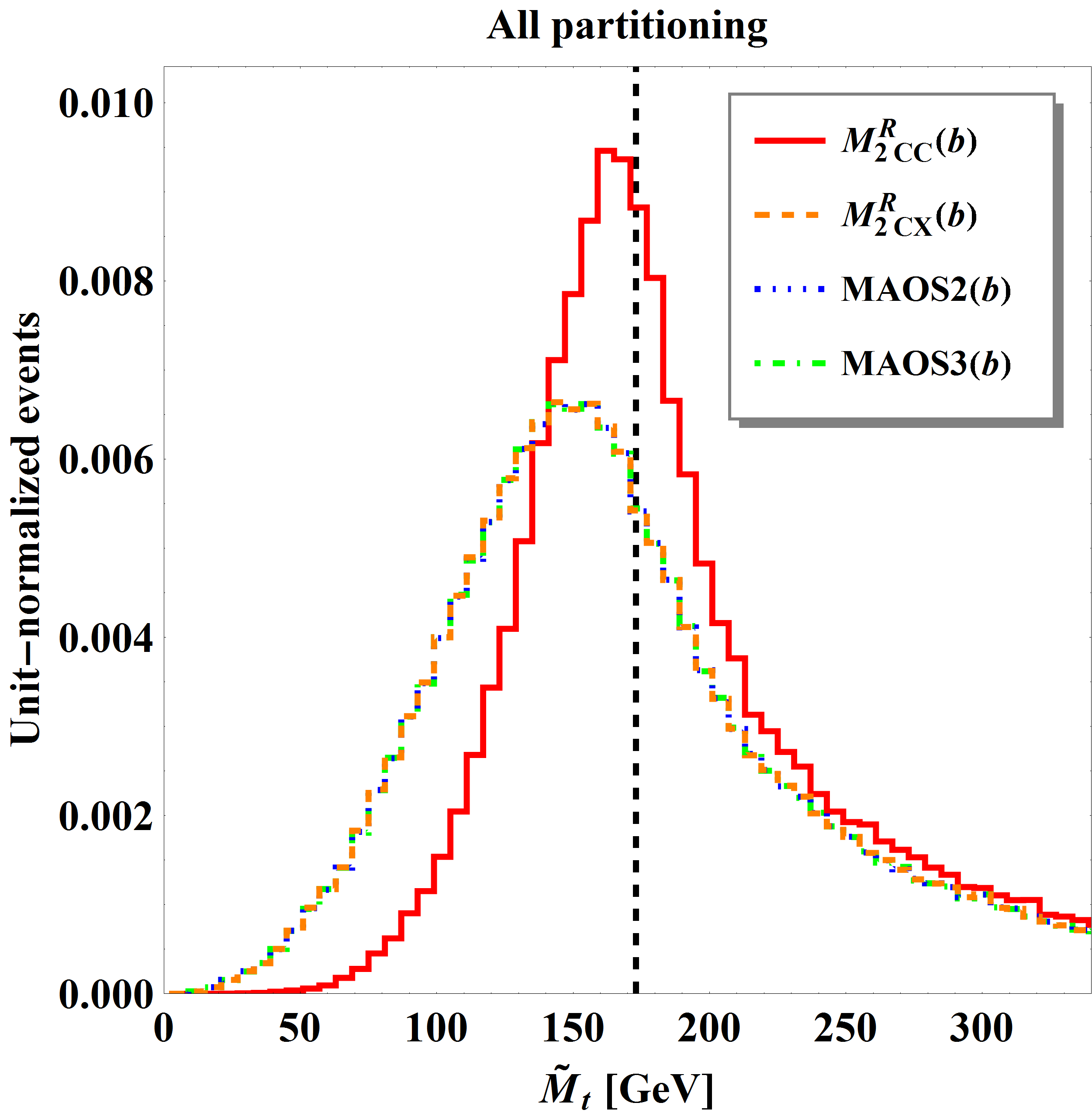}
\caption{\label{fig:reco_top}  
Comparison of the MAOS and $M_2$-assisted methods for top mass reconstruction from Figs.~\ref{fig:reco_maos_top}
and \ref{fig:reco_M2_top}. The left panel shows distributions of the reconstructed top mass $\tilde M_t$ 
with methods which use two mass inputs (the $W$-boson mass and the neutrino mass): the three MAOS methods from Fig.~\ref{fig:reco_maos_top},
MAOS4(ab) (blue solid line), MAOS1(b) (green dot-dashed line), and MAOS4(a) (cyan dotted line),
and the two $M_2$-based methods from Fig.~\ref{fig:reco_M2_top},
$M_{2CR}(ab)$ (red solid line) and $M_{2CR}(a)$ (orange dashed line).  
The right panel shows distributions of the reconstructed top mass $\tilde M_t$ 
with methods which use a single mass input (the neutrino mass):
MAOS2(b) (blue dotted line) and MAOS3(b) (green dot-dashed line) from Fig.~\ref{fig:reco_maos_top} 
and $M_{2CX}(b)$ (orange dashed line) and $M_{2CC}(b)$ (red solid line) from Fig.~\ref{fig:reco_M2_top}. 
}
\end{figure}

\begin{figure}[t]
\centering
\includegraphics[width=6.5cm]{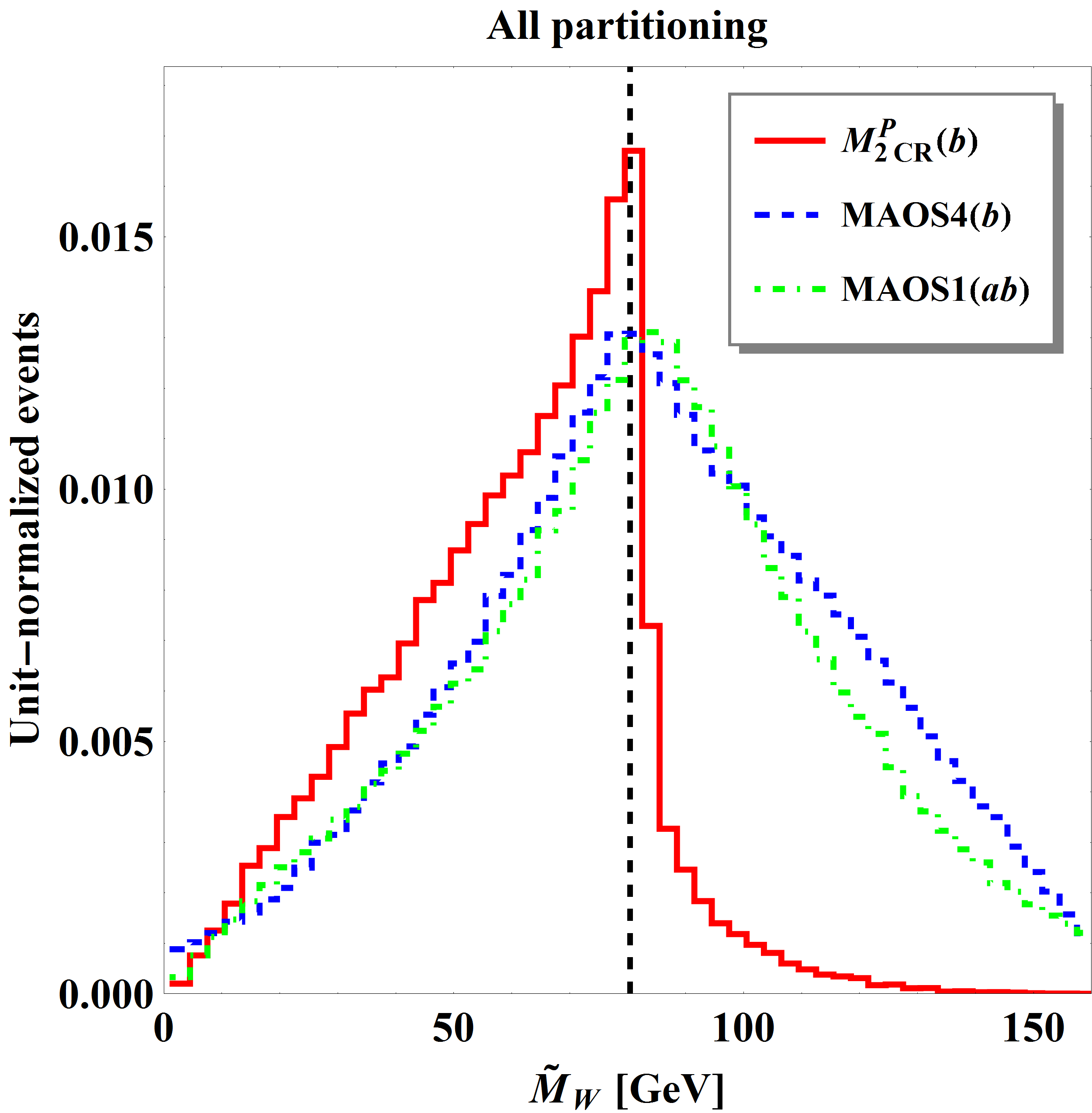}~~
\includegraphics[width=6.5cm]{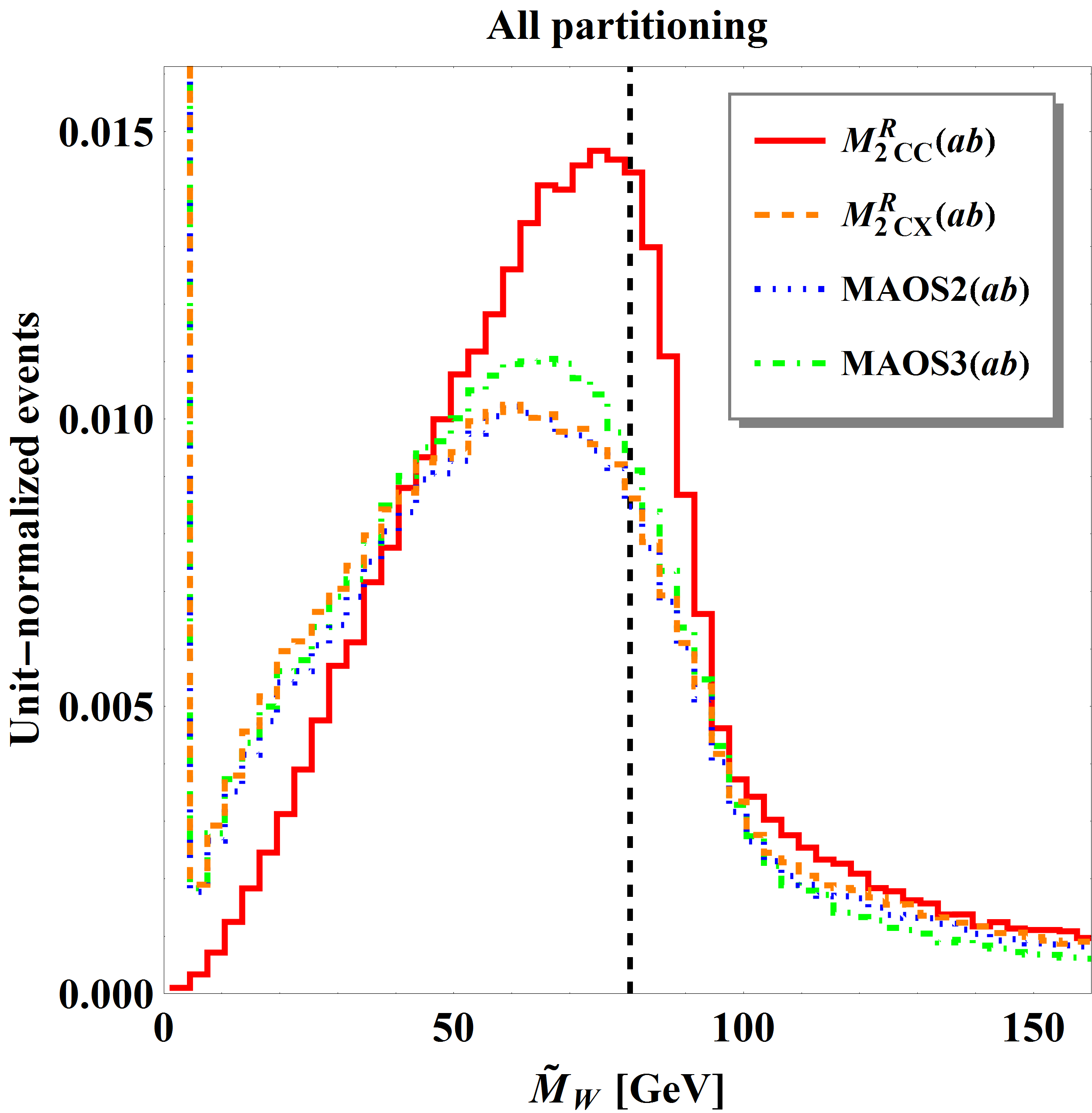}
\caption{\label{fig:reco_W} The same as Fig.~\ref{fig:reco_top}, but for the 
reconstructed mass $\tilde M_W$ of the $W$-boson. The left panel shows distributions of the reconstructed 
$W$-boson mass $\tilde M_W$ with methods which use two mass inputs (the top mass and the neutrino mass):
MAOS1(ab) (green dot-dashed line) and MAOS4(b) (blue dotted line) from Fig.~\ref{fig:reco_maos_W} 
and $M_{2CR}(b)$ (red solid line) from Fig.~\ref{fig:reco_M2_W}.  
The right panel shows distributions of the reconstructed $W$-boson mass $\tilde M_W$ 
with methods which use a single mass input (the neutrino mass):
MAOS2(ab) (blue dotted line) and MAOS3(ab) (green dot-dashed line) from Fig.~\ref{fig:reco_maos_W} 
and $M_{2CX}(ab)$ (orange dashed line) and $M_{2CC}(ab)$ (red solid line) from Fig.~\ref{fig:reco_M2_W}.   }
\end{figure}

\begin{figure}[t]
\centering
\includegraphics[width=6.5cm]{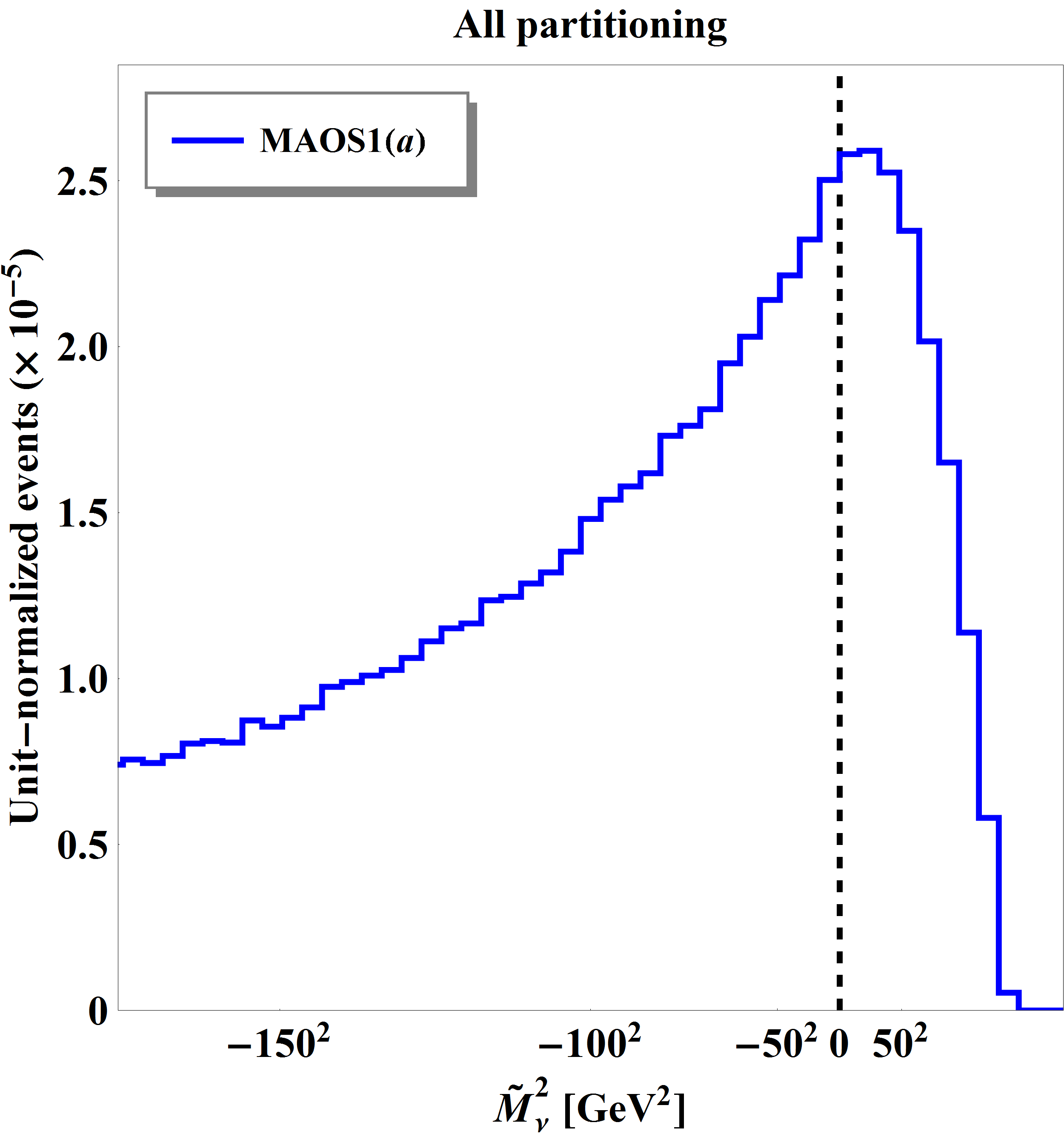}~~
\includegraphics[width=6.5cm]{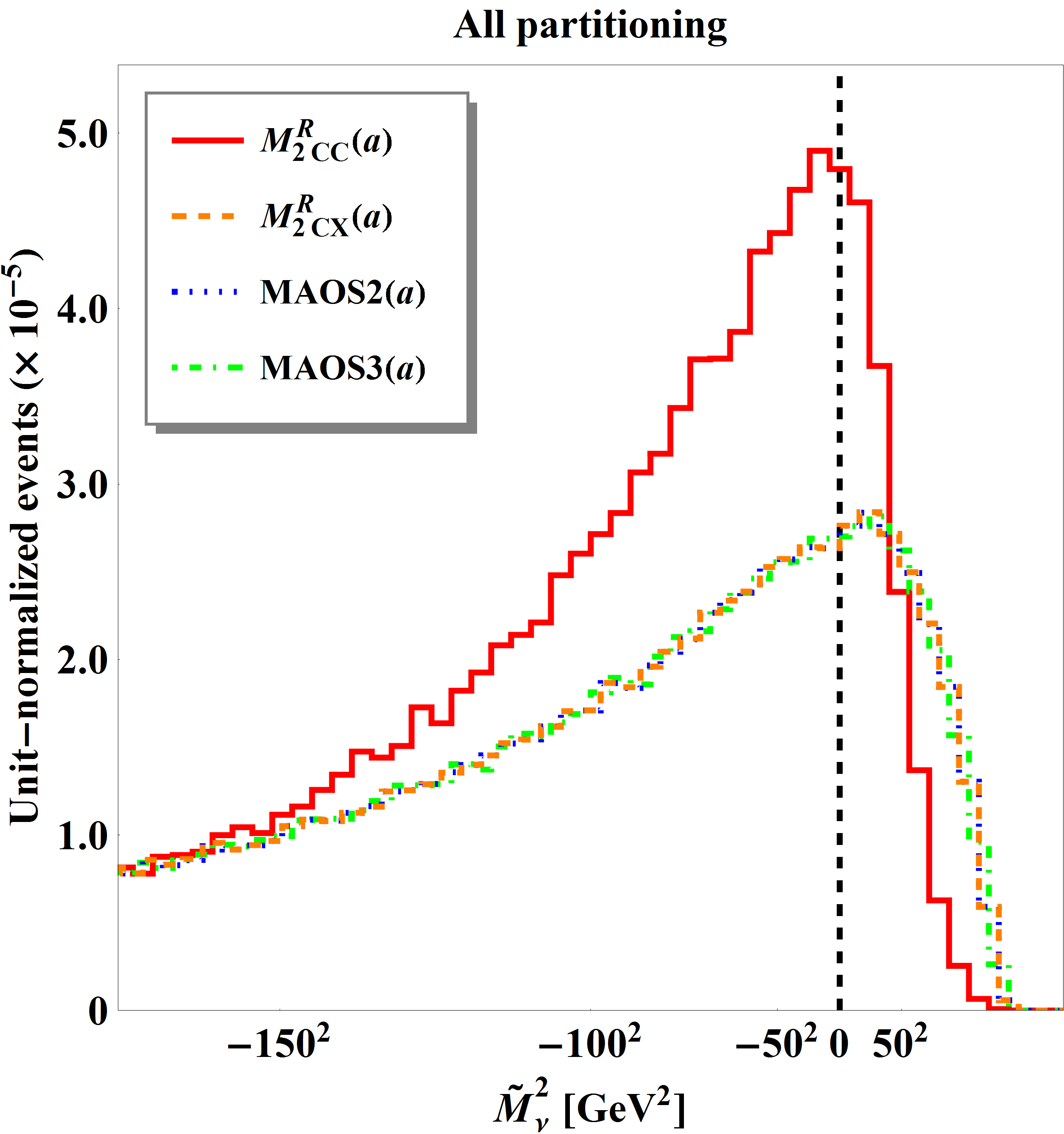}
\caption{\label{fig:reco_nu} 
The same as Fig.~\ref{fig:reco_top}, but for the reconstructed mass-squared $\tilde M^2_{\nu}$ 
of the neutrino. The left panel shows the distribution obtained with MAOS1(a) from Fig.~\ref{fig:reco_maos_nu}, 
which uses two mass inputs: the mass of the top quark and the mass of the $W$-boson.
The right panel shows distributions of the reconstructed neutrino mass squared $\tilde M^2_\nu$ 
with methods which use a single mass input (the $W$-boson mass):
MAOS2(a) (blue dotted line), and MAOS3(a) (green dot-dashed line) from Fig.~\ref{fig:reco_maos_nu} 
and $M_{2CX}(a)$ (orange dashed line) and $M_{2CC}(a)$ (red solid line) from Fig.~\ref{fig:reco_M2_nu}.
 }
\end{figure}

Having discussed the different versions of the more traditional MAOS method in Sec.~\ref{sec:reco_maos}
and the different options for $M_2$-assisted invisible momentum reconstruction in Sec.~\ref{sec:reco_M2},
we are now ready to contrast the two methods to each other. For this purpose, we reassemble the results
from the previous two subsections in Figs.~\ref{fig:reco_top}-\ref{fig:reco_nu}, so that only methods using 
the same number of theoretical mass inputs are compared on each plot: the distributions shown on the left panels 
of these figures require two mass inputs, while the distributions in the right panels need only one. 
Since we already showed the effects of combinatorics in the previous two subsections (compare the left and middle panels of 
Figs.~\ref{fig:reco_maos_top}-\ref{fig:reco_M2_nu}), here for simplicity we plot only the combined distributions, 
which include both the correct and the wrong lepton-jet assignment. Naturally, the use of the extra mass input should
allow for a better measurement, thus one should expect the distributions in the left panels of Figs.~\ref{fig:reco_top}-\ref{fig:reco_nu} 
to be more sharply peaked than those in the corresponding right panels.

Fig.~\ref{fig:reco_top} summarizes our previous results from Figs.~\ref{fig:reco_maos_top} and \ref{fig:reco_M2_top}
for the reconstruction of the top mass $\tilde M_t$. The distributions shown in the left panel require prior knowledge 
of both the $W$-boson mass $m_W$ and the neutrino mass $m_{\nu}$, while for the distributions shown in the right panel
one only needs to know $m_\nu$. The left panel of Fig.~\ref{fig:reco_top} demonstrates that the two $M_2$ methods,
$M_{2CR}(ab)$ (the red solid line) and $M_{2CR}(a)$ (the orange dashed line) clearly outperform their MAOS counterparts,
MAOS4(ab) (blue solid line), MAOS1(b) (green dot-dashed line) and MAOS4(a) (cyan dotted line) --- the peaks reconstructed
by means of $M_2$ are significantly more narrow, which should lead to a more precise mass measurement. Regardless of this width
difference, in all five cases the peak of the distribution is correctly centered on the true top mass used in the simulations
(indicated by the vertical dashed line). The right panel of Fig.~\ref{fig:reco_top} leads to a very similar conclusion for the 
set of methods which rely on a single mass parameter input --- here the method of $M_{2CC}(b)$ (red solid line) is clearly the best, 
while the other three, MAOS2(b) (blue dotted line), MAOS3(b) (green dot-dashed line), and $M_{2CX}(b)$ (orange dashed line), 
are in a perfect tie, which is {\em not} a numerical coincidence, but rather expected theoretically. First, the procedures of 
MAOS2 and MAOS3 differ only for unbalanced events, of which there are none in subsystem $(b)$. 
Furthermore, it is known that the variables $M_{T2}$ and $M_{2CX}$ are identical in any subsystem \cite{Cho:2014naa}, 
see eq.~(\ref{M2XXMT2equality}). Therefore they would lead to the same invisible momentum reconstruction, 
which is indeed confirmed by the right panel in Fig.~\ref{fig:reco_top}.

Fig.~\ref{fig:reco_W} reassembles our previous results from Figs.~\ref{fig:reco_maos_W} and \ref{fig:reco_M2_W}
for the reconstructed $W$-boson mass $\tilde M_W$. The left panel shows the distributions which need two mass inputs,
the top mass $m_t$ and the neutrino mass $m_\nu$. All three distributions peak at the correct value of the $W$ mass
indicated with the vertical dashed line. However, the distribution obtained with $M_{2CR}(b)$ (the red solid line)
is slightly more narrow than the other two, corresponding to MAOS1(ab) (the green dot-dashed line) and MAOS4(b) (the blue dotted line).
The right panel in Fig.~\ref{fig:reco_W}  collects the distributions from Figs.~\ref{fig:reco_maos_W} and \ref{fig:reco_M2_W}
which require only the neutrino mass as an input. Here we notice that MAOS2(ab) (the blue dotted line) and 
MAOS3(ab) (the green dot-dashed line) give slightly different results, due to the presence of unbalanced events in subsystem $(ab)$. 
Once again, the theorem from \cite{Cho:2014naa} ensures that the distributions for MAOS2(ab) (the blue dotted line)
and $M_{2CX}(ab)$ (the orange dashed line) are the same\footnote{The careful reader might notice that in the right panel of 
Fig.~\ref{fig:reco_W}, the blue dotted line for MAOS2(ab) and the orange dashed line for $M_{2CX}(ab)$ are slightly different, 
in apparent violation of the theorem from \cite{Cho:2014naa}. The reason for this is somewhat technical 
and has to do with the different way in which we produce the plots for MAOS2(ab) and $M_{2CX}(ab)$. 
We have verified that for balanced events, the results are identical, as expected. However, for unbalanced events, 
the MAOS2(ab) prescription yields two possible values for the longitudinal momenta, both of which are available to us 
as the solutions to a simple quadratic equation. Then, when we produce plots for MAOS2, we enter
both solutions in the histogram, each with a weight 1/2. These two solutions correspond to the two equally deep global minima 
of the target function used to compute $M_{2CX}(ab)$ \cite{Cho:2014naa}. Since the $M_{2CX}(ab)$ minimization is done numerically 
via {\sc Optimass} \cite{Cho:2015laa}, its numerical algorithm will randomly pick and converge to one of these minima,
giving us only one of the two solutions, which we then plot with weight 1.}. 
Just like we saw in the right panel of Fig.~\ref{fig:reco_top}, the distribution obtained from the $M_{2CC}$-type variable, 
in this case $M_{2CC}(ab)$ (the red solid line), has the best properties:
its peak is relatively narrow and appears closest to the true $W$-boson mass. 

Finally, in Fig.~\ref{fig:reco_nu} we revisit our results from Figs.~\ref{fig:reco_maos_nu} and \ref{fig:reco_M2_nu}
for the reconstructed neutrino mass-squared $\tilde M^2_{\nu}$. This time there is only a single method, MAOS1(a),
which uses two mass inputs, and correspondingly, it is depicted in the left panel. The remaining four methods use a single
input, the $W$-boson mass, and are shown in the right panel. Once again, in accordance with the theorem from
\cite{Cho:2014naa}, the distributions for $M_{2CX}(a)$ (the orange dashed line) and MAOS2(a) (the blue dotted line)
are identical. The lack of unbalanced events in subsystem $(a)$ implies that the distributions corresponding to 
MAOS2(a) and MAOS3(a) are also the same. The remaining fourth distribution, based on $M_{2CC}(a)$ (the red solid line) 
is different, however, and appears to be the most promising for the purposes of a mass measurement of $\tilde M_{\nu}$.

In conclusion of this section, let us summarize our main result. We contrasted the MAOS and $M_2$ methods 
for invisible momentum reconstruction by examining their potential for a mass measurement of an unknown particle through a bump hunt.
We analyzed each of the three subsystems in the event topology of Fig.~\ref{fig:DecaySubsystem} and found that the invisible
momentum reconstruction offered by the $M_2$ class of variables is generally superior to MAOS --- the reconstructed 
invariant mass peaks are more narrow and better localized. An additional theoretical advantage of the $M_2$ approach 
is that it is less ambiguous, as it always provides a unique ansatz for balanced events. In the next section we shall continue
to investigate the $M_2$ approach in the most general case of the event topology from Fig.~\ref{fig:DecaySubsystem},
where $A$, $B$ and $C$ are arbitrary new physics particles.

\section{Applicability to BSM scenarios}
\label{sec:BSM}

Our discussion in the previous section was limited to the SM $t\bar{t}$ dilepton event topology. The dilepton $t\bar{t}$ example is 
appealing to an experimentalist mainly because we know it is present in the data and can be used as a toy playground for new physics 
searches \cite{Chatrchyan:2013boa,CMS:2016kgk,ATLAS:2012poa}. 
Given that the ultimate goal of the LHC is to discover new physics and measure the new particle mass spectrum, in this section we shall
abandon the $t\bar{t}$ example and instead consider the most general case of the event topology of Fig.~\ref{fig:DecaySubsystem}, 
where the mass spectrum $(m_A,m_B,m_C)$ is completely arbitrary, and {\em not} $(m_t,m_W,m_\nu)$, as in the previous section.
A concrete realization in SUSY is provided by the process (\ref{eq:stop}) of stop production, in which the masses of the top squark,
chargino and sneutrino are a priori unknown. 

The main goal of this section will be to revisit the bump hunting mass measurement technique discussed previously and investigate how well 
it does in the general mass parameter space $(m_A,m_B,m_C)$, away from our previous ``study point" $(m_t,m_W,m_\nu)$.
Since the exact nature of the new physics particles $A$, $B$ and $C$ is unknown, in the simulations of this section 
we shall decay particles $A$ and $B$ by pure phase space. For concreteness, we shall continue to assume that particles 
$A$ are colored fermions produced similarly to top quarks. For fairness in comparing the sensitivity at different points in mass parameter space,
it would be nice to fix the overall signal rate. An easy way to do this is to fix the mass (and hence the cross-section) of the heaviest particle $A$. 
In what follows we shall choose $m_A=500$ GeV; this has the additional benefit of reducing the dimensionality of the relevant mass parameter space to two. 
The masses of the two remaining particles will be varied as $m_B\in (0,m_A)$ and $m_C\in (0,m_B)$. We shall then investigate 
the sensitivity of the method as a function of $m_B$ and $m_C$.

\begin{figure}[t]
\centering
\includegraphics[height=4.7cm]{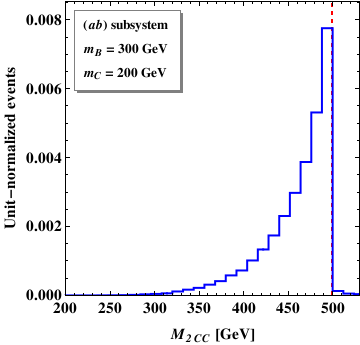}
\includegraphics[height=4.7cm]{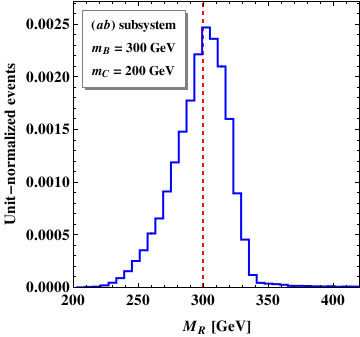}
\includegraphics[height=4.7cm]{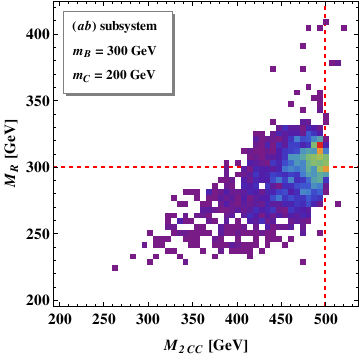}\\
\includegraphics[height=4.7cm]{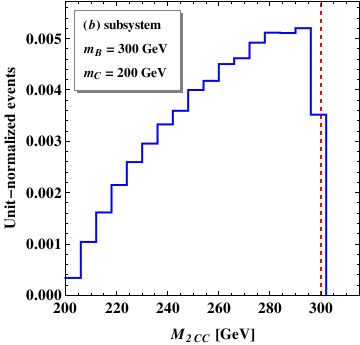}
\includegraphics[height=4.7cm]{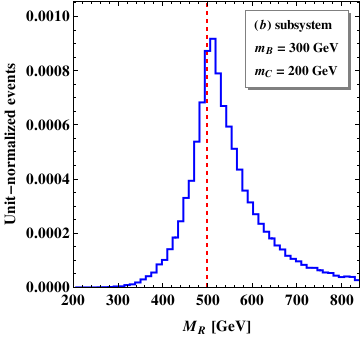}
\includegraphics[height=4.7cm]{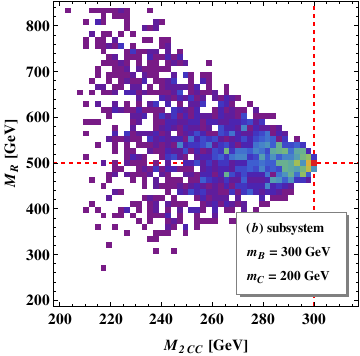}
\caption{\label{fig:works} An example of stop production (\ref{eq:stop}) which works rather well. The mass spectrum is 
$m_A=500$ GeV, $m_B=300$ GeV, $m_C=200$ GeV.  The top row shows the $M_{2CC}(ab)$ distribution (left panel), the corresponding
reconstruction of the relative particle $B$ in the (ab) subsystem (middle panel), and their correlation (right panel).
The bottom row shows the $M_{2CC}(b)$ distribution (left panel), the corresponding
reconstruction of the relative particle $A$ in the (b) subsystem (middle panel), and their correlation (right panel).
The dashed lines mark the true values of the particle masses in each case.
}
\end{figure}

In Fig.~\ref{fig:works} we revisit the main study point considered in \cite{Cho:2014naa}, namely 
$m_A=500$ GeV, $m_B=300$ GeV, $m_C=200$ GeV. In the left panels we show the distributions\footnote{For concreteness,
all plots in this section will be made with the correct lepton-jet pairing, thus avoiding the combinatorial issue.} 
of $M_{2CC}(ab)$ (upper row) and $M_{2CC}(b)$ (lower row), both made with the correct choice of $m_C$.
We observe that both distributions peak very nicely at their kinematic endpoint, allowing 
a measurement of the corresponding mass ($m_A$ for the case of $M_{2CC}(ab)$ and $m_B$ for the case of $M_{2CC}(b)$)
from either the peak of the distribution\footnote{Note the importance of adding the relative constraint (\ref{eq:relatives}). 
Without it, the distribution of $M_{2CX}(b)$ does {\em not} peak at the kinematic endpoint, but at lower values \cite{Cho:2014naa}.}, 
or the location of the kinematic endpoint. For our purposes, however, we are mostly interested in using the invisible momentum 
ansatz for reconstructing the mass of the relative particle, namely $m_B$ for the case of $M_{2CC}(ab)$ and $m_A$ 
for the case of $M_{2CC}(b)$. This reconstruction is shown in the two middle panels of Fig.~\ref{fig:works}.
We see that the reconstructed relative mass distributions have very sharp, well-defined peaks positioned very close to the
true values of the masses, which are indicated by the vertical dashed lines. We conclude that for the particular study point
shown in Fig.~\ref{fig:works}, the invisible momentum reconstruction is quite successful and the bump hunt measurement is very promising.
For future reference, the two right panels in Fig.~\ref{fig:works} then show the correlations between the two variables plotted 
in the left and middle panels of each row.

\begin{figure}[t]
\centering
\includegraphics[height=4.7cm]{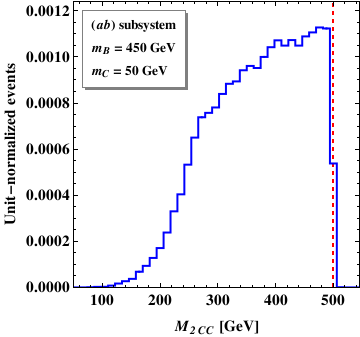}
\includegraphics[height=4.7cm]{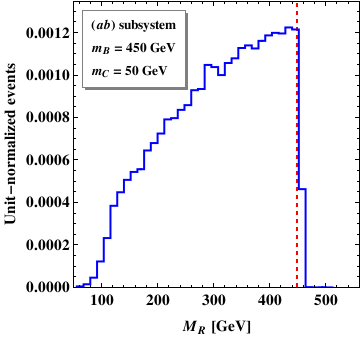}
\includegraphics[height=4.7cm]{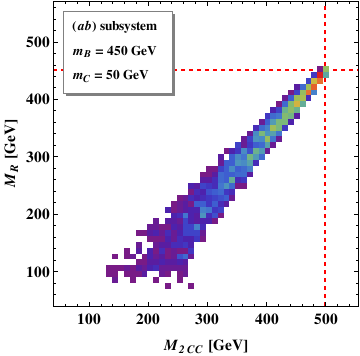}\\
\includegraphics[height=4.7cm]{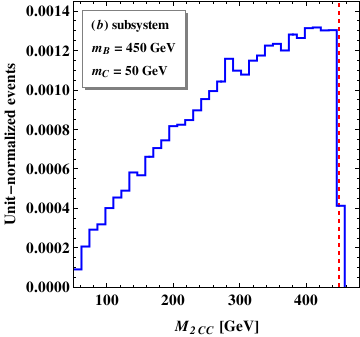}
\includegraphics[height=4.7cm]{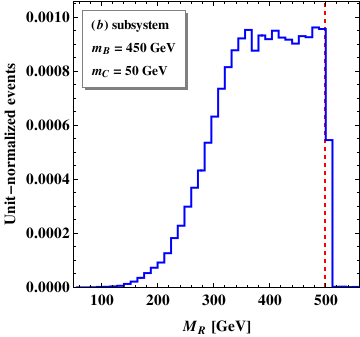}
\includegraphics[height=4.7cm]{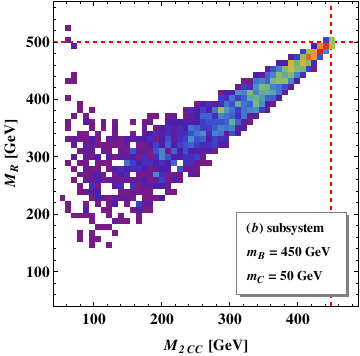}
\caption{\label{fig:doesnotwork1} The same as Fig.~\ref{fig:works}, but for a study point which does not work as well: 
$m_A=500$ GeV, $m_B=450$ GeV, $m_C=50$ GeV.  }
\end{figure}

Fig.~\ref{fig:doesnotwork1} presents the same results, but for a different study point, $m_A=500$ GeV, $m_B=450$ GeV, and $m_C=50$ GeV.
The mass spectrum was judiciously chosen so that the shapes of the relevant kinematic distributions are adversely affected.
For example, as shown in the upper left panel, the peak in the $M_{2CC}(ab)$ distribution is now much broader, extending 
significantly to the left (compare to the upper left panel in Fig.~\ref{fig:works}). Reconstructing the masses of the relative 
particles now appears to be a bit more problematic, as illustrated by the middle panels in Fig.~\ref{fig:doesnotwork1} ---
in the upper row, the peak in the distribution of $m_B$ reconstructed with the invisible momenta from $M_{2CC}(ab)$, 
is very asymmetric. The lower middle panel shows an even worse situation:
the distribution of $m_C$ reconstructed with the invisible momenta from $M_{2CC}(b)$ appears flat from $300$ GeV 
all the way to $500$ GeV, making the corresponding mass determination quite uncertain.

Fortunately, there exists a way to recover sensitivity. The basic idea can be understood from the correlation plots in the right panels
of Fig.~\ref{fig:doesnotwork1}. Notice that the most populated bins are situated very close to the true values of the masses,
$m_B=450$ GeV and $m_A=500$ GeV. The problem arises because of the appearance of the tail extending towards lower values of the 
reconstructed relative mass $M_R$, so that when we project this two-dimensional plot on the $y$-axis, the obtained distribution 
is skewed towards lower values of $M_R$ as well. This basic observation suggests the two possible solutions to the problem.
First, instead of bump hunting on a one-dimensional histogram, one may target directly the most populated bins in the two-dimensional
correlation plots shown in the right panels of Fig.~\ref{fig:doesnotwork1} (note that this method would have also worked 
on our previous example shown in Fig.~\ref{fig:works}). The use of such two-dimensional correlation plots was previously suggested 
in order to detect the kinematic boundaries of the available phase 
space \cite{Costanzo:2009mq,Burns:2009zi,Matchev:2009iw,Debnath:2015wra,Debnath:2016mwb,Debnath:2016gwz},
while here we propose to use them in order to find the location of the highest density.

\begin{figure}[t]
\centering
\includegraphics[width=4.9cm]{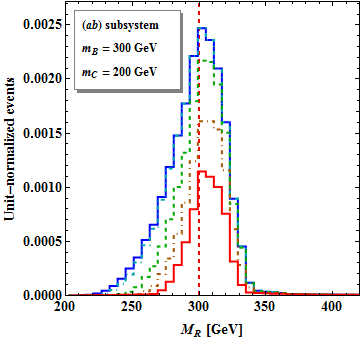}
\includegraphics[width=4.9cm]{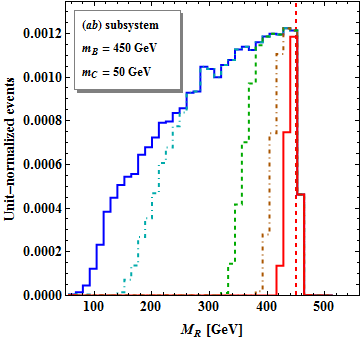}
\includegraphics[width=4.9cm]{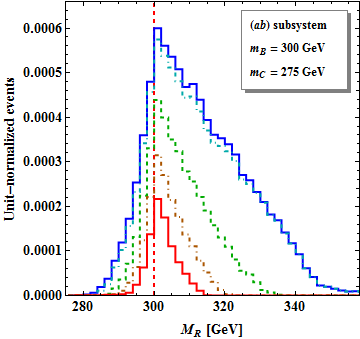}
\\
\includegraphics[width=4.9cm]{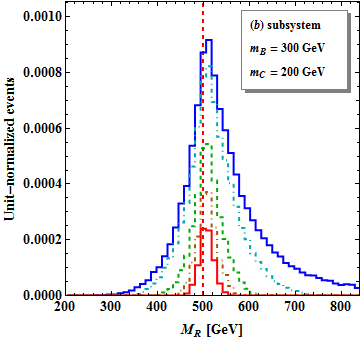}
\includegraphics[width=4.9cm]{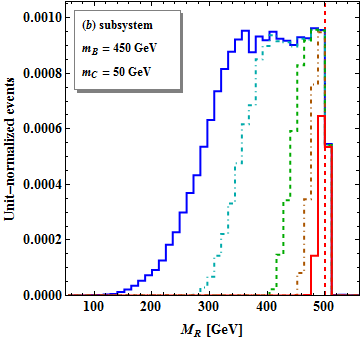}
\includegraphics[width=4.9cm]{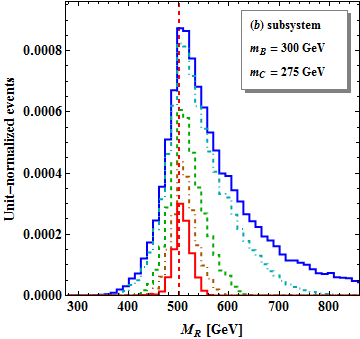}
\caption{\label{fig:withCut} The effect of a preselection cut on the reconstructed relative mass distributions shown in the 
middle panel plots from Figs.~\ref{fig:works}, \ref{fig:doesnotwork1} and \ref{fig:doesnotwork2}. The preselection cut is applied on the variable
$M_{2CC}(ab)$ (upper row) or $M_{2CC}(b)$ (lower row). The left, middle and right plots correspond to the study points from 
Fig.~\ref{fig:works}, Fig.~\ref{fig:doesnotwork1} and Fig.~\ref{fig:doesnotwork2}, respectively. Events are selected in the top 5\% (red), 
top 10\% (orange), top 20\% (green) and top 50\% of the allowed range for the $M_{2CC}$ variable.
}
\end{figure}
An alternative approach is based on the following observation. The right panels in Figs.~\ref{fig:works} and \ref{fig:doesnotwork1}
show that the correct value of the relative mass $M_R$ is obtained for events with extreme values of the $M_{2CC}$ 
kinematic variable plotted on the $x$-axis.
In other words, the ansatz for the invisible momenta tends to work best for events near the $M_{2CC}$ kinematic endpoint 
(this was first pointed out in the context of MAOS reconstruction \cite{Cho:2008tj}, and follows from the general principle that 
kinematic endpoints are attained at very special extreme momentum configurations \cite{Hinchliffe:1996iu,Kersting:2009ne,Matchev:2009fh}).
Thus the precision of the one-dimensional bump hunting method will be recovered, if we simply apply a preselection cut on 
$M_{2CC}$ to eliminate the effect of the tail. Ideally, one would like to select only events which sit right at the $M_{2CC}$ 
kinematic endpoint, but such a severe cut may cause too large of a loss of statistics. This trade-off is illustrated in Fig.~\ref{fig:withCut},
which shows the effect of the preselection cut on the reconstructed relative mass distributions shown in the 
middle panel plots from Fig.~\ref{fig:works} and \ref{fig:doesnotwork1}. The preselection cut is applied on the corresponding
$M_{2CC}$ variable from the $x$-axis of the scatter plots in the right panels of Figs.~\ref{fig:works} and \ref{fig:doesnotwork1},
$M_{2CC}(ab)$ (plots in the upper row) or $M_{2CC}(b)$ (plots in the lower row). The left (middle) plots correspond to the study point from 
Fig.~\ref{fig:works} (Fig.~\ref{fig:doesnotwork1}). The middle plots in Fig.~\ref{fig:withCut} nicely illustrate the benefit from the preselection cut ---
the unwanted events from the tail are removed and the mass bump is rendered more symmetric, and is now centered on the correct  
mass value for the relative particle. However, those benefits do come at a cost - the number of events in the mass bump is correspondingly 
reduced. (Note, however, the upper middle panel of Fig.~\ref{fig:withCut}, where the cut seems to cause no appreciable loss in statistics.)
On the other hand, the left plots in Fig.~\ref{fig:withCut} show that for our first study point from Fig.~\ref{fig:works}, the cut does not 
lead to a big improvement in the shape of the distribution, but this is because the shape was already very good to begin with, and thus
a preselection cut would be unnecessary.

\begin{figure}[t]
\centering
\includegraphics[width=4.9cm]{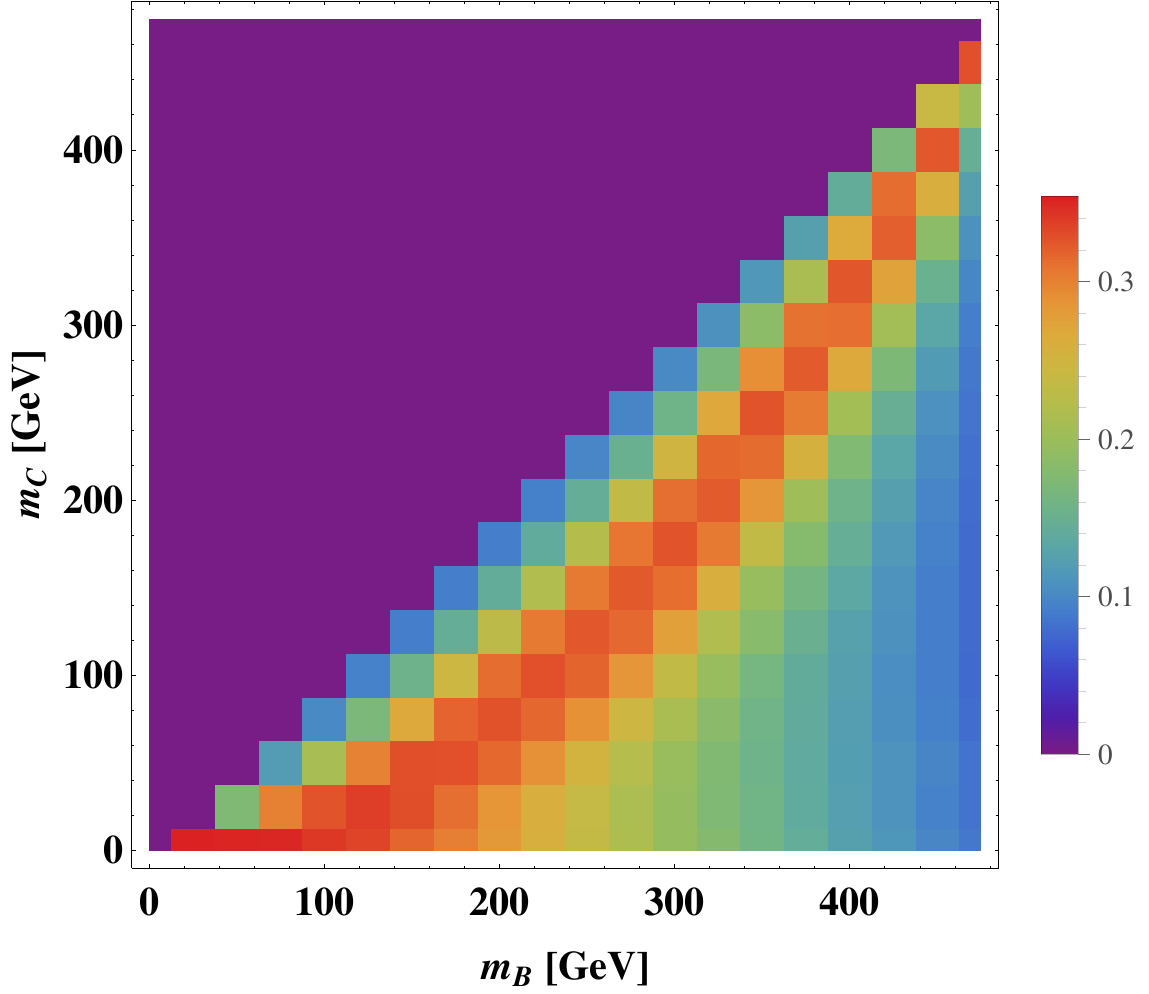}
\includegraphics[width=4.9cm]{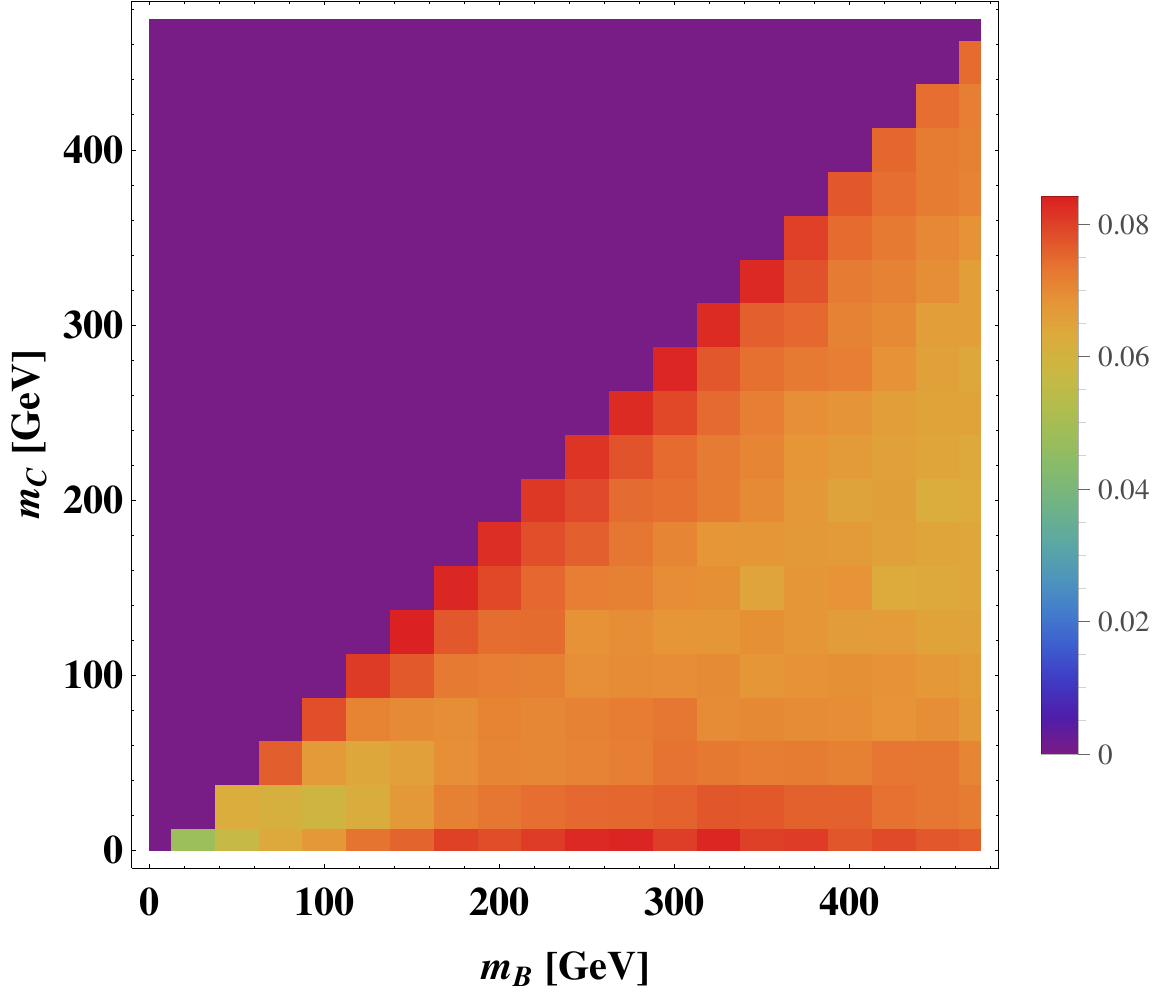}
\includegraphics[width=4.9cm]{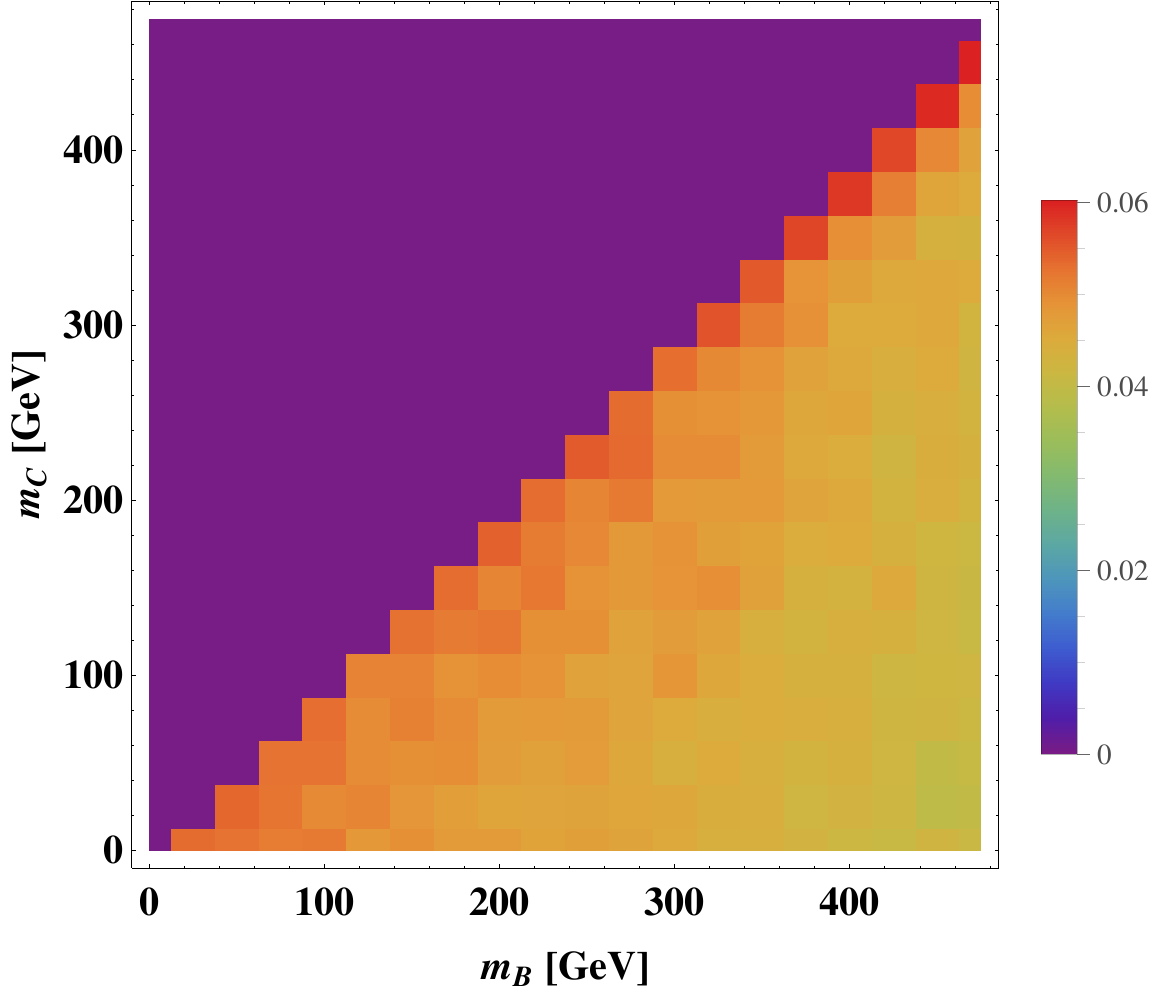}
\caption{\label{fig:endrate} The fraction of events falling within the rightmost 5\% of the allowed range for 
$M_{2CC}(ab)$ (left panel), $M_{2CC}(b)$ (middle panel) and $M_{2CC}(a)$ (right panel). The mass of $A$ is fixed at $m_A=500$ GeV,
while $m_B$ and $m_C$ are varied as $m_B\in (0,m_A)$ and $m_C\in (0,m_B)$. }
\end{figure}
The loss of statistics observed in Fig.~\ref{fig:withCut} as a result of the preselection cut suggests that a crucial issue 
for us to consider is the population of the $M_{2CC}$ bins near the upper kinematic endpoint. 
This is investigated in Fig.~\ref{fig:endrate} as a function of the general mass parameter space $(m_B,m_C)$ for a 
fixed $m_A=500$ GeV. The figure shows results for each of the three subsystems: $(ab)$ in the left panel, 
$(b)$ in the middle panel, and $(a)$ in the right panel. For any given choice of $m_B$ and $m_C$,
we take the allowed range for the corresponding $M_{2CC}$ variable, i.e., the difference between its 
upper and lower kinematic endpoints, and divide it into 20 equal size bins. Then the rainbow scale in
Fig.~\ref{fig:endrate} indicates the fraction of events which fell into the very last bin, i.e. in the upper 5\% of the $M_{2CC}$ 
range, close to the upper kinematic endpoint. 

Fig.~\ref{fig:endrate} reveals that throughout the whole mass parameter space, the rightmost bin is very well populated 
in the case of the $(ab)$ subsystem, and less so in the case of the $(b)$ and $(a)$ subsystems.
Given that invisible momentum reconstruction works best for events near the last bin, this suggests that 
variables based on the $(ab)$ subsystem have a certain advantage in terms of statistics and accuracy. 
Upon closer inspection of the left panel in Fig.~\ref{fig:endrate}, we find that the last bin is maximally populated  
if the mass spectrum satisfies the relation
\beq
m_C = \frac{m_B^2}{m_A},
\label{LucsFormula}
\eeq
whose physical meaning is the following --- in the rest frame of particle $A_i$, particle $C_i$ remains at rest, while the visible
particles $a_i$ and $b_i$ are back-to-back.
The relation (\ref{LucsFormula}) was approximately satisfied for the study point in Fig.~\ref{fig:works}, 
where we had $m_C=200$ GeV and $m_B^2/m_A=180$ GeV.
On the other hand, the study point in Fig.~\ref{fig:doesnotwork1} was characterized by
$m_C=50$ GeV and $m_B^2/m_A=405$ GeV, which significantly violated (\ref{LucsFormula})
by $m_C$ being too low. Note that the relation (\ref{LucsFormula}) is scale invariant, i.e., 
the result does not change if we inflate all masses by the same constant factor. We checked this with 
explicit simulations, and verified that much heavier spectra which satisfy (\ref{LucsFormula}), continue to 
exhibit nice reconstructed peaks and vice versa. 

\begin{figure}[t]
\centering
\includegraphics[height=4.7cm]{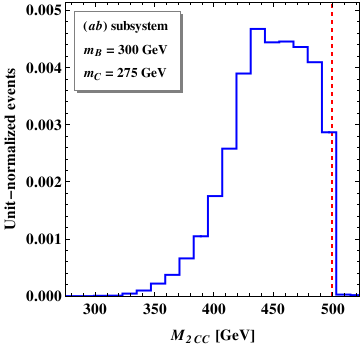}
\includegraphics[height=4.7cm]{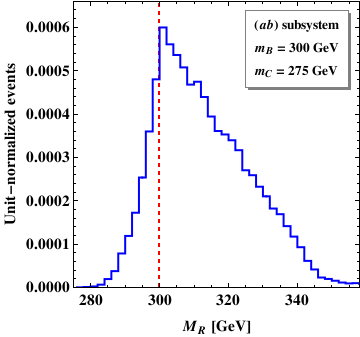}
\includegraphics[height=4.7cm]{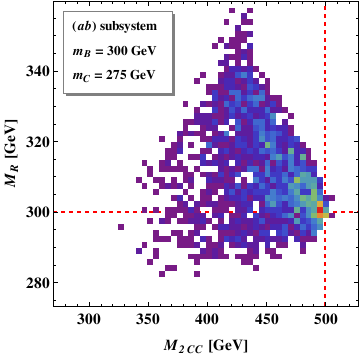}\\
\includegraphics[height=4.7cm]{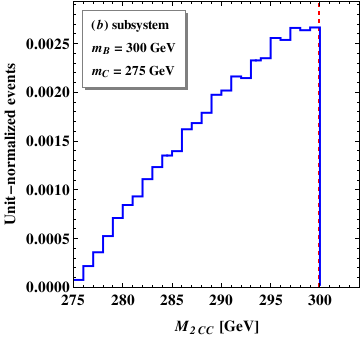}
\includegraphics[height=4.7cm]{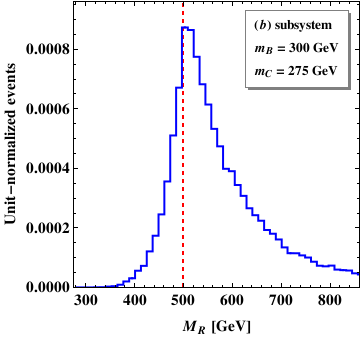}
\includegraphics[height=4.7cm]{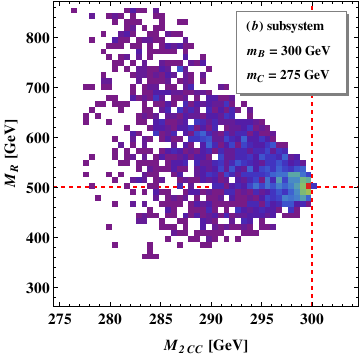}
\caption{\label{fig:doesnotwork2} The same as Fig.~\ref{fig:doesnotwork1}, but for an example with a relatively high value of $m_C$: 
$m_A=500$ GeV, $m_B=300$ GeV, $m_C=275$ GeV.  }
\end{figure}

In conclusion of this section, we shall test the prediction (\ref{LucsFormula}) by 
choosing a point for which $m_C$ is too high. Let us again take $m_A=500$ GeV and $m_B=300$ GeV, 
as in Fig.~\ref{fig:works}, only now increase the value of $m_C$ to $275$ GeV, well above the prediction from (\ref{LucsFormula}).
The result is shown in Fig.~\ref{fig:doesnotwork2}.
The upper left panel shows that, as designed, the events near the $M_{2CC}(ab)$ kinematic endpoint are depleted, and the 
peak of the $M_{2CC}(ab)$ distribution has now moved to lower values, away from the kinematic endpoint.
The right panels again exhibit tails, only this time the tails curl up towards higher values of the reconstructed relative mass,
leading to an overestimate of the mass --- the bulk of the $M_R$ distributions in the middle panels extend {\em above}
the nominal mass of the respective parent particle. However, these problems can be again overcome by the two techniques considered earlier --- 
applying a preselection cut on events near the $M_{2CC}(ab)$ kinematic endpoint (see the right panels in Fig.~\ref{fig:withCut}), 
or directly targeting the most populated bins in the two-dimensional correlation plots in the right panels of Fig.~\ref{fig:doesnotwork2}.

\section{Conclusions and outlook}
\label{sec:conclusions}

Understanding the kinematics of events with missing transverse momentum at hadron colliders like the LHC is 
an important task, since many new physics models have collider signatures with dark matter particles and/or neutrinos,
whose individual momenta and energies are not measured in the detector. The two traditional approaches for analyzing such events 
are 1) where available, use a sufficient number of on-shell constraints to solve for the invisible momenta exactly; and 
2) use variables which do not require any actual knowledge of the individual invisible momenta. However, recently 
several prescriptions for assigning {\em approximate} values to the individual invisible momenta have emerged.
The main goal of this paper was to advertise the existence of {\em a large number of} such ansatze (\ref{ansatz}) 
and demonstrate their usefulness for the purposes of a mass measurement through a bump hunt. Our specific 
points are the following.
\begin{itemize}
\item {\em Many different ansatze are possible.} Quite often, any given prescription for assigning invisible momenta 
has many different variations, as we demonstrated in Sections~\ref{sec:maos} and \ref{sec:M2}, where we defined
12 versions of the MAOS method and 18 versions of the $M_2$ method, respectively, in the case of the dilepton $t\bar{t}$ event topology. 
\item {\em The $M_2$ class of variables automatically provides ansatze for the longitudinal components of the invisible momenta.}
As discussed in Sections~\ref{sec:ansatze} and \ref{sec:m2aos}, the important advantage of the $M_2$ class of 
(3+1)-dimensional invariant mass variables is that they automatically provide values for the longitudinal components 
of the invisible momenta, without any need for additional mass inputs. In that sense, they are on the same theoretical 
footing as the MAOS2 and MAOS3 versions of the MAOS method, but significantly outperform them in the presence of the
relative mass constraint (\ref{eq:relatives}).
\item {\em The $M_2$-based reconstruction of invisible momenta is superior to the MAOS schemes.}
In this paper, we compared the performance of the two methods using the example of a bump hunt mass measurement.
Our results in Sec.~\ref{sec:reco_all} showed that the invisible momenta found by the $M_2$ class of variables
generally lead to a better determination of the new particle masses.
\item {\em Software support.} With the release of the public code {\sc Optimass} \cite{Cho:2015laa} which is capable of computing the 
on-shell constrained $M_2$ variables for general event topologies, the corresponding ansatze for the invisible momenta
are also readily available and can be used for phenomenological studies similar to the one in this paper. For example, one could
imagine spin measurements as in Refs.~\cite{Cho:2008tj,Cheng:2010yy,Guadagnoli:2013xia}, or designing procedures for
reducing the combinatorial background \cite{Rajaraman:2010hy,Baringer:2011nh,Choi:2011ys,us}.
\item {\em Sensitivity study throughout the mass parameter space.} In Section~\ref{sec:BSM} we investigated the 
precision of the invisible momentum reconstruction throughout the full mass parameter space, and identified the regions where 
sensitivity can be lost. We proposed to mitigate the problem by either studying the 2D correlations of the reconstructed 
mass and the $M_2$ variable, or by applying a preselection cut on the $M_2$ variable in order to only select events 
near its kinematic endpoint.  
\end{itemize}

\acknowledgments
We would like to thank Won Sang Cho and Sung Hak Lim for useful discussions. 
This work is supported in part by a US Department of Energy grant DE-SC0010296. 
DK is supported by the Korean Research Foundation (KRF) through the CERN-Korea Fellowship program.

\end{document}